\documentclass[manuscript,screen]{acmart}
\AtBeginDocument{%
  }

\setcopyright{acmlicensed}
\copyrightyear{2018}
\acmYear{2018}
\acmDOI{XXXXXXX.XXXXXXX}
\acmConference[Conference acronym 'XX]{Make sure to enter the correct
  conference title from your rights confirmation emai}{June 03--05,
  2018}{Woodstock, NY}
\acmISBN{978-1-4503-XXXX-X/18/06}

\usepackage{balance}  
\usepackage{amsmath}

\usepackage{gensymb}
\usepackage{makecell}
\usepackage{url}
\usepackage{subfigure}
\usepackage{diagbox}

\usepackage{multirow}
\usepackage{xspace}
\usepackage{mathtools}
\usepackage{enumerate}
\usepackage{float}
\usepackage{soul}

\usepackage{graphicx,epstopdf,url,color}
\usepackage{algpseudocode}
\usepackage{amssymb}
\usepackage[lined,linesnumbered,boxed,commentsnumbered, ruled,vlined]{algorithm2e}
\usepackage{float}
\usepackage{array}
\newfloat{algorithm2e}{t}{lop}
\newcolumntype{C}[1]{>{\centering\let\newline\\\arraybackslash\hspace{0pt}}m{#1}}
\usepackage{balance}
\usepackage{placeins}
\usepackage{amsthm}
\usepackage{threeparttable}
\newtheorem{theorem}{\textbf{Theorem}}
\newtheorem{definition}{\textbf{Definition}} 
\newtheorem{example}{\textbf{Example}}

\newtheorem{lemma}[theorem]{\textbf{Lemma}}

\newcommand{\ie}{\emph{i.e.,}\xspace}
\newcommand{\eg}{\emph{e.g.,}\xspace}
\newcommand{\etal}{\emph{et al.}\xspace}
\newcommand{\aka}{\emph{a.k.a.}\xspace}
\newcommand{\etc}{\emph{etc.}\xspace}

\begin{document}

\title{A Universal Scheme for Dynamic Partitioned Shortest Path Index: Survey, Improvement, and Experiments}

\author{Mengxuan Zhang}
\authornote{Both authors contributed equally to this research.}
\email{mengxuan.zhang@anu.edu.au}
\affiliation{%
  \institution{Australian National University}
  \country{Australia}
}

\author{Xinjie Zhou}
\authornotemark[1]
\email{xzhouby@connect.ust.hk}
\author{Lei Li}
\authornote{Lei Li is the corresponding author.}
\email{thorli@ust.hk}
\affiliation{%
  \institution{DSA Thrust, HKUST(GZ) and HKUST}
  \country{China}
}


\author{Ziyi Liu}
\affiliation{%
  \institution{HKUST}
  \country{China}
}
\email{zyl@ust.hk}

\author{Goce Trajcevski}
\affiliation{%
 \institution{Iowa State University}
 \country{USA}}
 \email{gocet25@iastate.edu}

\author{Yan Huang}
\affiliation{%
  \institution{University of North Texas}
  \country{USA}}
  \email{Yan.Huang@unt.edu}

\author{Xiaofang Zhou}
\affiliation{%
  \institution{HKUST and DSA Thrust, HKUST(GZ)}
  \country{China}}
\email{zxf@cse.ust.hk}

\renewcommand{\shortauthors}{Zhang et al.}

\begin{abstract}
Shortest Path (SP) computation is a fundamental operation in many real-life applications such as navigation on road networks, link analysis on social networks, etc. These networks tend to be massive, and graph partitioning is commonly leveraged to scale up the SP algorithms. However, the \textit{Partitioned Shortest Path (PSP)} index has never been systematically investigated. Moreover, few studies have explored its index maintenance in dynamic networks. In this paper, we survey the dynamic PSP index and propose a \textit{universal scheme} for its design and analysis. Specifically, we first review the SP algorithms and put forward a novel \textit{structure-based} partition method classification to facilitate the selection of partition methods. Furthermore, we summarize the existing \textit{Pre-boundary} PSP strategy and propose two novel strategies (\textit{No-boundary} and \textit{Post-boundary}) to improve its index performance. Lastly, we propose a \textit{universal scheme} with three dimensions (\textit{SP algorithm}, \textit{partition method}, and \textit{PSP strategy}) to facilitate the analysis and design of the PSP index. Based on this scheme, we put forward five new PSP indexes with a prominent query or update efficiency performance. Extensive experiments are conducted to evaluate the performance of the PSP index and the effectiveness of the proposed techniques, with valuable guidance on the PSP index design.
\end{abstract}




\received{20 February 2007}
\received[revised]{12 March 2009}
\received[accepted]{5 June 2009}

\maketitle
\section{Introduction}
\label{sec:intro}
Shortest Path (SP) query in networks is an essential building block for various daily life applications. Given an origin-destination pair such as $(s,t)$, the \textit{point-to-point shortest path query} aims to find the path with the minimum length from vertex $s$ to vertex $t$ on network $G$.
In real-life applications, this shortest path could be the path with minimum traveling time in a road network~\cite{ding2008finding,li2017minimal,li2018go,li2019time,li2020fastest, bast2016route,HTSP_zhou2025High}, the fastest connection in the web graph or Internet~\cite{boccaletti2006complex,scheideler2002models}, or the most intimate relationships in the social network~\cite{gong2016efficient,kolaczyk2009group, opsahl2010node}.
With the development of urban traffic systems and the evolvement of online interaction platforms, real-life networks tend to be massive, which brings great challenges to the scalability of the existing shortest-path algorithms with either heavy memory or expensive search overheads. 
For example, \textit{index-free} algorithms such as \textit{Dijkstra's}~\cite{dijkstra1959note} and \textit{A*}~\cite{Astar_hart1968formal} directly search on the graphs in a best-first manner for query processing, thus are time-consuming on large networks due to large search space. 
To this end, \textit{index-based} algorithms such as \textit{Contraction Hierarchy (CH)}~\cite{CH_geisberger2008contraction} and \textit{2-Hop Labeling (HL)}~\cite{cohen2003reachability} are proposed to expedite the query efficiency. In particular, \textit{2-hop labeling}~\cite{cohen2003reachability} such as \textit{Pruned Landmark Labeling (PLL)}~\cite{PLL_akiba2013fast} and \textit{Hierarchical 2-Hop labeling (H2H)}~\cite{H2H_ouyang2018hierarchy} has the state-of-the-art query efficiency but requires $O(nm^{1/2})$ space to store the index for a graph with $n$ vertices and $m$ edges. Moreover, since the networks are dynamic in nature with evolving edge weights or structures, extra efforts such as space and maintenance overheads are required to handle the updates for shortest path indexes, further deteriorating their scalability.

\begin{figure}[t]
	\centering
	\includegraphics[width=0.8\linewidth]{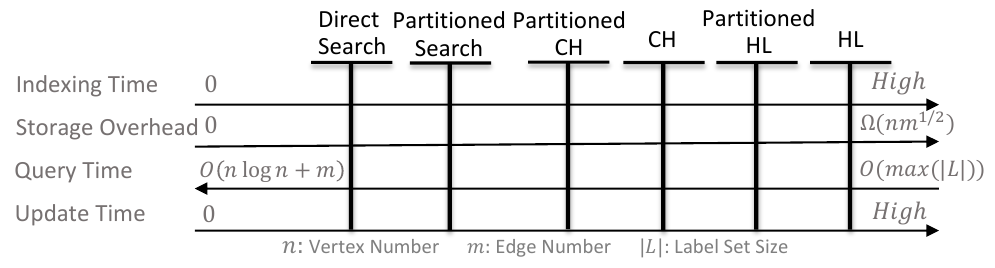}
	\caption{Performance Comparison of SP Algorithms}
	\label{fig:Comparison}
\end{figure}

Graph partitioning is often used to improve the scalability of shortest path algorithms and enable more complicated path problems like time-dependent \cite{li2019time,li2020fastest}, constraint \cite{wang2016effective,liu2021efficient,liu2022FHL}, and \textit{top-$k$} path enumeration \cite{yu2020distributed,yu2024distributed}.
We refer to the shortest path algorithms (indexes) adopting graph partitioning as the \textit{Partitioned Shortest Path (PSP)} algorithms (indexes)\footnote{The term ``index'' also refers to ``index-based algorithm''. We use the terms ``algorithm'' and ``index'' interchangeably when the context is clear.}. 
They generally decompose a large network into several smaller ones such that the query processing could be enhanced for index-free algorithms, or the construction time and index size can be reduced for index-based algorithms. 
As shown in Figure \ref{fig:Comparison}, we classify the existing SP algorithms into six categories and compare them briefly in terms of \textit{index construction time}, \textit{storage space}, \textit{query time}, and \textit{update time}. 
On one extreme are the \textit{direct search} algorithms \cite{dijkstra1959note,Astar_hart1968formal,li2020fast,zhang2020stream} that require no index and can work in dynamic environments directly but are slow for query answering. Then the \textit{partitioned search} adds various information to guide and reduce search space \cite{ParDiSP_chondrogiannis2016pardisp,mohring2007partitioning,bauer2010sharc} to improve query processing. On the other extreme, index-based solutions such as \textit{CH} \cite{CH_geisberger2008contraction} and \textit{HL} \cite{cohen2003reachability,PLL_akiba2013fast,PSL_li2019scaling,H2H_ouyang2018hierarchy} have larger index sizes, longer construction and maintenance time but faster query performance. Their partitioned versions \cite{CT_li2020scaling,li2019time,QbS_wang2021query,farhan2022batchhl,liu2021efficient,liu2023multi,liu2022FHL} are faster to construct with smaller sizes but longer query times.
As these works have better performance than their unpartitioned versions, they may create the following misconceptions regarding the benefits of partitions on pathfinding-related problems: 
1) Graph partitioning and shortest path algorithms are irrelevant since they are two different research branches. This is often phrased as ``\textit{graph partitioning is a well-studied problem and their methods could use any of the existing partition solutions}'' such that only one partition method is used without justification experimentally \cite{wang2016effective,li2019time,liu2021efficient,abraham2011hub,PbS_chondrogiannis2014exploring,yu2020distributed};
2) Applying graph partitioning is always better, which is the reason why these PSP indexes were proposed in the first place;
3) Different PSP indexes are irrelevant and very different from each other, as evidenced by the fact that they only compared with their non-partitioned counterparts \cite{ParDiSP_chondrogiannis2016pardisp,CT_li2020scaling,zheng2022workload,liu2022FHL,LG-Tree_dan2022lg} but not with each other.
However, these misconceptions are not always true.



\textbf{Motivation}.
Although the PSP indexes have been widely applied in the past few decades, they have not been studied systematically. 
Specifically, the existing solutions typically glue one partition method with one SP algorithm without discriminating their characteristics, figure out a way to query correctly, and then claim superiority. For instance, \textit{THop}~\cite{li2019time} utilizes \textit{PUNCH} \cite{PUNCH_delling2011graph} for partitions and \textit{PLL} \cite{PLL_akiba2013fast} for boundary and inner indexes, and processed the queries by concatenating inner and boundary indexes.
While \textit{G-tree}~\cite{Gtree_zhong2013g} uses \textit{METIS}~\cite{METIS_Karypis98MeTis} for hierarchical partitions and builds the \textit{distance matrices} for vertices on different levels, answering the queries by dynamic programming.
These two PSPs seem very different, with different partition methods and path indexes. 
However, they essentially share the same strategy for organizing index construction and query processing procedures. 
Besides, most PSP indexes only compare with the non-partitioned counterparts and claim their superiority without comparing with other PSP indexes. 
Consequently, there lacks a generalized scheme to organize, analyze, and compare the \textit{PSP indexes} insightfully and fairly. 
Moreover, how to extend the PSP indexes to dynamic networks is also unknown.
To this end, we aim to theoretically and systematically study the dynamic PSP index by conducting a comprehensive survey and proposing a universal scheme to guide its design for different scenarios.

\textbf{Challenges.} 
However, achieving this goal presents several significant challenges.

Firstly, graph partition methods are normally classified from different perspectives (like \textit{vertex-cut} / \textit{edge-cut}, \textit{in-memory} / \textit{streaming}, and \etc.) with various optimization objectives (such as \textit{cut size minimization}, \textit{balancing}, and \etc.), but these criteria are not path-oriented and thus hardly help to tell which partition method is suitable for certain path applications. Therefore, we propose a new partitioning classification from the perspective of \textit{partition structure} for easier partition method selection.
We also briefly review the shortest path algorithms used in the PSP index from the perspective of query and update efficiency to guide their selection.

Secondly, it is unclear how to construct, query, and update a new PSP index as all existing works only claim their unique structures can answer queries correctly and seldom involve index maintenance. 
Therefore, we dive into the \textit{PSP strategy}, which determines how to build the index, coordinate multiple partitions for correct query answering, and deal with index updates.
In particular, we summarize classic \textit{pre-boundary} PSP strategy and then propose two novel \textit{PSP strategies} (\textit{no-boundary} and \textit{post-boundary} strategies) and index update methods to improve its index performance.
Besides, a \textit{pruning-based overlay graph optimization} is also proposed to improve the index and update efficiency further.

Thirdly, the current PSP indexes are delicately designed structures with tightly coupled components and all claim certain superiority. However, since no effort has been put into the abstract PSP index itself, it is unclear what determines their performance, when to use particular indexes, and how to design a suitable new PSP index for a new scenario. 
Therefore, based on the rich literature, we propose an insightful and universal PSP index scheme with three critical dimensions, \ie \textit{shortest path algorithm}, \textit{partition method}, and \textit{PSP strategy}.
Based on this scheme, we propose five new PSP indexes to achieve better efficiency in either query or update operations by selecting the specific strategy or method on each dimension as per system requirements. 

\textbf{Contributions.} Our contributions and survey structure are listed as follows:
\begin{list}{$\bullet$}{\leftmargin=1em \itemindent=0em}
\item \textbf{[Literature Review]} 
We propose a novel \textit{structure-based} partitioning classification to facilitate the partition method selection for the PSP index, and provide comprehensive literature reviews of corresponding partitioning methods and PSP indexes. Besides, we also briefly review the shortest path algorithms used in PSP index. (Sections~\ref{sec:SPAlgorithms}-\ref{sec:partitionStructure})
\item \textbf{[Improvements]} 
We summarize the traditional PSP strategy as \textit{Pre-boundary} strategy and propose two novel PSP strategies called \textit{No-Boundary} and \textit{Post-Boundary} strategies to improve its index construction and update efficiency.
Furthermore, we investigate the index maintenance of these PSP strategies and design
a novel \textit{pruning-based overlay graph optimization} to prune the unnecessary computation. (Section~\ref{sec:PSPStrategy})
\item \textbf{[PSP Index Scheme and New PSP Indexes]} 
We propose a universal PSP index scheme by decoupling the PSP index into three dimensions:  \textit{shortest path algorithm}, \textit{partition method}, and \textit{PSP strategy}, such that the \textit{PSP index} can be analyzed and designed systematically.
Based on the scheme, we identify and design five new PSP indexes suitable for specific network structures and scenarios to validate the effectiveness of our scheme. (Section~\ref{sec:universalScheme})
\item \textbf{[Experimental Evaluation]} 
We implement comprehensive experimental studies to explore the performance of various PSP indexes under our proposed universal scheme and provide guidelines for designing PSP indexes with different application scenarios. (Section~\ref{sec:Experiment})

\end{list}

\section{Preliminary}
\label{sec:preliminary}

In this paper, we focus on the dynamic weighted network $G(V,E)$ with $V$ denoting the vertex set and $E$ denoting the edge set. We denote the number of vertices and edges as $n=|V|$ and $m=|E|$. Each edge $e\in E$ is assigned a non-negative weight $w=|e|$, which can \textit{increase} or \textit{decrease} in ad-hoc. 
Note that we focus on edge weight change updates since the structural update (\ie \textit{edge/vertex deletion/insertion}) can be converted to edge weight updates and leverage their update mechanisms for structural updates~\cite{DCH_ouyang2020efficient,DH2H_zhang2021dynamic,DPSL_zhang2021efficient,zhang2021experimental}, as will be discussed in
Section~\ref{subsec:IndexUpdate}.
For each $v\in V$, we represent its neighbors as $N(v)=\{u,|(u,v)\in E\}$.
We associate each vertex $v\in V$ with order $r(v)$ indicating its importance in $G$.
In this paper, we use \textit{boundary-first ordering}~\cite{liu2022FHL} to generate the order for the PSP indexes.
For non-partitioned SP indexes, we leverage \textit{degree} to decide the order of small-world networks while the \textit{minimum degree elimination}-based ordering~\cite{MDE_berry2003minimum} for the road networks.
A path $p=\left<v_0,v_1,\dots,v_k\right> ((v_i,v_{i+1})\in E,0\le i<k)$ is a sequence of adjacent vertices with length of $l(p)=\sum_{i=0}^{k-1}w(v_i,v_{i+1})$, where $w(v_i,v_{i+1})$ is the edge length from $v_i$ to $v_{i+1}$. Given an origin $s$ and destination $t$ vertex (OD pair), the shortest path query $Q(s,t)$ calculates the path with the minimum length between them, \ie $p_G(s,t)$ with the shortest distance $d_G(s,t)$. 
We omit the notation $G$ when the context is clear.
Note that we consider $G$ as an undirected graph, and our techniques can be easily extended to directed graphs.
Graph partitioning generally divides a graph $G$ into multiple disjoint subgraphs $\{G_i\}$ $(i=1,2,\dots,k)$ with\footnote{In this paper, we generally refer to graph partitioning as edge-cut partitioning for easier illustration.} $V_i\cap V_j=\emptyset$ and $\bigcup V_i=V$.
$\forall v\in G_i$, we say $v$ is a \textit{boundary vertex} if there exists a neighbor of $v$ in the another subgraph, that is $\exists u\in N(v), u\in G_j (i\neq j)$. Otherwise, $v$ is an \textit{inner vertex}. We represent the boundary vertex set of $G_i$ as $B_i$ and that of $G$ as $B=\bigcup B_i$. 
The PSP indexes then build an \textit{overlay graph} $\tilde{G}$ on all boundary vertices to connect different partitions and preserve the global distances for all \textit{cross-partition queries} (\ie $s\in G_i,t\notin G_j,i\neq j$).
For $(u,v)\in E$, we say $(u,v)$ is an \textit{inter-edge} if both its two endpoints $u$ and $v$ are boundary vertices from different subgraphs (\ie $u\in B_i, v\in B_j, i\neq j$). Otherwise, it is an \textit{intra-edge}. The corresponding edge sets are denoted as $E_{inter}$ and $E_{intra}$, respectively. 

\begin{figure}
	\centering
	\includegraphics[width=0.8\linewidth]{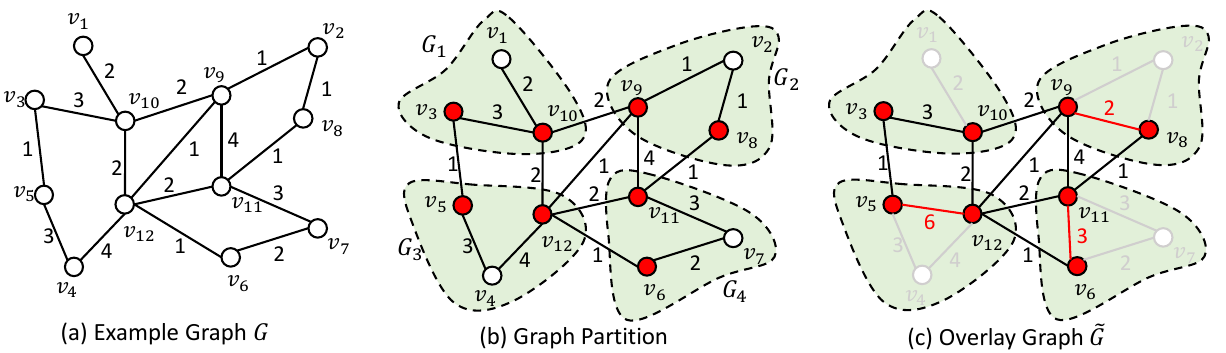}
	\caption{Example Graph $G$ and Overlay Graph $\tilde{G}$}
	\label{fig:exampleGraph}
\end{figure}

\begin{example} \label{example:graphExample}
Figure \ref{fig:exampleGraph} presents an example graph $G$ which is partitioned into four subgraphs $G_1$ to $G_4$, with the red vertices denoting the boundary vertices (\textit{e.g.,} $B_1=\{v_3, v_{10}\}$) and the white vertices representing the inner vertices (\textit{e.g.,} $v_1$ and $v_2$). Edges outside partitions like $e(v_3,v_5)$ are inter-edges, and $e(v_3,v_{10})$ is an intra-edge.
Figure~\ref{fig:exampleGraph}-(c) illustrates an overlay graph $\tilde{G}$ built on boundary vertices to connect all subgraphs while preserving the global distances (\textit{e.g.,} $|e_{\tilde{G}}(v_8,v_9)|=2$).
\end{example}

It is worth noting that \textit{boundary vertex} plays an important role in the shortest path computation with its \textbf{cut property}: given a shortest path $p(s,t)$ with its endpoints in two different subgraphs $s\in G_i, t\notin G_i$, there is at least one vertex $v$ which is in the boundary vertex set of $G_i$ on the path, that is $\exists v\in p(s,t), v\in B_i$. 
Therefore, \textit{boundary vertices} determines the correctness and complexity of partitioned shortest path computation, as will be discussed in Section~\ref{sec:PSPStrategy}.
Various partition methods are involved in the existing PSP methods such as 
\textit{METIS} \cite{METIS_Karypis98MeTis} in 
\textit{CRP} \cite{CRP_delling2017customizable} and \cite{holzer2009engineering, ParDiSP_chondrogiannis2016pardisp, PbS_chondrogiannis2014exploring}; 
\textit{PUNCH} \cite{PUNCH_delling2011graph} in \textit{COLA} \cite{wang2016effective}, \textit{FHL} \cite{liu2021efficient, liu2023multi}, \textit{FHL-Cube} \cite{liu2022FHL} and \textit{THop} \cite{li2019time, li2020fastest};
\textit{core-tree decomposition} \cite{CT_li2020scaling} in \textit{Core-Tree} \cite{CT_li2020scaling, zheng2022workload};
\textit{multi-level partitioning} \cite{METIS_Karypis98MeTis} in \textit{G-tree} \cite{Gtree_zhong2013g,zhong2015g}, \textit{HiTi} \cite{jung1996hiti}, and \textit{CRP} \cite{CRP_delling2017customizable}.
However, their selections are performed in an ad-hoc manner without comparing each other and discerning the SP computation requirements.

\section{Shortest Path Algorithms in PSP Index}
\label{sec:SPAlgorithms}
In this Section, we briefly review the shortest path algorithms used in the PSP indexes and roughly compare their query efficiency and update efficiency to facilitate the selection of shortest path algorithms.
It is worth noting that there are several surveys regarding the shortest path algorithms, such as SP algorithms in static networks~\cite{sommer2014shortest,zhang2022shortest} and SP algorithms in dynamic networks~\cite{zhang2021experimental}. 
The interested reader could refer to \cite{sommer2014shortest,zhang2022shortest,zhang2021experimental} for more details.

\textbf{Direct Search.}
Direct search algorithms such as \textit{Dijkstra}'s \cite{dijkstra1959note}, $A^*$ \cite{Astar_hart1968formal}, or \textit{ALT}~\cite{ALT_goldberg2005computing} search the graph in a best-first manner. They are generally used when the PSP index cannot be constructed due to the large graph size (\textit{QbS} \cite{QbS_wang2021query}) or complicated problem \cite{wang2016effective,MCSPs_zhou2024efficient}. Direct search algorithms take no time in index maintenance on dynamic networks, but their query processing is the slowest due to a large search space. The PSP indexes adopting direct search include \textit{G-tree}~\cite{Gtree_zhong2013g}, \textit{CRP}~\cite{CRP_delling2017customizable}, \textit{PbS}~\cite{PbS_chondrogiannis2014exploring}, \textit{BMHPS}~\cite{MCSPs_zhou2024efficient}, and so on.

\textbf{Contraction Hierarchy (CH)}. CH \cite{CH_geisberger2008contraction} is a widely used lightweight index that contracts the vertices in a pre-defined order and preserves the shortest path information by adding shortcuts among the contracted vertex's neighbors. 
The \textit{search-based CH} \cite{CH_geisberger2008contraction} has a small index size but takes longer to build and maintain \cite{zhang2021experimental,geisberger2012exact} while \textit{concatenation-based CH} \cite{DCH_ouyang2020efficient,wei2020architecture} is much faster to build and maintain and is more suitable for the graph with small \textit{treewidth}~\cite{zhang2021experimental,H2H_ouyang2018hierarchy,CT_li2020scaling,DCT_zhou2024Scalable}.
\textit{CH}'s query performance is generally around 10$\times$ faster than \textit{Dijkstra's} but much slower than the \textit{2-hop labeling}~\cite{cohen2003reachability}. 
The \textit{Dynamic CH (DCH)}~\cite{DCH_ouyang2020efficient} extends the CH index to dynamic networks and achieves almost the fastest index update efficiency among index-based solutions. Surprisingly, no PSP index has used \textit{CH} as the underlying index so far.

\textbf{Pruned Landmark Labeling (PLL)}. Although many \textit{2-hop labeling} methods have been proposed in the past decade, only two are widely used: \textit{PLL}~\cite{PLL_akiba2013fast,PSL_li2019scaling} and \textit{Tree Decomposition (TD)}~\cite{H2H_ouyang2018hierarchy}. Built by either \textit{pruned search} \cite{PLL_akiba2013fast} or \textit{propagation} \cite{PSL_li2019scaling,jin2020parallelizing,DPSL_zhang2021efficient,zhang2023parallel}, \textit{PLL}~\cite{PLL_akiba2013fast} and its parallel indexing version \textit{PSL}~\cite{PSL_li2019scaling} is the only index that can work on graphs with large treewidth, such as large social networks, web graphs, and \etc 
The query efficiency of \textit{PLL} is much higher than \textit{CH} but lower than \textit{TD}. 
The index maintenance of \textit{PLL} can be divided into \textit{search-based}~\cite{IncPLL_akiba2014dynamic,DecPLL_d2019fully,IUPLL_qin2017efficient} or \textit{propagation-based}~\cite{DPSL_zhang2021efficient,DCT_zhou2024Scalable}. The \textit{search-based} solutions involve time-consuming \textit{Dijktra's} searches during the index update. By contrast, \textit{propagation-based} solutions rely on propagation to rectify the incorrect labels, thus achieving much faster efficiency. Nevertheless, the update efficiency of \textit{PLL} is still slower than both \textit{CH} and \textit{TD}.
The PSP indexes adopting \textit{PLL} include \textit{THop}~\cite{li2019time} that uses it to build the overlay and partition indexes, \textit{QbS} \cite{QbS_wang2021query,farhan2022batchhl} that leverages it for building labels among landmarks for billion-scale small-world networks, and \textit{Core-tree} \cite{CT_li2020scaling,zheng2022workload,DCT_zhou2024Scalable} that uses it to build the core index.

\textbf{Tree Decomposition (TD)}.  
Combining the 2-hop labeling with hierarchy, the query processing and index update of \textit{TD} (\aka \textit{Hierarchical 2-Hop labeling (H2H)})~\cite{H2H_ouyang2018hierarchy,chen2021p2h} are much faster than \textit{PLL} for graphs with smaller treewidth~\cite{zhang2021experimental}, such as road networks. However, due to the vertex contraction mechanism, it cannot scale to large-treewidth networks.
The PSP indexes adopting \textit{TD} includes \textit{Core-tree} \cite{CT_li2020scaling,zheng2022workload,DCT_zhou2024Scalable} that leverage it to build the tree index, and \textit{FHL} \cite{liu2021efficient,liu2022FHL} that uses it to build the forest index, and \etc

\textbf{All-pair Table}. The pre-computation of all-pair distances among different vertices enables the fastest query but incurs time-consuming index construction and large index storage, which is infeasible for large graphs. Nevertheless, PSP indexes such as \textit{G-Tree} \cite{Gtree_zhong2013g,zhong2015g} and \textit{ROAD} \cite{lee2009fast,lee2010road} store the distance between boundary vertices in each partition to expedite the query processing.



\section{Structure-based Partition Method Classification}
\label{sec:partitionStructure}
An important step for designing a PSP index is to choose a suitable graph partition method.
However, it is non-trivial to choose a partition method fulfilling our preferred performance because 1) Numerous partition methods with different characteristics were proposed in the past decades; 2) They are classified under different criteria \cite{bulucc2016recent}, for example, from the perspectives of \textit{partition manners} (spectral, flow, graph glowing, contraction and multi-level, \etc), \textit{partition objectives} (balance and minimal cut), \textit{computation manner} (in-memory, distributed and streaming), and \textit{cut type} (edge-cut and vertex-cut), but it is not clear which criterion is beneficial to PSP index; 3) Their relationship with (or the influence on) PSP index is unknown and has never been studied. Based on these doubts, in this Section, we propose a novel \textit{structure-based} classification of partition methods to facilitate the partition method selection. In particular, we categorize existing partition methods into three types: \textit{Planar Partitioning}, \textit{Core-Periphery Partitioning}, and \textit{Hierarchical Partitioning}, and provide comprehensive reviews about corresponding partition methods and PSP indexes in Section~\ref{subsec:planarP}-\ref{subsec:hierarchicalP}. An insightful discussion about these partition structures are also provide in Section~\ref{subsec:partitionDiscussion}.

\subsection{Planar Partitioning}\label{subsec:planarP}
\textit{Planar Partitioning (PP)} treats all partitions equally on one level. We provide its formal definition as follows.

\begin{definition}\label{def:planar}
	\textbf{(Planar Partitioning).} Given a graph $G$, planar partitioning decomposes $G$ into multiple equally-important subgraphs $\{G_i|1\leq i\leq k\} (k\in [2,\infty])$ with
    1)
    $\bigcup_{i\in [1,k]} V(G_i)=V$, $V(G_i)\cap V(G_j)=\emptyset, \forall i,j\in [1,k],i\neq j$
    , or \footnote{We also consider the vertex-cut scenario of graph partition to generalize the definition of planar partitioning.} 2) $\bigcup_{i\in [1,k]} E(G_i)=E$, $E(G_i)\cap E(G_j)=\emptyset, \forall i,j\in [1,k],i\neq j$.
\end{definition}


\subsubsection{Planar Partitioning Algorithms}
\label{subsubsec:planarP_Algorithm}
Sharing almost the same constraint with edge-cut (or vertex-cut) partitioning, the planar partitioning highlights that all partitions are treated equally.
Representative methods include \textit{spectral partitioning} \cite{Pothen90SB,barnard1994fast}, \textit{growing-based partitioning} \cite{Bubble_diekmann2000shape}, \textit{flow-based partitioning} \cite{sanders2011engineering,PUNCH_delling2011graph}, \textit{geometric partitioning}~\cite{BB87RCB,simon1991partitioning}, \textit{node-swapping} \cite{KernighanLin70,FiducciaMattheyses82}, \textit{multilevel graph partitioning}~\cite{METIS_Karypis98MeTis,Scotch18}, etc.
We next briefly introduce these partition methods and corresponding PSP indexes.

\textit{\textbf{Spectral Partitioning.}}
\textit{Spectral Bisection (SB)}~\cite{Pothen90SB} is a powerful but expensive algorithm that can be applied in various applications. Its basic idea is to calculate the eigenvector $z_2$ corresponding to the second smallest eigenvalue (\ie \textit{Fiedler vector}) of the graph \textit{Laplacian matrix} $L$ and utilizes the median value $\overline{m}$ of it to bisect the graph, for the purpose of equal size. It assigns the vertices with an entry larger than $\overline{m}$ to one partition and all other vertices to the other. The computation of $z_2$ is usually done by Lanczos algorithm~\cite{lanczos1950iteration} which is quite time-consuming. Barnard and Simon \cite{barnard1994fast} propose a multi-level \textit{Recursive Spectral Bisection (RSB)} method to improve the efficiency of SB and achieved about an order-of-magnitude improvement without any loss in the quality of the edge-cut. Leete \etal extend RSB method to handle an arbitrary number of partitions and propose a parallel implementation of it to improve efficiency. Hendrickson and Leland~\cite{hendrickson1995improved} extend the bisection problem to $k$-way partitioning by using multiple eigenvalues, which can produce better results than recursive bisection when $k$ is 4 or 8.
Overall, although \textit{spectral partitioning} is a powerful method that is applied in many applications, it is very time-consuming and cannot scale to large-scale graphs.

\textit{\textbf{Growing-based Partitioning.}}
The bisection version of \textit{graph growing-based partitioning} bisects a graph by picking a vertex and growing the region centered at it in a breath-first fashion until half of the vertices are included~\cite{bulucc2016recent}. Similarly, the extension to $k$ partitions conducts the $k$ breadth-first searches from $k$ seed vertices until all vertices are traversed~\cite{goehring1994heuristic}. The partition quality of graph-growing algorithms is sensitive to the initial seed vertices. Therefore, some methods such as using nodes from a two-way partitioning, and finding the ``farthest'' vertex from current seeds are explored to find the initial seed vertices~\cite{goehring1994heuristic}.
Diekmann \etal \cite{Bubble_diekmann2000shape} extend the graph growing algorithm to its iterative version \textit{Bubble}, which tries to adjust the seed vertices to the centers of their regions and restart the growing process in each iteration. Such a procedure continues until the seed vertices are close to the centers for all partitions, thus producing higher partition quality. However, \textit{Bubble} cannot guarantee balanced partitions. Distance measures-based improvements for \textit{Bubble} are proposed by \cite{schamberger2004partitioning,meyerhenke2006accelerating} to improve the partition quality.
Due to the complexity caused by iterations, the graph growing partitioning is time-consuming when the graph is large. 

\textit{\textbf{Flow-based Partitioning.}}
\textit{Flow-based Partitioning} is based on the well-known \textit{max-flow min-cut} theorem~\cite{ford1956maximal}, which tries to find a minimum cut between two parts by computing the maximum flow. It is obvious that such an approach cannot ensure the balance requirement of partitioning. Therefore, flow-based partitioning is often used as a building block for local improvement in many advanced partitioning algorithms. 
For example, as a famous partitioning algorithm for road networks, \textit{PUNCH}~\cite{PUNCH_delling2011graph} leverages flow-based partitioning to improve the partition quality. It utilizes the max-flow computation to find \textit{natural cuts} (\ie a sparse cut separating a local region from the rest of the graph) which are geographically formed by road networks. To be specific, it first randomly picks a vertex $v$ as a \textit{center} from a certain area and finds the \textit{ring} and \textit{core} of $v$ by growing a BFS from $v$. Then, we contract the \textit{ring} and \textit{core} into two vertices $s$ and $t$, and compute the $s$-$t$ cut algorithm such as the push-relabel method~\cite{goldberg1988new} to get the min-cut (\ie \textit{natural cut}) of this area. 
Apart from natural cut detection, other techniques such as contraction and re-balancing are also leveraged to get the final partition result.
\textit{PUNCH} achieves very high partition quality on road networks because of the preservation of the \textit{natural cuts} and the contraction of dense regions. However, it is slower than many \textit{multilevel graph partitioning} algorithms such as \textit{METIS}~\cite{METIS_Karypis98MeTis}. Besides, it is spatial-aware and thus is only applicable on road networks.

\textit{\textbf{Geometric Partitioning.}}
\textit{Geometric partitioning} solely leverages the coordinate information of the graph to generate the partition, targeting at minimizing some related concepts such as the size of subdomain boundary~\cite{schloegel2000graph}.
\textit{Recursive Coordinate Bisection (RCB)}~\cite{BB87RCB,simon1991partitioning} is one of the basic algorithms to solve load-balancing problems in multi-dimensional space by recursively collapsing each dimension. In each recursion, \textit{RCB} divides the computational domain into two partitions by a cutting plane perpendicularly to the coordinate direction of the longest expansion of the domain.
Many theoretical works~\cite{teng1991unified,miller1993automatic,miller1998geometric,gilbert1998geometric} provide a performance guarantee of partitioning on geometrically defined graphs. 
\textit{Space-filling curve}~\cite{pilkington1994partitioning,sagan2012space} is another representative geometric partitioning method that maps a $d$-dimensional space to a $1$-dimensional line to preserve the locality of vertices.
Overall, \textit{geometric partitioning}  is only suitable for road networks and is conducive to applications that are often defined geometrically.

\textit{\textbf{Node-Swapping Heuristic.}}
\textit{Kernighan–Lin (KL)}~\cite{KernighanLin70} is one of the most famous heuristic algorithms that iteratively swaps nodes of different partitions to improve the partition quality. It first defines the \textit{edge-cut gain} of exchanging two vertices from two partitions. Based on the \textit{edge-cut gain} function, it greedily chooses a set of the pairs to maximize the improvement of the partitions in each iteration until no further gain can be achieved. The complexity of \textit{KL} algorithm is $O(n^2\log n)$. \textit{Fiduccia-Mattheyses (FM)}~\cite{FiducciaMattheyses82} algorithm improves the expensive complexity of \textit{KL} algorithm to $O(m)$ by several optimizations. Specifically, instead of swapping pairs of vertices, the FM algorithm moves one vertex with the greatest gain from one partition to another at each time and locks it for no further relocation. Moreover, the elaborately-designed data structures are utilized to improve efficiency. 
In general, the node-swapping heuristic usually serves as a local improvement for \textit{multilevel graph partitioning}. For example, \textit{METIS}~\cite{METIS_Karypis98MeTis} adopts the improved \textit{KL/FM} algorithm by confining the scope of gain re-computation to the boundary vertices to improve efficiency.


\textit{\textbf{Multilevel Graph Partitioning.}}
\textit{Multilevel Graph Partitioning (MGP)} scheme~\cite{barnard1994fast, HL95} is one of the most powerful approaches that can produce high-quality partitions efficiently. It generally includes three phases: \textit{coarsening phase}, \textit{partitioning phase}, and \textit{uncoarsening phase}. The \textit{coarsening} phase sequentially collapses the graph into a coarser one with fewer vertices (normally ends with hundreds of nodes). The second phase partitions the coarser graph by utilizing the algorithms mentioned above while the last phase projects the partition of the coarser graph back to the original graph with refinement during each projection step. Classic multilevel algorithms include \textit{Scotch}~\cite{pellegrini1996scotch}, \textit{METIS}~\cite{METIS_Karypis98MeTis}, \textit{KaPPa}~\cite{kappa2010}, \textit{PaToH}~\cite{atalyrek2011PaToH} and so on.
For example, 
\textit{METIS}~\cite{METIS_Karypis98MeTis} adopts heavy edge matching in the coarsening phase, improved \textit{KL/FM} algorithm in the partitioning phase, and boundary KL refinement in the uncoarsening phase. In particular, the heavy edge matching matches the vertex $v$ for $u$ such that $|e(u,v)|$ is maximum over all the adjacent edges of $u$. The intuition behind heavy edge matching is that incorporating adjacent vertices with heavy edge weight results in lightweight edges being the potential edge cuts to decrease the cut size. In the partitioning phase, \textit{METIS} improves the efficiency of \textit{KL} algorithm by early termination condition, \ie we stop exchanging vertices between partitions when the edge-cut does not decrease for several iterations. Due to the high efficiency and reasonable partition quality, \textit{METIS} is widely used as a building block for many downstream applications. 
Overall, \textit{MGP} can not only generate partitions with high quality but also have significant efficiency, thus they are widely used in many applications.

\textbf{\textit{In-Memory and Streaming Edge Partitioning.}}
Over the past decade, the \textit{vertex-cut partitioning (\aka edge partitioning)} that cuts the vertex and targets minimizing the \textit{replication factor} is drawing much attention as it is more effective than \textit{edge-cut partitioning} in reducing the communication volume of distributed query processing on skewed power-law graphs~\cite{HEP_mayer2021hybrid,bourse2014balanced}.
We divide edge partitioning methods into two categories: \textit{in-memory} and \textit{online streaming} algorithms. 
Intuitively, \textit{in-memory algorithms} need to load the whole graph into the main memory and hence can produce higher partition quality than online streaming methods as global information is available~\cite{zhang2017graph,HEP_mayer2021hybrid}. In contrast, \textit{online streaming algorithms} ingest edge stream and perform on-the-fly partitioning based on partial information of the graph and thus generate relatively low-quality partitions~\cite{HEP_mayer2021hybrid,CLUGP_kong2022clustering}.
One of the representatives of the \textit{in-memory} algorithms is the \textit{Neighbor Expansion (NE)} algorithm proposed by Zhang \etal \cite{zhang2017graph}, which aims to decrease the replication factor by greedily maximizing edge locality. 
\textit{Hybrid Edge Partitioner (HEP)}~\cite{HEP_mayer2021hybrid} improves the efficiency and memory overhead of NE by pruned graph representation and lazy edge removal strategy. In addition, it combines NE with streaming algorithm \textit{HDRF}~\cite{petroni2015hdrf} in a framework to take advantage of them.
\textit{Streaming algorithms} aim to improve partitioning efficiency and reduce the memory overhead.
The simplest streaming algorithm is \textit{Random}~\cite{gonzalez2012powergraph} which randomly assigns edges to a partition based on hashing, thus having limited partition quality. 
Therefore, heuristic-based methods such as \textit{HDRF}~\cite{petroni2015hdrf} and \textit{CLUGP}~\cite{CLUGP_kong2022clustering} are exploited to improve the partition quality.

\subsubsection{PSP Indexes with Planar Partitioning}
\label{subsubsec:planarP_PSP}
Typical PSP indexes adopting planar partitioning include \textit{Arc-Flag}~\cite{ArcFlag_ich2006extremely,ArcFlag_hilger2009fast}, \textit{TNR}~\cite{TNR_bast2007transit}, \textit{PbS}~\cite{PbS_chondrogiannis2014exploring}, \textit{ParDiSP}~\cite{ParDiSP_chondrogiannis2016pardisp}, and \textit{BMHPS}~\cite{MCSPs_zhou2024efficient}, \etc. We next briefly review some of them.

\textbf{\textit{Arc-Flag.}} \textit{Arc-flag}~\cite{ArcFlag_ich2006extremely} proposed by Lauther is a partition-based arc labeling that utilizes a goal-directed heuristic to speed up Dijkstra's algorithm. It divides a road network into several regions $P=\{p_j|j=1,...,k\}$ by rectangular geographic partition and precompute the arc flags whether an edge $e\in E$ is on any shortest path into the region $p_j\in P, j=1,...,k$. During the query phase, if the target vertex $t$ locates in partition $p_t$, then only the edge with the true arc flag of $p_t$ will be searched, thereby narrowing down the search space of Dijkstra's algorithm. The query efficiency can be further improved by leveraging the bidirectional technique. 
\cite{ArcFlag_hilger2009fast} explores the effect of graph partitioning methods including \textit{grid}, \textit{quadtree}, \textit{$kd$ tree}, and \textit{METIS} in \textit{Arc-Flag}. It turns out that \textit{$kd$ tree} and \textit{METIS} are better than the others.

\textit{\textbf{TNR.}} \textit{Transit Node Routing (TNR)}~\cite{TNR_bast2007transit} is a famous framework for processing long-distance shortest path queries on road networks. It is based on the observation that when traveling to a distant location, you will depart from your current location through one of the limited and important traffic junctions. In other words, for every vertex $v$, there are a small set of vertices that $v$ has to pass through one of them to access a distant vertex. We call such a small set of vertices as the \textit{access node} of $v$, denoted as $A(v)$. The union of all access nodes forms the \textit{transit node} $\mathcal{T}$ of graph $G$. The idea of \textit{TNR} is to leverage grid-based geometric partitioning to obtain subnetworks and then precompute distances between $v$ and $u\in A(v)$, and the distances between all transit nodes. Given a query $Q(s,t)$, a locality filter is used to decide whether such a query is a local query or not. If it is a long-distance query then we can answer it by $Q(s,t)=\min_{u\in A(s), w\in A(t)}\{d(s,u)+d(u,w)+d(w,t)\}$. Otherwise, $Q(s,t)$ is answered by another shortest path algorithm. \textit{TNR} demonstrates a good performance on long-distance queries on large road networks. Nonetheless, its preprocessing is very expensive.

\textbf{\textit{PbS and ParDiSP.}}
\textit{PbS}~\cite{PbS_chondrogiannis2014exploring} partitions the road network into several parts by \textit{METIS}~\cite{METIS_Karypis98MeTis} and constructs an overlay graph among boundary vertices for query processing. In particular, it \textit{PbS} builds in-component shortcuts for each vertex $v\in G_i$ (\ie the shortcuts from $v$ to its boundary vertices in $B_i$). Based on the in-component shortcuts, an overlay graph $H$ comprising a complete clique for each partition is constructed. 
To respond to a query about an OD pair $(s, t)$, the modified bidirectional search is performed from $s$ to $t$ if they belong to the same partition. In cases where they do not, $Q(s, t)$ is calculated as $\min_{b_i\in B_s, b_j\in B_t}\{d_{G_s}(s,b_i)+d_{H}(b_i,b_j)+d_{G_t}(t,b_j)\}$, where $B_s$ and $B_t$ represent the boundary vertex set of $G_s$ and $G_t$, respectively.
\textit{ParDiSP}~\cite{ParDiSP_chondrogiannis2016pardisp} improves the query efficiency of \textit{PbS} by modifying the overlay graph and the query processing method. Specifically, after partitioning the graph $G$ by \textit{METIS}, it builds an \textit{extended component} for each partition $G_i$. These components denoted as $G^*_i$, include all the vertices and edges in $G_i$ as well as those located on the shortest path between two boundary vertices of $G_i$. The overlay graph of \textit{ParDiSP} consists of all shortest paths between all boundary vertices of each partition $G_i$, along with the inter-edges. To answer the query $Q(s,t)$, the \textit{ALT} algorithm~\cite{ALT_goldberg2005computing} is employed on the extended component $G^*_i$ if $s,t\in G_i$. Additionally, the boundary vertices of $G_i$ are utilized as landmarks for \textit{ALT}. For cross-partition queries, \textit{ParDiSP} adopts the same query strategy as \textit{PbS}.

In summary, \textit{planar PSP indexes} leverage graph partitioning for shortest path computation in small-treewidth graphs such as road networks, which can achieve good query performance. 
Additionally, 
graph partitioning is the only solution that enables the more complicated shortest path problems like \textit{time-dependent} \cite{li2019time,li2020fastest}, \textit{constraint paths} \cite{liu2021efficient,liu2022FHL,liu2023multi}, and \textit{multi-criteria shortest paths}~\cite{MCSPs_zhou2024efficient} to construct their indexes in large networks.

\begin{figure}[t]
	\centering
	\includegraphics[width=0.8\linewidth]{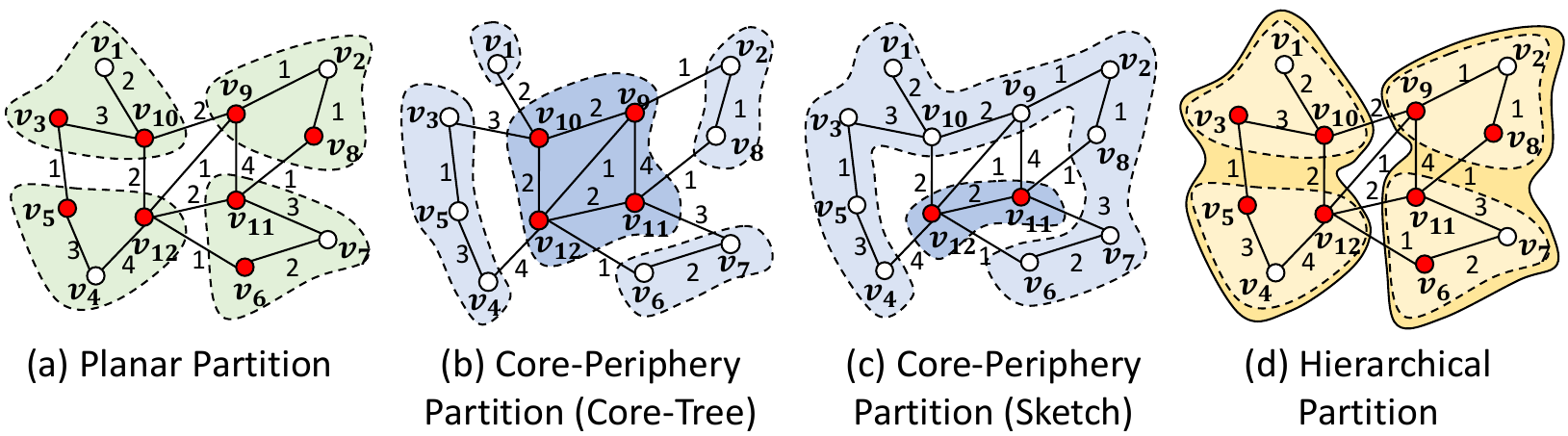}
	\caption{Different Partition Structures}
	\label{fig:partition_example}
\end{figure}

\subsection{Core-Periphery Partitioning}\label{subsec:corePeripheryP}
\textit{Core-periphery Partition (CP)} \cite{borgatti2000models,rombach2014core,elliott2020core,CTD_maehara2014computing} treats the partitions discriminately by taking some important vertices as ``Core'' and the remaining ones as ``Peripheries'' with the formal definition as follows.

\begin{definition}
	\textbf{(Core-Periphery Partitioning).} Given a graph $G$, core-periphery partitioning decomposes $G$ into two distinct parts: a core subgraph $G_c$ that contains the most important vertices, and $k-1$ peripheral subgraphs $\{G_i|1\leq i\leq k-1\} (k\in [2,\infty])$, satisfying $\bigcup_{i\in [1,k-1]} V(G_i) \cup V(G_c)=V, V(G_c)\cap V(G_i)=\emptyset (\forall 1\le i\le k-1)$ and $V(G_i)\cap V(G_j)=\emptyset (\forall i,j\in[1,k-1],i\neq j)$.
\end{definition}	

There are two big streams depending on how the core is formed: \textit{Core-Tree Decomposition} \cite{CTD_maehara2014computing,maehara2014computing,CT_li2020scaling,zheng2022workload,DCT_zhou2024Scalable} and \textit{Sketch} \cite{potamias2009fast,das2010sketch,gubichev2010fast,tretyakov2011fast,qiao2012approximate,qi2013toward,farhan2018highly,QbS_wang2021query,farhan2022batchhl}. It is worth noting that these two core-periphery partitioning is tightly integrated with corresponding PSP indexes. As such, it is hard for the core-periphery PSP index to choose other partition methods (\eg \textit{planar partitioning}) for graph partitioning.

\textbf{\textit{Core-Tree Decomposition}}~\cite{CTD_maehara2014computing} generates the core through tree decomposition \cite{robertson1984graph}, and the resulting periphery part is a set of small-width trees. Specifically, it leverages \textit{minimum degree elimination} \cite{MDETD_xu2005tree,MDE_berry2003minimum} to contract vertices of lower degree first such that the graph is contracted from the periphery towards the center, generating a set of growing trees (``Peripheries'') around a shrinking while denser graph (``Core'') formed by those non-contracted vertices. The contraction terminates when the width of one tree reaches the previously-set bandwidth threshold. 

A classic PSP index adopting core-tree decomposition is \textit{Core-tree (CT)} index~\cite{CT_li2020scaling,DCT_zhou2024Scalable}, which aims to scale up the distance labeling on graphs with core-periphery properties such as small-world networks. It leverage core-tree decomposition to divide the graph $G$ into a ``core'' (\ie a large partition of high-degree vertices) and a set of ``trees'' (\ie a number of small partitions with bounded treewidth).
To achieve efficient shortest path computation, a 2-hop labeling index (\ie \textit{PLL}~\cite{PLL_akiba2013fast}) is built on the core, while a tree decomposition-based index (\ie \textit{H2H}~\cite{H2H_ouyang2018hierarchy}) is built on the ``trees''. There are several reasons behind such a differentiation strategy. Firstly, lower-degree vertices tend to have larger label sizes compared to higher-degree ones since the high-degree vertices are more likely to be passed through by the shortest path of a random source-target query. Therefore, by removing them from the core, the index size of 2-hop labeling can be significantly reduced. Secondly, \textit{H2H} is unsuitable for large-treewidth graphs (the core) as the tree decomposition method cannot handle these graphs. However, it is very efficient on small-treewidth networks, which makes it an ideal SP index for the peripheries. \textit{DCT}~\cite{DCT_zhou2024Scalable} extends \textit{CT} to dynamic networks by efficient index maintenance algorithms.

\textbf{\textit{Sketch}} \cite{potamias2009fast,das2010sketch,gubichev2010fast,tretyakov2011fast,qiao2012approximate,qi2013toward,farhan2018highly,QbS_wang2021query,farhan2022batchhl} selects a set of vertices as \textit{landmarks}, which can be regarded as the core and treats the remaining parts as a periphery. A classic PSP index adopting Sketch is \textit{QbS}~\cite{QbS_wang2021query} which picks a set of landmarks as \textit{meta-graph} (\ie the \textit{core}) while the other part as \textit{sparsified graph}. The 2-hop labeling is computed for the meta-graph and a \textit{sketch} is obtained based on the meta-graph to summarize the important structure of shortest paths for the query pair. The shortest path computation on the sparsified graph is based on a bidirectional search and guided by the \textit{sketch}. 
Similarly, \textit{BatchHL}~\cite{farhan2018highly,farhan2022batchhl} also selects a set of landmarks and then builds a highway index on these landmarks. Besides, a highway cover labeling is built to connect the periphery vertices to at least one landmark to prune the search space during query processing. \cite{farhan2022batchhl} investigates the batch updates for \textit{BatchHL} in unweighted networks.

In summary, the identification of a small yet crucial set of vertices as the ``core'' of the network, with all other vertices being the ``periphery'', allows \textit{core-periphery PSP indexes} to achieve exceptional scalability in computing shortest paths on small-world networks. \textit{Core-Tree}~\cite{CT_li2020scaling} index improves the scalability of state-of-the-art non-partitioned \textit{PSL}~\cite{PSL_li2019scaling} with a negligible performance drop in the query time, while \textit{Sketch-based methods}~\cite{farhan2018highly,QbS_wang2021query} have the best scalability for shortest path computation but sacrificing the query efficiency due to the search on periphery.

\subsection{Hierarchical Partitioning}\label{subsec:hierarchicalP}
\textit{Hierarchical Partitioning (HP)} organizes the network partitions hierarchically and each level is equivalent to a \textit{planar partitioning}. 
We formally define it as follows.

\begin{definition}
	\textbf{(Hierarchical Partitioning).} Given a graph $G$, hierarchical partitioning method organizes $G$ hierarchically within $H$ levels where each level $h$ is a planar partitioning $\{G^h_i\}$ of $G$ and $\exists G^{h-1}_j\supset G^h_i, \forall G^h_i (1\le h\le H)$. 
\end{definition}

\textit{HP} aims to organize the partition results in multiple layers so that each layer only needs to focus on the search/index of this level. Such a level-by-level approach is effective in pruning the search space for faster query processing. 
Since each level of hierarchical partitioning is equivalent to planar partitioning, we further divide it into two categories according to different cut types: \textit{edge-cut based HP} that cuts edges and \textit{vertex-separator based HP} that cuts vertices. 
A general way to obtain hierarchical partitioning is to leverage \textit{multi-level graph partition} methods such as \textit{METIS}~\cite{METIS_Karypis98MeTis} since they naturally produce multiple levels of partitions during the partitioning process.
Classic PSP indexes adopting \textit{edge-cut based HP} include \textit{G-Tree}~\cite{zhong2015g}, \textit{G*-Tree}~\cite{Gstar_tree_li2019g}, \textit{CRP}~\cite{CRP_delling2017customizable}, \textit{HiTi}~\cite{jung1996hiti}, and \textit{SHARC}~\cite{bauer2010sharc}, etc.
While PSP indexes adopting \textit{vertex-separator based HP} include \textit{SMOG} \cite{holzer2009engineering} and \textit{ROAD}~\cite{lee2010road}, etc. We next briefly introduce some classic ones.

\textit{\textbf{G-Tree.}} \textit{G-Tree} \cite{Gtree_zhong2013g} is a balanced search tree for location-based queries on road networks (\eg point-to-point shortest path query, $k$ nearest neighbor ($k$NN) query, and keyword-based $k$NN query) by hierarchically organizing the partitions produced by \textit{METIS}~\cite{METIS_Karypis98MeTis}. Specifically, they first partition graph $G$ into $f$ subgraphs with roughly equal size and take them as the root's children nodes, and then recursively partition them into another $f$ children nodes until each leaf node's subgraph has no more than $\tau$ vertices. By precomputing the shortest path distance of the border-border pairs in the non-leaf nodes and border-vertex pairs in the leaf nodes, an \textit{assembly-based method} is proposed for efficiently retrieving the shortest distance between two vertices from \textit{G-tree}. 
In particular, when $s$ and $t$ are in the same leaf node (\ie $s,t\in G_i$), there are two subcases for $sp(s,t)$: (1) $sp(s,t)$ does not contain a vertex outside the leaf node $G_i$. In this case, we can obtain $d(s,t)$ by performing Dijkstra's algorithm in $G_i$. (2) $sp(s,t)$ contains a vertex outside the leaf node $G_i$. In this case, $d(s,t)$ can be computed by $\min_{b_1,b_2\in B_i}\{d(s,b_1)+d(b_1,b_2)+d(b_2,t)\}$. 
For the case that $s$ and $t$ belong to different leaf nodes (\ie $s\in G_s$ and $t\in G_t$), we implement a dynamic programming algorithm that searches from $G_s$ to $G_{LCA}$ and then to $G_t$, where $G_{LCA}$ is the \textit{lowest common ancestor (LCA)} of $G_s$ and $G_t$.
\textit{G*-Tree}~\cite{Gstar_tree_li2019g} further improve the query efficiency of \textit{G-Tree} by adding shortcuts among the lowest level partitions.

\textbf{\textit{CRP.}} To answer the shortest path queries on the road networks with arbitrary metrics, Delling \etal propose a novel partition-based method named \textit{Customizable Route Planning (CRP)}~\cite{CRP_delling2011customizable,CRP_delling2017customizable} to support real-time traffic updates and personalized optimization functions. The general idea of \textit{CRP} is to decouple \textit{topological} and \textit{metric} properties of the road network by graph partitioning. In particular, \textit{CRP} has three main stages: \textit{metric-independent preprocessing}, \textit{metric customization}, and \textit{query stage}. \textit{Metric-independent preprocessing} decomposes graph $G$ into multi-level partitions by \textit{PUNCH}~\cite{PUNCH_delling2011graph} in top-down fashion. We first generate top-level partitions, while the lower-level partitions are obtained by running \textit{PUNCH} on each subgraphs of the higher level.
Suppose the top-level partitions are $\{G_i|i=1,...,k\}$, then in \textit{metric customization} stage, a shortest-distance \textit{clique} between the boundary vertices $B_i$, denoted as $C_i$, is precomputed for each partition $G_i$. 
These boundary distances are computed in a bottom-up fashion for all levels.
The \textit{overlay graph} on all boundary vertices is also constructed.
It is worth mentioning that the \textit{overlay graph} will be further optimized by pruning techniques such as \textit{edge reduction} and \textit{skeletons}, resulting in a much-sparsified graph called \textit{skeleton graph}.
Finally, bidirectional search is used for query processing in the \textit{query stage}. 


In summary, \textit{hierarchical PSP indexes} organize the partitions in a tree style hierarchically, allowing for the exploration of pathfinding in a layer-by-layer manner. As the state-of-the-art method in this line of research, \textit{G-Tree}~\cite{Gtree_zhong2013g,Gstar_tree_li2019g} achieves good shortest path query performance on road networks. Nonetheless, it cannot outperform tree decomposition-based methods such as \textit{H2H}~\cite{H2H_ouyang2018hierarchy} and is unsuitable for small-world networks.

\begin{table}[]
\setlength\tabcolsep{1pt}

\caption{Structure-Based Partition Method Classification.} 
\label{tab:summary}
\footnotesize
\begin{tabular}{c|rl|cc|c|c}
\hline
\multirow{2}{*}{\textbf{Structure}} 
& \multicolumn{2}{c|}{\textbf{Partition Method}}          & \multicolumn{2}{c|}{\textbf{Objectives}}      & \multirow{2}{*}{\textbf{Spatial-Aware}\begin{tabular}[c]{@{}c@{}}\end{tabular}} & \multirow{2}{*}{\textbf{Cut Type}\begin{tabular}[c]{@{}c@{}}\end{tabular}} \\ \cline{2-5}
    & \multicolumn{1}{r|}{\textbf{Category}}         & \textbf{Representative Methods}        & \multicolumn{1}{c|}{\textbf{Balance}} & \textbf{Minimal Cut} &        &      \\ \hline
\multirow{9}{*}{\textbf{PP}}       & \multicolumn{1}{r|}{Spectral partition}                  & SP~\cite{Pothen90SB}, RSB~\cite{barnard1994fast}       & \multicolumn{1}{c|}{$\checkmark$}   & $\checkmark$    & $\times$  & Edge    \\
   & \multicolumn{1}{r|}{Growing-based partition}             & Bubble~\cite{Bubble_diekmann2000shape}          & \multicolumn{1}{c|}{$\times$}         & $\checkmark$    & $\times$     &     Edge    \\   
   & \multicolumn{1}{r|}{Flow-based partition}                & PUNCH\cite{PUNCH_delling2011graph}         & \multicolumn{1}{c|}{$\checkmark$}     & $\checkmark$    & $\checkmark$    & Edge   \\           
   & \multicolumn{1}{r|}{Geometric partition}                 & RCB~\cite{simon1991partitioning,huang1996effective}                  & \multicolumn{1}{c|}{$\checkmark$}     & $\checkmark$    & $\checkmark$               & Edge             \\
   & \multicolumn{1}{r|}{Node-swapping heuristic}             & KL~\cite{KernighanLin70}, FM~\cite{FiducciaMattheyses82}           & \multicolumn{1}{c|}{$\checkmark$}     & $\checkmark$    & $\times$                            & Edge      \\
   & \multicolumn{1}{r|}{Minimum $k$-cut}           & $k$-cut~\cite{Goldschmidt94}                   & \multicolumn{1}{c|}{$\times$}         & $\checkmark$    & $\times$       & Edge         \\
    & \multicolumn{1}{r|}{Multilevel graph partition}        & SCOTCH~\cite{pellegrini1996scotch}, METIS~\cite{METIS_Karypis98MeTis}, KaHyPar~\cite{kahypar_ahss2017alenex}               & \multicolumn{1}{c|}{$\checkmark$}     & $\checkmark$    & $\times$   & Edge       \\  
    & \multicolumn{1}{r|}{Streaming edge partition}            & HDRF~\cite{petroni2015hdrf,HEP_mayer2021hybrid}, CLUGP~\cite{CLUGP_kong2022clustering}          & \multicolumn{1}{c|}{$\checkmark$}     & $\checkmark$    & $\times$     & Vertex    \\                 
    & \multicolumn{1}{r|}{In-Memory edge partition}            & NE\cite{zhang2017graph}, HEP\cite{HEP_mayer2021hybrid}              & \multicolumn{1}{c|}{$\checkmark$}     & $\checkmark$    & $\times$                  & Vertex        
    \\\hline
\multirow{2}{*}{\textbf{\begin{tabular}[c]{@{}c@{}}CP\end{tabular}}}       & \multicolumn{1}{r|}{Core-tree decomposition}                 & Core-Tree\cite{maehara2014computing,CT_li2020scaling,zheng2022workload,elliott2020core,DCT_zhou2024Scalable}     & \multicolumn{1}{c|}{$\times$}         & $\times$        & $\times$                       & Edge  \\
    & \multicolumn{1}{r|}{Sketch}          & QbS~\cite{QbS_wang2021query}, BatchHL~\cite{farhan2018highly,farhan2022batchhl} & \multicolumn{1}{c|}{$\times$}         & $\times$        & $\times$                                                         & Edge  \\ \hline
    \multirow{3}{*}{\textbf{HP}}     & \multicolumn{1}{r|}{\multirow{2}{*}{Edge-cut based HP}} & HiTi~\cite{jung1996hiti}       & \multicolumn{1}{c|}{$\times$}         & $\times$        & $\checkmark$                                                     & Edge           \\
& \multicolumn{1}{r|}{}                          & SHARC~\cite{bauer2010sharc}, G-Tree \cite{zhong2015g}, CRP \cite{CRP_delling2017customizable}                                      & \multicolumn{1}{c|}{$\checkmark$}     & $\checkmark$    & $\times$  & Edge  \\
& \multicolumn{1}{r|}{Vertex-Separator based HP}          & SMOG \cite{holzer2009engineering}, ROAD ~\cite{lee2009fast,lee2010road}             & \multicolumn{1}{c|}{$\times$}         & $\times$        & $\times$           & Vertex \\ \hline
\end{tabular}

\raggedright\textbf{PP} is Planar Partitioning, \textbf{CP} is Core-Periphery Partitioning, and \textbf{HP} is Hierarchical Partitioning. 
\textbf{Balance} denotes if it targets at balanced partitioning, \textbf{Minimum Cut} denotes if it targets at minimize the cut size, \textbf{Spatial-Aware} denotes if it needs geographic coordinates, and \textbf{Cut Type} is the cut type with Edge denotes edge-cut and Vertex denotes vertex-cut. 
\end{table}

\subsection{Summary and Discussion}\label{subsec:partitionDiscussion}

We present the partition results of different partition structures in Figure~\ref{fig:partition_example}, where the red vertices represent the corresponding boundary vertices. 
Specifically, \textit{planar partitioning} generates four partitions with equivalent vertex size while \textit{core-periphery partitioning} produces both ``Core'' and ``Periphery'' (the periphery of Core-Tree decomposition is a set of small-width trees while Sketch treats the non-core part as a periphery). \textit{Hierarchical partitioning} organizes partitions hierarchically, and the leaf nodes (the lowest level partition result) may have the same partition result as planar partitioning if the same MGP method is used. 

We also summarize and categorize existing partition methods under our structure-based classification, as shown in Table \ref{tab:summary}. Note that for \textit{CP} and \textit{HP}, we use the corresponding PSP indexes as representative methods, given the absence of specific partitioning methods associated with them. Almost all planar partition methods adopt minimal cut size as the optimization objective because a smaller cut size could facilitate cross-partition communication. Some partition methods also consider the size balance of different partitions for a more balanced workload distribution. Some of them also require spatial coordinates, confining their applications to road networks. 

We next analyze the requirement of the PSP index and discuss the pros and cons of different partition structures.
For the PSP index, the boundary vertex number is a crucial factor in determining the index construction time, query time, and update time.
However, the planar and hierarchical partitions have no limit to the boundary number, even though reducing the cut size is one of their optimization goals. By contrast, the core-periphery partition limits the boundary number by pre-defined bandwidth or limited landmark number such that their performance would not deteriorate by the large border number. Therefore, the planar and hierarchical partitions are better for small-treewidth networks (\eg road networks) such that balanced partitions and fewer boundaries can be achieved together; while core-periphery is better for large-treewidth networks (\eg social networks) as it limits the boundary vertex number by bandwidth or landmark number.

\section{Novel Partitioned Shortest Path Strategies}
\label{sec:PSPStrategy}
Apart from the shortest path algorithms and partition methods, another key problem for the PSP index is how to organize and coordinate the shortest path indexes of different partitions for correct index construction, query processing, and index maintenance. We call the strategy for solving the above problem as \textit{PSP Strategy}.
In this section, we first summarize the classic \textit{pre-boundary} PSP strategy and then propose two novel PSP strategies (\textit{no-boundary} and \textit{post-boundary} strategy) to improve the index construction and maintenance efficiency of existing solution with non-trivial correctness guarantee. In addition, we also propose a \textit{pruning-based overlay optimization} to further enhance these strategies. 
Note that we use planar PSP index for illustrating the PSP strategy in this section for simplicity.


\subsection{Traditional Pre-Boundary Strategy}
\label{subsec:Preliminary_Strategy}
The PSP index $L$ typically consists of two components: the \textit{partition indexes} $\{L_i\}$ for each subgraph (partition) $G_i$, and the \textit{overlay index} $\tilde{L}$ for the overlay graph which is composed of the boundary vertices of all partitions, \ie $L=\{L_i\}\cup \tilde{L}$. 
Regardless of the specific SP algorithm, almost all the existing PSP indexes \cite{jung2002efficient, CRP_delling2017customizable, holzer2009engineering, PbS_chondrogiannis2014exploring,ParDiSP_chondrogiannis2016pardisp, Gtree_zhong2013g, ta2017efficient, wang2019querying, wang2016effective,liu2021efficient,liu2022FHL, li2019time,li2020fastest,QbS_wang2021query,farhan2022batchhl, gubichev2010fast,qiao2012approximate} construct the PSP index and process the queries under the following procedure. 

\begin{figure}[t!]
    \centering
    \includegraphics[width=0.8\linewidth]{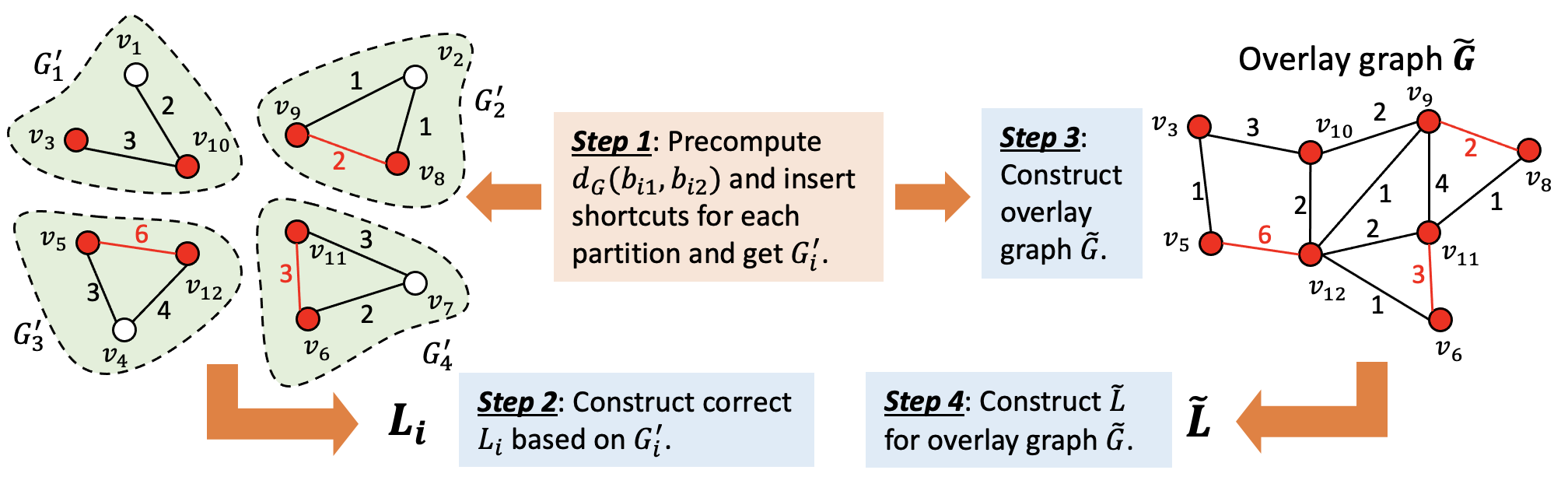}
    \caption{Traditional PSP Strategy (Pre-boundary Strategy)}
    \label{fig:Pre_Boundary_Example}
\end{figure}

\textbf{Index Construction.} 
It first leverages a partition method to divide the graph $G$ into multiple subgraphs $\{G_i\}$ and then builds the PSP index $L$ by the following four steps, as illustrated in Figure \ref{fig:Pre_Boundary_Example}.

\underline{\textit{Step 1}} precomputes the global distance between all boundary vertex pairs $(b_{i1},b_{i2}), (b_{i1},b_{i2}\in B_i)$ for each partition $G_i$ and insert shortcuts $e(b_{i1},b_{i2})=d_{G}(b_{i1},b_{i2})$ into $G_i$ to get $G'_i$.
For instance, the global distance between boundary vertex pairs $(v_3,v_{10})\in G_1$, $(v_8,v_9)\in G_2$, $(v_5,v_{12})\in G_3$, $(v_6,v_{11})\in G_4$ are calculated and inserted into their corresponding subgraphs to form new subgraphs $G'_1, G'_2, G'_3, G'_4$, respectively;

\underline{\textit{Step 2}} constructs the partition index $L_i$ based on $G'_i$. The index can be \textit{CH} \cite{CH_geisberger2008contraction}, \textit{PLL} \cite{PLL_akiba2013fast} or \textit{H2H} \cite{H2H_ouyang2018hierarchy}, and the index construction procedure among multiple partitions can be parallelized as it takes each partition as input independently;

\underline{\textit{Step 3}} constructs the overlay graph $\tilde{G}$ based on the precomputed shortcuts in \textit{Step 1}. Specifically, $\tilde{G}$ is composed of shortcuts between boundary vertex pairs $\{B_i\times B_i\}$ 
and the inter-edges,
that is $V_{\tilde{G}}=\{B_i\}, E_{\tilde{G}}=\{B_i\times B_i\}\cup E_{inter}$;

\underline{\textit{Step 4}} constructs the overlay index $\tilde{L}$ on $\tilde{G}$. Similar to $L_i$, $\tilde{L}$ can be any index type.

Note that the construction of $\tilde{L}$ (\textit{Step 3, Step 4}) can be parallelized with $L_i$ (\textit{Step 2}) since they are independent and both rely on \textit{Step 1}. 
For Step 1, we need $|B|$ times of \textit{Dijkstra}'s with time complexity \textit{$O(|B|\cdot(n\log n + m))$}. Then each partition's label $L_i$ can be constructed in parallel, and $\tilde{L}$'s construction is also independent of them, so its complexity is the worst case of them: $max\{O_c(G_i), O_c(\tilde{G})\}$, where $O_c$ is the complexity of underlying index's construction time as our discussion is not fixed to any specific index type. 

\textbf{Query Processing.} 
The shortest path queries can be divided into two categories and processed by utilizing $L$:
\begin{align*}
	&\textit{Case 1: Same-Partition}, \ie \forall s,t\in G_i, Q(s,t)=d_{L_i}(s,t);\\
	&\textit{Case 2: Cross-Partition}, \ie \forall s\in G_i, t\in G_j (i\neq j), Q(s,t)=
 \begin{footnotesize}
    \begin{cases}
	d_{\tilde{L}}(s,t) & s,t\in B\\
	\min\limits_{b_{q}\in B_j}\{d_{\tilde{L}}(s, b_{q})+d_{L_j}(b_{q}, t)\} & s\in B, t\notin B\\
	\min\limits_{b_{p}\in B_i}\{d_{L_i}(s, b_{p})+d_{\tilde{L}}(b_{p}, t)\} & s\notin B, t\in B\\
	\min\limits_{b_{p}\in B_i,b_{q}\in B_j}\{d_{L_i}(s, b_{p})+d_{\tilde{L}}(b_{p}, b_{q})+d_{L_j}(b_{q}, t)\} & s\notin B, t\notin B\\
	\end{cases}
	 \end{footnotesize}
\end{align*}


In summary, when $s$ and $t$ are in the same partition, we can use $L_i$ to answer $d_G(s,t)$; otherwise, we have to use $L_i, L_j$ and $\tilde{L}$. The same-partition query complexity is $O_q(L_i)$, where $O_q$ is the index's query complexity. The cross-partition query is made up of three parallel query sets, and the complexity is the worst of them: $max\{|B_i|\cdot O_q(L_i),|B_i|\cdot |B_j| \cdot O_q(\tilde{L}), |B_j|\cdot  O_q(L_j)\}$. 
The proof of the query processing is in the Appendix of our extended version \cite{FullVersion}.
Since the traditional partitioned index starts by precomputing the all-pair distance among boundary vertices, we call this approach \textbf{\textit{Pre-Boundary Strategy}}.

\subsection{Novel No-Boundary and Post-Boundary Strategies}
\label{subsec:Scheme_NoBoundary}
The pre-boundary strategy with the theoretical guarantee for correct partitioned shortest-path processing could handle all the queries well. However, the Step 1 of the \textit{Pre-Boundary} could be very time-consuming because only the index-free SP algorithms like \textit{Dijkstra's} \cite{dijkstra1959note} or \textit{A$^*$} \cite{Astar_hart1968formal} could be utilized. As a result, the index construction and maintenance efficiency suffer when the graph has numerous boundary vertex pairs. 
However, it appears that pre-computing the all-pair boundary distance is an essential procedure for constructing the ``correct'' \textit{PSP} index. Otherwise, the correctness of the subgraph index $L_i$ cannot be guaranteed because the shortest distance between boundary vertices within one partition could pass through another partition. For instance, as shown in Figure \ref{fig:Pre_Boundary_Example}, $sp(v_6,v_{11})=\left<v_6,v_{12},v_{11}\right>$ goes outside $G_4$ and passes through $v_{12}\in G_3$. As a result, $d(v_6,v_{11})$ cannot be answered correctly only with $G_3$ and the global shortest distance $d(v_6,v_{11})$ calculation is essential before constructing $L_3$. 
Then, a question arises naturally: do we really require precomputing the boundary all-pair distance first? In other words, is there a chance to keep index correctness by dropping this time-consuming step? Based on this brainstorm, we break the traditional misconception and propose a novel \textbf{\textit{No-Boundary Strategy}} as follows, which significantly reduces the index construction/maintenance time by skipping the time-consuming pre-computation step.

\begin{figure}
    \centering
    \includegraphics[width=0.8\linewidth]{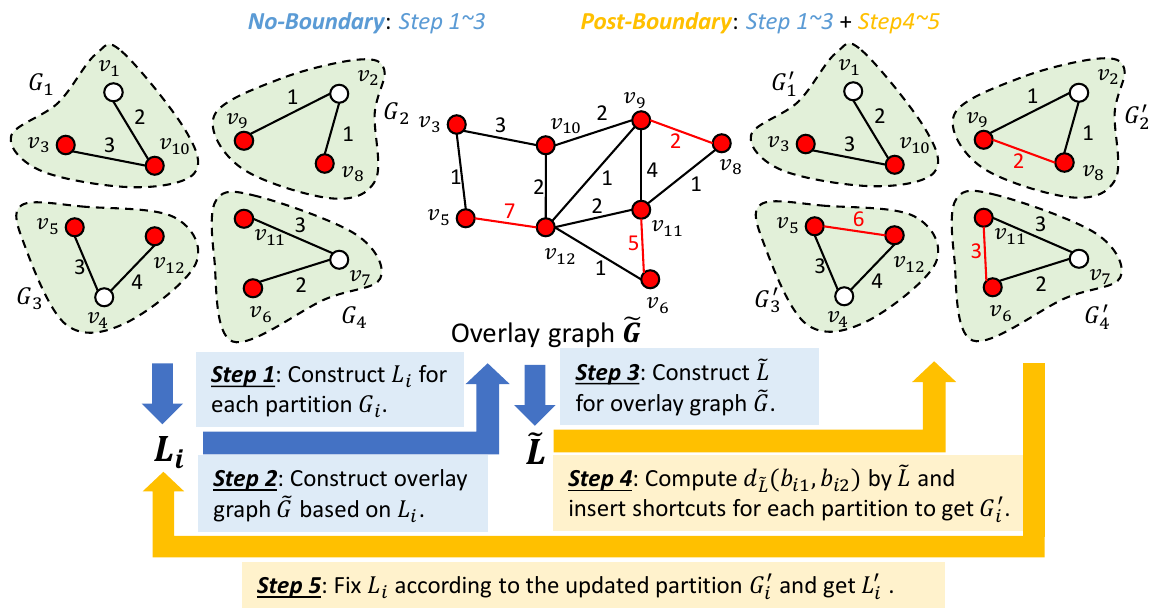}
    \caption{No-Boundary and Post-Boundary Strategies}
    \label{fig:no_boundary_index_construction}
\end{figure}

\textbf{Index Construction.}
It contains three steps as shown in Figure \ref{fig:no_boundary_index_construction}. With a graph partitioned into subgraphs $\{G_i\}$, we start by constructing the partition index $L_i$ for each subgraph $G_i$ parallelly in \underline{\textit{Step 1}}; Then \underline{\textit{Step 2}} construct the overlay graph $\tilde{G}$ based on $\{L_i\}$. For instance, $\tilde{G}$ is made up of boundary all-pair edges $(v_3,v_{10})$, $(v_8,v_9)$, $(v_5,v_{12})$, $(v_6,v_{11})$ (with their weight calculated based on $L_1, L_2, L_3$, and $L_4$) and inter-edges. It should be noted that although the overlay graph $\tilde{G}$ has the same structure as that in the \textit{Pre-boundary Strategy}, their edge weights differ as $\{L_i\}$ of them are constructed from different subgraphs; \underline{\textit{Step 3}} constructs the overlay index $\tilde{L}$ for $\tilde{G}$. Since the \textit{partition indexes} $\{L_i\}$ are constructed in parallel first followed by the construction of $\tilde{L}$, the \textit{No-Boundary} takes $max\{O_c(G_i)\}+O_c(\tilde{G})$ time in index construction.


\textbf{Query Processing}. 
Now that $L_i$ cannot answer $G_i$'s query correctly as the global distance between boundary all-pair is not covered, then how can \textit{No-Boundary} answer query correctly? 
Before revealing it, we first use the following theorem to prove that the correctness of $\tilde{L}$ still holds even though it is built upon the incorrect $\{L_i\}$, \ie the queries between boundary vertices can be correctly processed with $\tilde{L}$. 


\begin{theorem}
	\label{the:BoundaryLabel}
	$\forall s,t\in B, d_G(s,t)=d_{\tilde{L}}(s,t)$.
\end{theorem}

\begin{figure}
	\centering
	\includegraphics[width=0.6\linewidth]{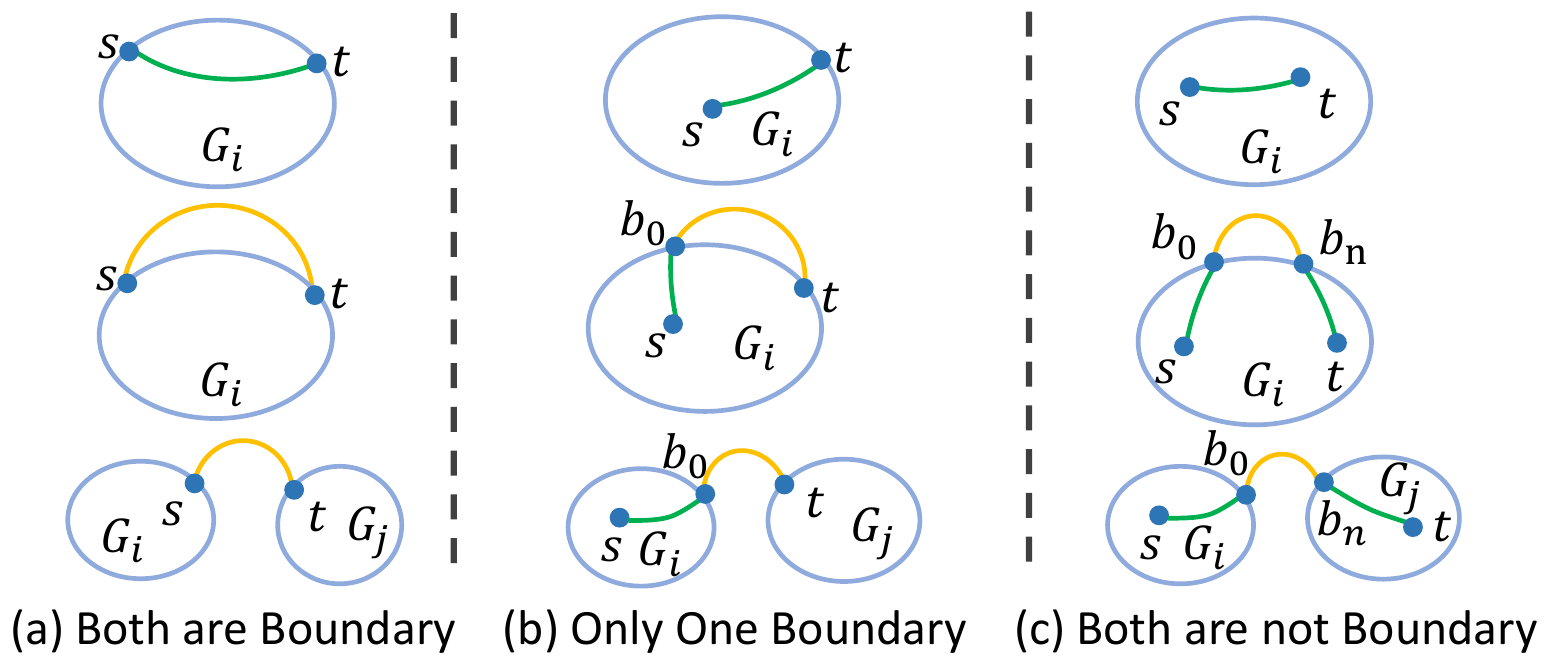}
	\caption{Different Categories of OD Distribution}
	\label{fig:querycorrectnessproof}
\end{figure}

\begin{proof}
We 
divide 
all the scenarios into three cases as shown in Figure \ref{fig:querycorrectnessproof}-(a): 1) $s$ and $t\in G_i$ and $p_G(s,t)$ only passes through the interior of $G_i$, then it is obvious that $d_G(s,t)=d_{G_i}(s,t)$ holds. Since $\tilde{w}(s,t)=d_{G_i}(s,t)$ is leveraged for $\tilde{G}$'s construction, we can obtain that $d_G(s,t)=d_{\tilde{G}}(s,t)=d_{\tilde{L}}(s,t)$; 2) $s$ and $t\in G_i$ with $p_G(s,t)$ going outside of $G_i$; 3) $s\in G_i$ and $t\in G_j (i\neq j)$. In the latter two cases, we take the concise form of $p_G(s,t)$ by extracting only the boundary vertices as $p_c=\left<s,b_0,b_1,\dots,b_n,t\right>$ ($b_i\in B, 0\le i\le n$). For two adjacent vertices $b_i,b_j\in p_c$, if 
$b_i$ and $b_j$ are in the same partition,
then $\tilde{w}(b_i,b_j)$ can be correctly obtained in $\tilde{G}$ 
via case 1);
otherwise, $(b_i,b_j)$ is an inter-edge with $\tilde{w}(b_i,b_j)=w(b_i,b_j)$ naturally correct. 
Thus, 
the shortest distance among boundary vertices can be correctly calculated by accumulating the edge weights on $\tilde{G}$. 
\end{proof}







Based on the above Theorem, we answer the shortest distance queries by discussing all scenarios in two cases:

\textit{Case 1: Same Partition} $s,t\in G_i$, we report $Q(s,t)$ by distance concatenation as shown in the following theorem:
\begin{theorem}
	\label{theorem:NoBoundary_SamePartition}
	$\forall s, t\in G_i$, $d_G(s,t)$ $=\min\{d_{L_i}(s,t),$ $\min\{$ $d_{L_i}(s,b_{i1})+d_{\tilde{L}}(b_{i1},b_{i2})+d_{L_i}(b_{i2},t)\}\}$, where $b_{i1}$ and $b_{i2}\in B_i$.
\end{theorem}
\begin{proof}
We denote $d_{L_i}(s,t)$ as $d_2$, $\min\{d_{L_i}(s,b_{i1})+d_{\tilde{L}}(b_{i1},b_{i2})+d_{L_i}(b_{i2},t)\}$ as $d_4$, and consider two subcases:

\textit{Subcase 1}: $p_G(s,t)$ does not go outside of $G_i$, as shown in the first row of Figure~\ref{fig:querycorrectnessproof}. No matter $s$ and $t$ are borders or not, $d_{L_i}(s,t)$ (\ie $d_2$) is enough to answer $d_G(s,t)$ as $L_i$ is built based on $G_i$ which contains all necessary information for finding the shortest path;

\textit{Subcase 2}: $p_G(s,t)$ passes outside of $G_i$, as shown in the second row of Figure~\ref{fig:querycorrectnessproof}. If $s$ and $t$ are both boundary vertices, $d_G(s,t)=d_{\tilde{L}}(s,t)$ holds by referring to Theorem~\ref{the:BoundaryLabel}. If $s$ and $t$ are both non-boundary vertices, we take the concise form of $p_G(s,t)$ by extracting only the boundary vertices as $p_c=\left<s,b_0,b_1,\dots,b_n,t\right>$ ($b_i\in B, 0\le i\le n$). Therefore, $d_4$ can correctly handle this case as the  $d_G(b_0,b_n)$ can be answered by $d_{\tilde{L}}(b_0,b_n)$ as per Theorem~\ref{the:BoundaryLabel}, while $d_G(s,b_0)$ and $d_G(b_n,t)$ can be answered by $d_{L_i}(s,b_0)$ and $d_{L_i}(b_n,t)$ by referring to Case 1. If either $s$ or $t$ is a non-boundary vertex, its distance is the special case of $d_4$ and can be easily proved.
\end{proof}	

\textit{Case 2: Cross-Partitions}, \ie $s\in G_i, t\in G_j, i\neq j$, we process them in the same manner as \textit{Pre-Boundary Strategy}. 

\begin{lemma}\label{lemma:noBoundaryCrossPartiQuery}
The cross-partition queries can be correctly processed in the No-Boundary Strategy.
\end{lemma}
\begin{proof}
We prove and discuss it with three subcases as shown in Figure~\ref{fig:querycorrectnessproof}-(c).
	
\textit{Subcase 1}: $s,t\in B$, then $d_G(s,t)$ can be correctly answered according to Theorem~\ref{the:BoundaryLabel};

\textit{Subcase 2}: $s\in B$ or $t\in B$. As shown in the second illustration in Figure~\ref{fig:querycorrectnessproof}-(c), suppose $s$ is an inner vertex of $G_i$ and $t$ is a boundary vertex, we take the concise $p_{s,t}$ by extracting the boundary vertices as $p_c=\left<s,b_0,\dots,b_n,t\right>$. Then $b_0\in B_i$ and $p_{s,t}$ can be treated as concatenated by two sub-paths $p_{s, b_0}\oplus p_{b_0,t}$. Specifically, $d_G(s,b_0)=d_{L_i}(s,b_0)$ by referring to the Subcase 1 of Case 1, and $d_G(b_0,t)=d_{\tilde{L}}(b_0,t)$ by referring to Theorem~\ref{the:BoundaryLabel};

\textit{Subcase 3}: $s\notin B, t\notin B$, which is the extended case of Subcase 2, so $d_G(s,t)$ can also be correctly answered.
\end{proof}

For cross-partition queries, we can use $\tilde{L}$ to answer $d_G(s,t)$ when $s$ and $t$ are both boundary vertices; otherwise, we have to use $L_i,L_j$ and $\tilde{L}$. The intra-query complexity is $max\{O_q(L_i), B_i\times B_j\times O_q( \tilde{L}),B_j\times O_q(L_j)\}$, while the inter-query takes $O_q( \tilde{L})$ time when $s$ and $t$ are both boundary vertices, and the complexity of other case is the worst of them: $max\{B_i\times O_q(L_i),B_i\times B_j \times O_q(\tilde{L}),B_j\times  O_q(L_j)\}$. 
Therefore, we make a bold attempt to enhance the performance of PSP index by skipping the heavy pre-computation, such that the index construction time is largely reduced so as the index maintenance (as will introduced in Section \ref{subsec:IndexUpdate}). Surprisingly, with incorrect distance value in the \textit{No-boundary index}, we prove theoretically that it can support the query answering with correctness guarantee.

By comparing the query processing procedure of the traditional \textit{Pre-boundary} strategy and our proposed \textit{No-boundary} strategy, we find that it is more complex for \textit{No-boundary} to process those queries with endpoints in the same partition because of the incorrectness of $\{L_i\}$. We repair this weakness by fixing the incorrect boundary all-pair distance in each partition $G_i$ and then transforming $\{L_i\}$ to their correct version. Specifically, we calculate the boundary all-pair distance $d(b_{i1},b_{i2})$ within each partition $G_i$ by leveraging $\tilde{L}$ and insert its corresponding edge $e(b_{i1},b_{i2})$ into $G_i$ to get $G'_i$ (\underline{\textit{Step 4}}). For instance, $d(v_3,v_{10})$, $d(v_8,v_9)$, $d(v_5,v_{12})$ and $d(v_6,v_{11})$ are calculated based on $\tilde{L}$ and these edges (as colored red in Figure \ref{fig:no_boundary_index_construction}) are inserted into their corresponding partitions with new subgraphs $\{G'_i\}$ formed. Followed by, we treat those newly inserted edges as graph updates and refresh $\{L_i\}$ to $\{L'_i\}$ by invoking the index maintenance algorithms \cite{DCH_ouyang2020efficient,DH2H_zhang2021dynamic,zhang2022relative,farhan2022batchhl,DPSL_zhang2021efficient,DCT_zhou2024Scalable} in each partition parallelly (\underline{\textit{Step 5}}). As these procedures are implemented after the \textit{No-boundary}, we call this strategy as \textbf{\textit{Post-boundary}}. By referring to Theorem \ref{the:BoundaryLabel}, the correct global distance information is contained in $\{G'_i\}$, so $\{L'_i\}$ can support the correct query processing of $Q(s,t)$ with $s,t\in G_i$. 
With the consistent query procedure as that of \textit{Pre-boundary}, the query efficiency of \textit{Post-boundary} reaches the state-of-the-art level.

\subsection{Index Update for Different PSP Strategies}
\label{subsec:IndexUpdate}
The PSP algorithms have been widely used on static networks, however, how to maintain the PSP index to support its application in dynamic environments has never been discussed.
In this section, we propose universal index update procedures for different PSP strategies to guarantee the correctness of the partitioned index in dynamic scenarios.

\begin{figure}
    \centering
    \includegraphics[width=\linewidth]{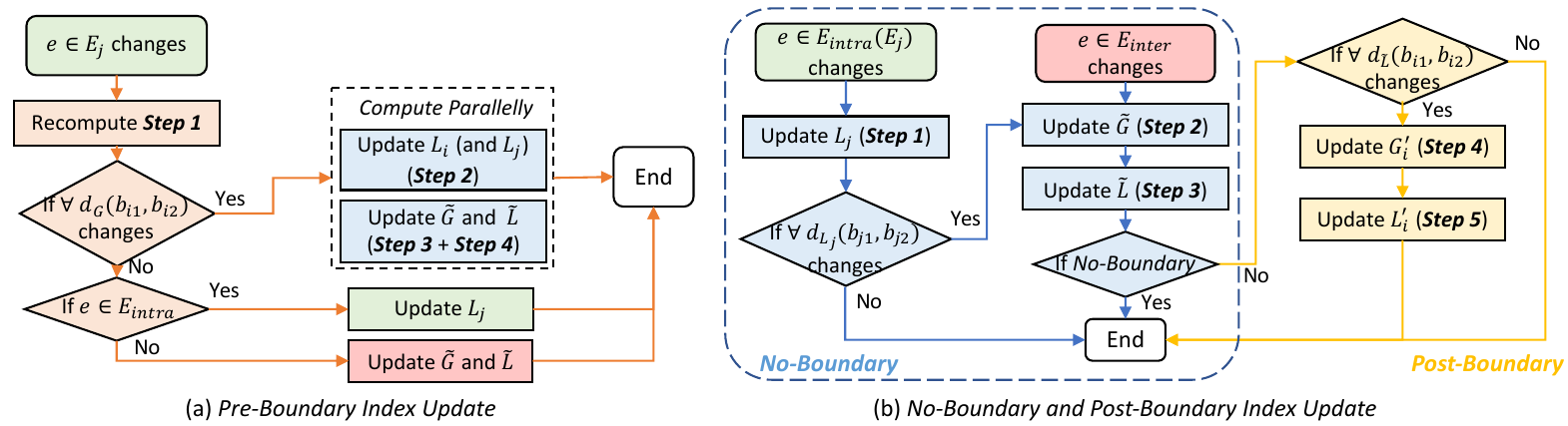}
    \caption{Index Update of Different PSP Strategies}
    \label{fig:indexupdate}
\end{figure}

\textbf{Index Update for Pre-boundary Strategy.}  
As shown in Figure \ref{fig:indexupdate}-(a), when the weight of edge $e\in G$ changes, we first recalculate the boundary-all pair distance (as the Step 1 in index construction) and identify the changed weight $e(b_{i1},b_{i2})$ between boundary vertices in each partition $G_i$. Then we need to update the corresponding partition index $L_i$ and overlay index $\tilde{L}$.
Step 1 takes $O(|B|\cdot(n\log n+m))$ time, while the partition and $\tilde{G}$ index takes $max\{O_u(G_i),O_u(\tilde{G})\}$ time to update in parallel, where $O_u$ is the update complexity for different indexes.

\begin{lemma}
	\label{Lemma:ProofPreBoundary}
    The update procedures of \textit{Pre-Boundary Strategy} can be correctly deal with any edge weight updates. 
\end{lemma}	
\begin{proof}
	First of all, we need to recalculate \textit{Step 1} to identify the affected edges $e(b_{i1},b_{i2})$ between boundary vertex pairs in each partition. Even though this step is time-consuming, 
	it cannot be skipped since it would be hard to identify the affected edges. For example, suppose the shortest path between the boundary pair $(b_{j1},b_{j2})$ in $G_j$ passes through an edge $e\in G_i$ with $d_{\tilde{G}}(b_{j1},b_{j2})=d_0$. When $e$ increases, we could update $L_i$ and then $\tilde{L}$. But $\tilde{L}$ cannot be correctly updated since it could be that $d_{\tilde{G}}(b_{j1},b_{j2})>d_0$, such that $d_{\tilde{G}}(b_{j1},b_{j2})$ cannot be refreshed to the correct value. It is because $d_0$ contains the old smaller edge weight while cannot be identified, since the shortest distance index always takes the smallest distance value. Then we could select those affected edges $e(b_{i1},b_{i2})$ by comparing their old and new weights. Lastly, we update their corresponding partition index $L_i$ and $\tilde{L}$ in parallel.	
\end{proof}

\textbf{Index Update for No-Boundary Strategy.}
Since the weight change of inter-edges does not affect the index for each subgraph, we divide index updates into two scenarios as shown in Figure~\ref{fig:indexupdate}-(b). Scenario 1: Inter-edge weight change. When  $e\in E_{inter}$ changes, only $\tilde{L}$ needs an update; Scenario 2: Intra-edge weight change. When $e\in E_{intra}$ ($e\in E_j$) changes, we first update $L_j$ and compare the old and new weights of $e(b_{j1},b_{j2})$ between boundary in $G_j$. If there is an edge weight update, we need to further update $\tilde{L}$. The update complexity is $max\{O_u(G_i)\}+max\{|B_i|^2\cdot O_q(L_i)\}+O_u(\tilde{G})$.

\begin{lemma}
	\label{Lemma:ProofNoBoundary}
    The update procedures of \textit{No-Boundary Strategy} can be correctly deal with any edge weight updates. 
\end{lemma}	
\begin{proof}
	In the \textit{inter-edge update} case, since $e\in \tilde{G}, e\notin G_i, \forall e\in E_{inter}$, the weight change of $e$ could only affect the correctness of $\tilde{L}$. So only $\tilde{L}$ should be checked and updated. In the \textit{Intra-edge update} case, since $e\in G_i$, its weight change will firstly affect $L_i$. Then it could affect $\tilde{G}$ as $e_{\tilde{G}}(b_{il_1},b_{il_2})=d_{L_i}(b_{il_1},b_{il_2})$. So $\tilde{L}$ also needs update if $e_{\tilde{G}}(b_{il_1},b_{il_2})$ changes after checked.
\end{proof}

\textbf{Index Update for Post-Boundary Strategy.}
It is similar to \textit{No-Boundary} with an additional judgment and processing shown in Figure~\ref{fig:indexupdate}-(b). Scenario 1: Intra-edge weight change. Suppose $e\in E_i$ changes, we update $G_i, L_i$ and then update $\tilde{G}, \tilde{L}$ if any $d_{L_j}(b_{j1},b_{j2})$ changes. Then we update $\{G'_i\}, \{L'_i\}$ if $d_{G_i'}(d_{i1},d_{i2})$ and $d_{\tilde{L}}(d_{i1},d_{i2})$ are different; Scenario 2: Inter-edge weight change. Suppose $e\in E_{inter}$ changes, we update $\tilde{G}, \tilde{L}$ and then update $\{G_i'\}, \{L_i'\}$.
Therefore, the update time complexity is 
$max\{O_u(G_i)\}+max\{|B_i|^2\cdot O_q(L_i)\}+O_u(\tilde{G})+max\{|B_i|^2\cdot O_q(\tilde{L})\}+max\{O_u(G_i)\}$.


\begin{lemma}
	\label{Lemma:ProofPostBoundary}
	The update procedures of \textit{Post-Boundary Strategy} can be correctly deal with any edge weight updates. 
\end{lemma}	
\begin{proof}
	Firstly, we prove the necessity to keep both $\{G_i\}, \{L_i\}$ and $\{G_i'\}, \{L_i'\}$. Similarly to the \textit{Pre-Boundary Strategy}, those boundary edges in $\{G_i'\}$ would keep the old smaller value such that the index could not be correctly updated as explained in Lemma \ref{Lemma:ProofPreBoundary}. So keeping $\{G_i\}, \{L_i\}$ gives us a chance to update $\tilde{L}, \{L_i\}$ correctly as proved in Theorem \ref{theorem:NoBoundary_SamePartition}.
	Then, following the \textit{No-Boundary Strategy} update, we recompute the shortest distance between the all-pair boundaries leveraging $\tilde{L}$ and compare their values on $L_i'$, then update $G_i', L_i'$.
\end{proof}

\textbf{Extension for Structural Updates.}
We next discuss how to support structural updates (edge/vertex insertion/deletion). 
For edge deletion, it is equivalent to increasing the edge weight to $+\infty$. 
Vertex deletion can be regarded as deleting all its adjacent edges ~\cite{DCH_ouyang2020efficient,DH2H_zhang2021dynamic,DPSL_zhang2021efficient}. 
Regarding the edge insertion, it is equivalent to decreasing an edge weight from $+\infty$ to a certain value~\cite{DCH_ouyang2020efficient,DH2H_zhang2021dynamic,DPSL_zhang2021efficient} and thus we can utilize the edge weight decrease update. In particular, there are three cases for a newly inserted edge $e(u,v)$: 1) $u$ and $v\in G_i$ belong to the same partition; 2) $u,v$ belong to different partitions but they are both boundaries; 3) $u,v$ belong to different partitions and at least one of them is not boundary vertex.
For case 1), we insert $e$ to $G_i$ and leverage edge decrease update to maintain $L_i$ (and $\tilde{L}$). For case 2), we update the overlay graph $\tilde{G}$ (and the partition index $L_i$ if the endpoints of the new edge belong to the same partition $G_i$) and overlay index $\tilde{L}$ by edge weight decrease update. For case 3), supposing $u\notin B$, this case results in a new boundary vertex $u$ and it affects the partition result, so index reconstruction is needed.
The vertex insertion is equivalent to a set of edge insertions~\cite{DCH_ouyang2020efficient,DH2H_zhang2021dynamic,DPSL_zhang2021efficient}.

\textbf{Extension for Batch Updates.}
Our techniques also support batch updates. Since CH and TD initially supports the batch updates, we directly leverage the corresponding techniques~\cite{DCH_ouyang2020efficient,DH2H_zhang2021dynamic} for the overlay and partition index maintenance. While for PLL that only supports streaming updates, a direct approach to process the batch
weight update is to leverage the techniques designed for the streaming update. Nevertheless, the all-pair distances computation among boundary vertices could be shared among one batch update for faster index maintenance.


\begin{table}[]
\caption{Time Complexity of Different PSP Strategies}
\label{table:Scheme}
\footnotesize
\centering 
\begin{tabular}{|c|cc|c|c|}
\hline
                                                & \multicolumn{2}{c|}{\textbf{Opeartions}}                              & \textbf{Procedure}                                                     & \textbf{Complexity}                                                                  \\ \hline
\multirow{4}{*}{\rotatebox{90}{\makecell{\textbf{Pre-}\\\textbf{Boundary}}}} & \multicolumn{2}{c|}{\textbf{Construction}}                            & $\{B_i\times B_i\}\rightarrow \{L_i\}, \tilde{L}$                    & $O(|B|\cdot (n\log n + m))+max\{O_c(G_i)+O_c(\tilde{G})\}$                                          \\ \cline{2-5} 
                                                & \multicolumn{1}{c|}{\multirow{2}{*}{\textbf{Query}}} & \textbf{Intra} & $L_i$                                                                  & $O_q(L_i)$                                                                           \\ \cline{3-5} 
                                                & \multicolumn{1}{c|}{}                                & \textbf{Inter} & $L_i\oplus \tilde{L}\oplus L_j$                                                & \makecell{$max\{|B_i|\cdot O_q(L_i), |B_i|\cdot |B_j| \cdot O_q(\tilde{L}),|B_j|\cdot  O_q(L_j)\}$} \\ \cline{2-5} 
                                                & \multicolumn{2}{c|}{\textbf{Update}}                                  & $\{B_i\times B_i\}\rightarrow \{L_i\}, L$                            & $O(|B|\cdot (n\log n + m))+max\{O_u(G_i)+O_u(\tilde{G})\}$                                          \\ \hline

\multirow{4}{*}{\rotatebox{90}{\makecell{\textbf{No-}\\\textbf{Boundary}}}}           & \multicolumn{2}{c|}{\textbf{Construction}}                            & $\{L_i\}\rightarrow \tilde{G}\rightarrow \tilde{L}$                                                 & $max\{O_c(G_i)\}+max\{|B_i|^2\cdot O_q(L_i)\}+O_c(\tilde{G})$                                                     \\ \cline{2-5} 
                                                & \multicolumn{1}{c|}{\multirow{2}{*}{\textbf{Query}}} & \textbf{Intra} & $L_i\oplus \tilde{L}\oplus L_i$                                                & $max\{|B_i|\cdot O_q(L_i), |B_i|^2 \cdot O_q(\tilde{L}))\}$                    \\ \cline{3-5} 
                                                & \multicolumn{1}{c|}{}                                & \textbf{Inter} & $L_i\oplus \tilde{L}\oplus L_j$                                                & \makecell{$max\{|B_i|\cdot O_q(L_i), |B_i|\cdot |B_j|\cdot O_q(\tilde{L}),|B_j|\cdot O_q(L_j)\}$} \\ \cline{2-5} 
                                                & \multicolumn{2}{c|}{\textbf{Update}}                                  & $\{L_i\}\rightarrow\tilde{G}\rightarrow \tilde{L}$                                                 & $max\{O_u(G_i)\}+max\{|B_i|^2\cdot O_q(L_i)\}+O_u(\tilde{G})$                                                     \\ \hline
\multirow{4}{*}{\rotatebox{90}{\makecell{\textbf{Post-}\\\textbf{Boundary}}}}         & \multicolumn{2}{c|}{\textbf{Construction}}                            & \makecell{$\{L_i\}\rightarrow \tilde{G}\rightarrow \tilde{L}$\\$\rightarrow\{B_i\times B_i\}\rightarrow \{L_i\}$} & \makecell{$max\{O_c(G_i)\}+max\{|B_i|^2\cdot O_q(L_i)\}+O_c(\tilde{G})$\\ $+ max\{|B_i|^2\cdot O_q(\tilde{G})\}+ max\{O_u(G_i)\}$}                    \\ \cline{2-5} 
                                                & \multicolumn{1}{c|}{\multirow{2}{*}{\textbf{Query}}} & \textbf{Intra} & $L_i$                                                                  & $O_q(L_i)$                                                                           \\ \cline{3-5} 
                                                & \multicolumn{1}{c|}{}                                & \textbf{Inter} & $L_i\oplus \tilde{L}\oplus L_j$                                                & \makecell{$max\{|B_i|\cdot O_q(L_i), |B_i|\cdot |B_j| \cdot O_q(\tilde{L}),|B_j|\cdot O_q(L_j)\}$} \\ \cline{2-5} 
                                                & \multicolumn{2}{c|}{\textbf{Update}}                                  & \makecell{$\{L_i\}\rightarrow \tilde{G}\rightarrow\tilde{L}$\\$\rightarrow\{B_i\times B_i\}\rightarrow \{L_i\}$} & \makecell{$max\{O_u(G_i)\}+max\{|B_i|^2\cdot O_q(L_i)\}+O_u(\tilde{G})$\\$+max\{|B_i|^2\cdot O_q(\tilde{G})\}+max\{O_u(G_i)\}$}                      \\ \hline
\end{tabular}

\raggedright
Because different SP indexes have different complexities but share the same logical procedure, we use the following notations to represent the logical complexities: $O_c$ is the index construction complexity, $O_q$ is the query complexity, and $O_u$ is the index update complexity. $max$ returns the maximum processing time among all partitions.
\end{table}

We analyze and summarize the time complexity of PSP strategies in terms of index construction, query processing, and index update in Table \ref{table:Scheme}.
Notably, the boundary vertex numbers $|B|$ and $|B_i|$ are two crucial factors of the index performance. 
The advantage of the no-boundary and post-boundary strategies over pre-boundary strategy source from the replacement of the time-consuming Dijkstra's searches ($O(|B|\cdot(m+n\log n)\})$) with index-based all-pair boundary query processing ($O(max\{|B_i|^2\cdot O_q(L_i)\})$). 
Therefore, such an optimization generally results in much better performance if the underlying partition index is 2-hop labeling (in which case $O_q(L_i)=w(G_i)$, where $w(G_i)$ is the treewidth of $G_i$). However, if the query processing is conducted in a search-based manner (\eg direct search or CH) and the boundary number $|B_i|$ is very large, the no-boundary strategy's performance may dramatically deteriorate. Such a deterioration is even worse for the post-boundary strategy. Nevertheless, no-boundary and post-boundary strategies generally have much better performance than the pre-boundary strategy in most cases.

\subsection{Pruning-based Overlay Graph Optimization}
\label{subsec:Scheme_Prune}
Though it is flexible to select the suitable boundary strategy according to application scenarios, we observe that all boundary strategies are hindered by the extremely dense overlay graph. The density of overlay graph $\tilde{G}$ increases dramatically as the all-pair boundary vertices in each partition are connected during its construction, which badly affects the index performance by slowing down the index construction, query processing, and index update \cite{zhang2021experimental}. Then is it possible to improve index performance and decrease the density of $\tilde{G}$ by deleting some unnecessary edges in $\{B_i\times B_i\}$? 
We explore this question by categorizing the boundary vertices as follows:
\begin{definition}[\textbf{Half / Full-Connected Boundary Vertex}] 
	\label{defition:ConnectVertex}
	$b\in B_i$ is a half-connected boundary vertex if $\exists u\in N(b), u\in G_i, u\notin B_i$ and we denote them as $B_i^H$ with $B^H=\{B_i^H\}$; otherwise, $b$ is a full-connected boundary vertex and denoted as $B_i^F$ with $B^F=\{B_i^F\}$.
\end{definition}

\begin{figure}
    \centering
    \includegraphics[width=0.7\linewidth]{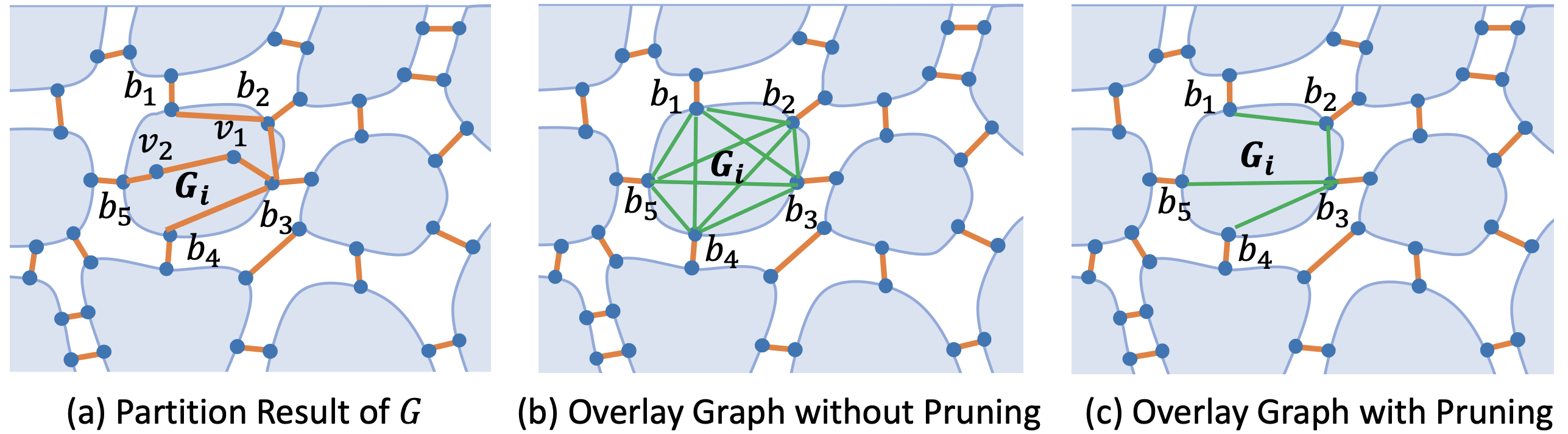}
    \caption{Example of Overlay Graph Simplification}
    \label{fig:pruning}
\end{figure}

As illustrated in Figure \ref{fig:pruning}-(a), suppose partition $G_i$ has five boundary vertices, where $b_1$, $b_2$, and $b_4$ are full-connected boundary vertices since their neighbors are all boundary vertices while $b_3$ and $b_5$ are half-connected boundary vertex since they have neighbors that are non-boundary vertices ($v_1$ and $v_2$).
We observe that the shortest paths from full-connected boundary vertices to other boundary vertices always pass through at least one of their boundary neighbors.
Let us consider the shortest path $p$ from $b_4\in B_i^F$ to any other boundary vertex $b\in B$, then the second vertex $v$ along $p$ must be a boundary vertex with $v\in N(b_4)$. Therefore, the global distance information of $p$ will still be kept in $\tilde{G}$ if we delete edges $\{(b_4, b_i)\}$ with $b_i\in B_i, b_i\notin N(b_4)$, since it is impossible for $p$ to pass through these edges.
By contrast, for the shortest path starting and ending with the half-connected boundary vertices $b_s,b_t\in B_i^H$, it is necessary to insert an edge $(b_s,b_t)$ into $\tilde{G}$ to preserve its distance information, as the second vertex on this path could be a non-boundary vertex which is not contained in the overlay graph. 
As analyzed above, we can shrink $\tilde{G}$ ($\{B_i\times B_i\}\cup E_{inter}$) to its slimmer version $\tilde{G}'$ (composed of $\{B_i^H\times B_i^H\}$, $E_{inter}$ and $B_i^F$'s adjacent edges). 
Let us only take $G_i$ as an example for the clear presentation, the original overlay graph as shown in Figure \ref{fig:pruning}-(b) will be shrunk to a sparser one in Figure \ref{fig:pruning}-(c) after the pruning technique, reducing 6 redundant shortcuts.
In the following Lemma, we prove the global distance can still be preserved in the overlay graph after pruning.
\begin{lemma}
\label{lemma:BoundaryShrinking}
	$\forall b_s, b_t\in B, d_{G}(b_s,b_t)=d_{\tilde{G'}}(b_s,b_t)$.
\end{lemma}
\begin{proof}
We denote the concise form of $p_G(b_s, b_t)$ by only taking the borders as $\{b_s=b_0,\dots,b_t=b_h\},(0<j<h)$.

When $h=1$, if either $b_s\in B^F$ or $b_t\in B^F$, then $d_{G}(b_s,b_t)=w_{G}(b_s,b_t)$ holds with $b_s\in N(b_t)$ by referring to definition \ref{defition:ConnectVertex}. Since $(b_s,b_t)\in \tilde{G}'$, $d_{G}(b_s,b_t)=d_{\tilde{G'}}(b_s,b_t)$ holds. The same applies when $b_s, b_t\in B^H$ with two endpoints in different subgraphs $b_s\in B_i, b_t\in B_j (i\neq j)$. If $b_s,b_t\in B^H$ with $b_s, b_t\in B_i$, edge $(b_s,b_t)$ are inserted into $\tilde{G}'$ with their global shortest distance in both \textit{Pre-Boundary} and \textit{Post-Boundary} strategies, so $d_G(b_s, b_t)=d_{\tilde{G}'}(b_s, b_t)$ naturally holds. In the \textit{No-Boundary} strategy, though only the local shortest distance is inserted in the overlay graph with $d_{G_i}(b_s,b_t)=d_{\tilde{G}'}(b_s,b_t)$, it holds that $d_{G}(b_s,b_t)=d_{G_i}(b_s,b_t)$ since there exists no other boundary vertex besides two endpoints in $p_{G}(b_s,b_t)$ and it only pass through $G_i$, so $d_{G}(b_s,b_t)=d_{\tilde{G'}}(b_s,b_t)$ holds.

When $h>1$, the shortest path distance is accumulated as $d_{G}(b_s,b_t)=\sum_{j=0}^{h-1}d_G(b_{j},b_{j+1})$. Since $d_G(b_{j},b_{j+1})=d_{\tilde{G}'}(b_{j},b_{j+1})$ holds for $0\le j<h$ by referring to the case when $h$=1, we can get that $d_G(b_s,b_t)=d_{\tilde{G}'}(b_s,b_t)$. So we prove that $d_{G}(b_s,b_t)=d_{\tilde{G'}}(b_s,b_t)$ holds for all scenarios and boundary strategies.
\end{proof}	
Therefore, we enhance the index performance by utilizing the sparser $\tilde{G}'$.
Note that Lemma~\ref{lemma:BoundaryShrinking} also holds with edge weight updates since this optimization only depends on the structure of $\tilde{G}$.
\section{Universal PSP Index Scheme and Newly Generated PSP Indexes}
\label{sec:universalScheme} 
In this section, we propose a universal PSP index scheme and then put forward five representative PSP indexes for query-oriented or update-oriented applications based on this scheme.

\setlength{\belowcaptionskip}{-4pt}
\begin{figure*}
	\centering
	\includegraphics[width=1\linewidth]{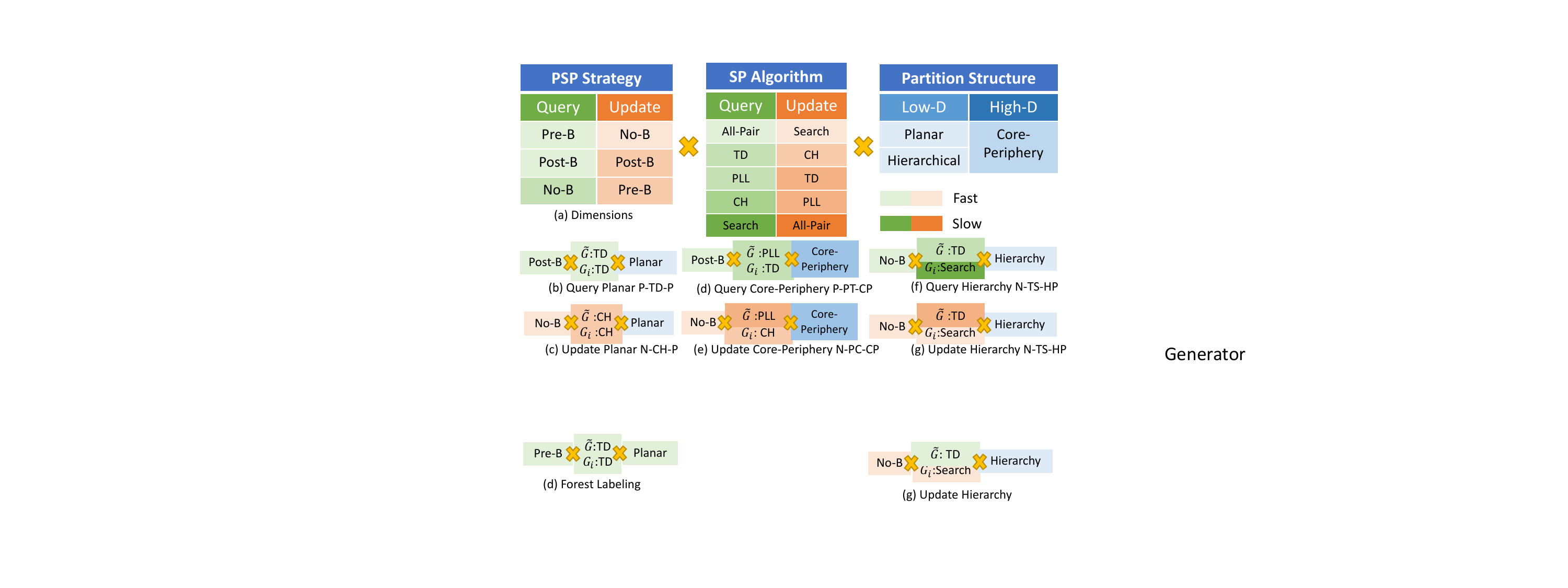}
	\caption{PSP Index Generator with PSP Index Scheme and Representative Generated PSP Indexes}
	\label{fig:Coupling}
\end{figure*}

 \subsection{PSP Index Dimensions and Universal Scheme}
\label{subsec:PSP}
Based on the comprehensive literature review and analysis of the PSP index in previous Sections, we identify three critical dimensions of the PSP index as follows.

\textbf{1) \textit{Shortest Path Algorithm}} that decides the underlying methods for the overlay graph and partitions. The performance of different SP algorithms on dynamic networks has been thoroughly studied in our previous work \cite{zhang2021experimental}; 

\textbf{2) \textit{Graph Partition Method}}, which divides the graph into multiple subgraphs and is the key difference from the standalone SP indexes. The existing partition methods are not designed for path indexes, so how they affect the PSP index is unknown and deserves study; 

\textbf{3) \textit{PSP Strategy}}, which provides a procedure to organize the index construction and update for different subgraphs and coordinate those indexes such that the query correctness is guaranteed. This dimension is crucial because it determines the complexity of both the query and update. However, only straightforward pre-boundary strategy exists currently. Thus we further propose two novel strategies to improve its index construction and maintenance performance.

Given a dynamic network, these three dimensions provide a roadmap to construct a PSP index by selecting a specific \textit{graph partition method}, \textit{SP algorithm} and \textit{PSP strategy} such that the application requirement (query / update efficiency) can be satisfied. 
Inspired by this, we propose a universal \textit{\textbf{PSP Index Scheme}} as shown in Figure \ref{fig:Coupling}. The main part is the \textit{PSP index generator}, which is formed by coupling those three components. For easy selection of the PSP strategy and SP algorithm, we sort all of their choices roughly in terms of query and update efficiency. 
As shown in Figure~\ref{fig:Coupling}, the lighter color signifies faster efficiency for PSP strategy and SP algorithm. Meanwhile, planar and hierarchical partitions are more suitable for low-treewidth networks, while core-periphery can deal with high-treewidth graphs.

There are two benefits of our PSP index scheme. Firstly, the existing \textit{PSP indexes} can find their positions in our scheme and we can compare them theoretically and fairly. 
For exmaple, 1) \underline{Pre-Boundary + Search/All-Pair + Hierarchy/ Planar}: \textit{HiTi} \cite{jung2002efficient}, \textit{CRP} \cite{CRP_delling2017customizable}, \textit{PbS} \cite{PbS_chondrogiannis2014exploring}, \textit{ParDiSP}~\cite{ParDiSP_chondrogiannis2016pardisp} pre-compute  shortest distances between boundaries to guide the search. In a broad sense, \textit{Arc-flag} \cite{mohring2007partitioning} and \textit{SHARC} \cite{bauer2010sharc} also belong to this category. \textit{G-tree} \cite{Gtree_zhong2013g,Gstar_tree_li2019g} uses dynamic programming to compute the distance between layer's all-pair information to replace searching and it is used in \textit{kNN} \cite{zhong2015g}, \textit{ride-sharing} \cite{ta2017efficient}, \textit{time-dependent} \cite{wang2019querying}, and \textit{distance embedding} \cite{huang2021learning}. They are slow to construct due to \textit{Pre-B} and inefficient to query due to direct search; 
2) \underline{Pre-Boundary+Search/PLL+Planar}: 
\textit{COLA} \cite{wang2016effective} builds labels for the \textit{skyline shortest path} on the overlay graph for the constrained shortest path. Its construction suffers from \textit{Pre-B} and query suffers from searching; 
3) \underline{Pre-Boundary+TD/TD+Planar}: \textit{FHL} \cite{liu2021efficient,liu2022FHL} builds the \textit{TD} both within and between partitions for multi-dimensional skyline paths; 
4) \ul{Pre-Boundary+PLL/PLL+Planar}: \textit{T2Hop} \cite{li2019time,li2020fastest} utilizes two layers of \textit{PLL} to reduce the complexity of long-range time-dependent paths, and this structure's performance is limited by \textit{PLL}; 
5) \ul{Pre-Boundary+Search/All-Pair/PLL+Sketch}: This category works on huge graphs where index is impossible so only a small number of landmarks are selected to prune the search \cite{das2010sketch,QbS_wang2021query,farhan2022batchhl} or approximate the result \cite{gubichev2010fast,qiao2012approximate}.

Secondly, given the specific application requirements in terms of computation efficiency, space consumption, or partition type, our scheme could provide one PSP index with a specified PSP strategy, SP algorithm, and partition structure through the \textit{PSP index generator}.
Especially when all the existing ones are query-oriented, there is a vast space of combinations for new ones. Specifically, the generated PSP index needs to choose the components from each of the three dimensions, and its name also has three parts corresponding to the three dimensions. 
Although we can enumerate them all, we choose to introduce five representative ones from the perspectives of \textit{query-} and \textit{update-}oriented under different partition structures, as shown in Figure~\ref{fig:Coupling}.
The details of these five PSP indexes will be elaborated in Section~\ref{subsec:universalScheme_Index}.

 \subsection{Representative Generated New PSP Indexes}
 \label{subsec:universalScheme_Index} 

\textbf{1. Query-Oriented Planar PSP Index: P-TD-P.}
We introduce how to construct a planar PSP index that is efficient in query processing.
Firstly, in terms of PSP strategy, since both \textit{Pre-Boundary} and \textit{Post-Boundary} are efficient in query answering and our proposed \textit{Post-Boundary} is generally faster than \textit{Pre-Boundary} in index maintenance, we use it as the PSP strategy.
Secondly, in terms of the SP algorithm, we could choose \textit{TD} as the index for both partitions and overlay graph. Although \textit{all-pair} is faster, its space consumption is intolerable. Its structure is shown in Figure \ref{fig:Coupling}-(a). Specifically, the index construction takes $O(V_{max}\cdot (\log V_{max}+h_{max}\cdot w_{max}))$ time for the partition \textit{TD}, $O(V_{\Tilde{G}}(\log V_{\Tilde{G}}+h_{\Tilde{G}}\cdot w_{\Tilde{G}}))$ for the overlay \textit{TD}, $O(B^2_{max}w_{\Tilde{G}})$ for boundary correction, and $O(w_{\Tilde{G}}+w^2_{max}\cdot \delta)$ for the partition \textit{TD} refresh, where $\delta$ is the affected shortcut number and $max$ are the corresponding max values in the partitions. 
For the query processing, the intra-query takes $O(w_{max})$ as partition index is correct, and the inter-query takes $O(max\{B_{max}w_{max},B^2{max}w_{\Tilde{G}}\})$ for the partition and overlay query and $O(B^2_{max})$ for the combinations. For the index update, it takes $O(w^2_{max}\delta)$ to update partition \textit{TD}, $O(w^2_{\Tilde{G}}\delta)$ to update overlay \textit{TD}, and $O(B^2_{max}w_{\Tilde{G}})$ to check boundaries.

\textbf{2. Update-Oriented Planar PSP Index: N-CH-P.}
In terms of the PSP strategy, we use \textit{No-Boundary} as it requires the least effort to update. In terms of SP algorithm, we can choose \textit{CH} as the underlying index as it is fast to update while the query processing is better than direct search (Figure \ref{fig:Coupling}-(b)). Its index construction is faster with $O(V_{max}\cdot w^2_{max}\cdot \log V_{max})$ for partition \textit{CH} and  $O(V_{\Tilde{G}}\cdot w^2_{\Tilde{G}}\cdot \log V_{\Tilde{G}})$ for overlay \textit{CH}; the query time is longer with $O(max\{B_{max}\cdot w_{max}\log V_{max},$ $B^2_{max} \cdot w_{\Tilde{G}}\log V_{\Tilde{G}}\})$ for the intra- and overlay searching, and $O(B^2_{max})$ for the combinations; the index update is faster with
$O(\delta w_{max})$ for partition \textit{CH} maintenance and $O(\delta w_{\Tilde{G}})$ for overlay \textit{CH} maintenance.

\textbf{3. Query-Oriented Core-Tree PSP Index: P-PT-CP.}
The core-periphery partition index comprises the core index $L_c$ and the periphery index $\{L_i\}$. 
Although the core does not belong to any partition, they are connected to the partitions, and we treat the core as the overlay graph. 
Different from the previous planar partition, the core here usually has a large degree so its index is limited to \textit{PLL}. As for \textit{sketch}, we omit it here because its \textit{PLL} core + pruned direct search seems to be the only solution for huge networks. Next, we discuss the remaining parts.
For the PSP strategy, \textit{Post-Boundary} is utilized as the queries within the periphery can be handled without the core index. The periphery index uses \textit{TD} because the periphery usually has a small degree. This structure is shown in Figure \ref{fig:Coupling}-(c). In terms of index construction, because the periphery is constructed through contraction, we regard it as a by-product of the partition phase and do not construct their labels in the first step. Then it takes $O(w_c E_c \log V_c + w^2_c V_c \log^3 V_c)$ for the core \textit{PLL}, $O(B^2_{max}w_c \log V_c)$ for boundary correction, and $O(V_{max}\cdot w^2_{max}\cdot \log V_{max})$ for periphery label. In terms of query processing, the intra-query takes $O(w_{max})$ time. The inter-query takes $O(max\{B_{max}w_{max},B^2_{max}w_c\log V_c\})$ for the periphery and core, and $O(B^2_{max})$ for the combinations. In terms of index update, it takes $O(w^2_{max}\cdot\delta)$ for the periphery \textit{TD}, $O(w_cE_c\log V_c)$ for the core \textit{PLL}, and $O(B^2_{max}w_c \log V_c)$ for boundary correction.

\textbf{4. Update-Oriented Core-Tree PSP Index: N-PC-CP.}
As shown in Figure \ref{fig:Coupling}-(d), N-PC-CP uses \textit{CH} for faster periphery update while \textit{PLL} is for the core. It adopts \textit{No-Boundary} strategy for faster update. In index construction, only core needs $O(w_c E_c \log V_c + w^2_c V_c \log^3 V_c)$ time. The query time is longer with $O(max\{B_{max}\allowbreak w_{max}\log V_{max},$ $B^2_{max}w_c\log V_c\})$ for intra and core, and $O(B^2_{max})$ for combinations.
The index update is fast with $O(\delta w_{max})$ for periphery shortcuts and $O(w_cE_c\log V_c)$ for the core.

\textbf{5. Update / Query-Oriented Hierarchical PSP Index: N-TS-HP.}
This category organizes $L$ levels of partitions hierarchically with several lower partitions forming a larger partition on the higher level. We use $L^l_i$ to denote the index of partition $G^l_i$ on level $l$. Different from the SOTA \textit{G-Tree} stream of indexes which uses all-pair in their hierarchical overlay graph, we replace it with the hierarchical labels for better query and update performance. Specifically, for vertices in each layer, we store their distance to vertices in their upper layers. As this is essentially 2-hop labeling, we use \textit{TD} to implement it with orderings corresponding to the boundary vertex hierarchy. Such a replacement in the overlay index could answer queries and update much faster than the original dynamic programming-based layer all-pair index. 
As for the partitions, this structure tends to generate small partitions so we inherit the original search for fast query processing. Consequently, no partition index leads to inevitable boundary all-pair searches. Fortunately, our \textit{No-Boundary} restricts the search space to the small partition compared with \textit{G-Tree}'s whole graph \textit{Pre-Boundary} (Figure \ref{fig:Coupling}-(e) and (f)). To construct the index, the partition boundary all-pair takes $O(B_{max}\cdot V_{max}(logV_{max}+E_{max}))$, and its by-product boundary-to-partition can be cached for faster inter-query. 
Then the overlay \textit{TD} takes $O(V_{\Tilde{G}}(logV_{\Tilde{G}} + h_{\Tilde{G}}w_{\Tilde{G}}))$ time. In query processing, the intra-query takes $O(B_{max}V_{max}(logV_{max}+E_{max})$ for the direct search (very rare as the partitions are small), while the inter-query 
The query takes $O(B_{max}w_{\Tilde{G}}\})$ for intra- (constant time with cache) and hierarchical query, and $O(B^2_{max})$ for combination. As for index update, it takes $O(B_{max}\cdot V_{max}(logV_{max}+E_{max}))$ to update the partition all-pairs and $O(w^2_{\Tilde{G}}\cdot \delta)$ to update the overlay graph.

In summary, these five PSP indexes are novel structures, each with unique technical challenges, especially in updating. 
Nevertheless, we believe the PSP scheme holds greater significance than these new indexes, as it can guide new PSP index designs. 
We omit their details due to limited space, but their source codes are provided \cite{SourceCode}. Besides, we take the N-CH-P index as an example and elaborate on its details in the extended version \cite{FullVersion}.
\section{Experimental Evaluation}\label{sec:Experiment}
In this section, we evaluate the effectiveness of the proposed methods and provide a systematic evaluation of PSP indexes. All the algorithms are implemented in C++ with full optimization on a server with 4 Xeon Gold 6248 2.6GHz CPUs (total 80 cores / 160 threads) and 1.5TB memory. The default thread number is set to 150. 






\setlength{\belowcaptionskip}{-10pt}
\begin{table}[t]
	\caption{Dataset Description}
	\label{table:Dataset}
	\begin{threeparttable}[]
	\centering
	\small
 \setlength\tabcolsep{2pt}
	\begin{tabular}{l|l|l|r|r|l}
		\hline
		\textbf{Category}                             & \textbf{Name}   & \textbf{Dataset}        & \textbf{$|V|$}        & \textbf{$|E|$}  & \textbf{Type}   \\ \hline
		\multirow{4}{*}{Road Network}    & NY \tnote{1}     & New York City  & 264,346    & 730,100  &  road   \\ \cline{2-6} 
		& FL \tnote{1}    & Florida        & 1,070,376  & 2,687,902    &  road    \\ \cline{2-6} 
		& W \tnote{1}     & Western USA    & 6,262,104  & 15,119,284   &  road    \\  \cline{2-6}
		& US \tnote{1}    & Full USA       & 23,947,347 & 57,708,624  &  road     \\ \hline
		\multirow{4}{*}{Complex Network} & GO \tnote{2} & Google         & 855,802    & 8,582,704    &  web  \\ \cline{2-6} 
		& SK \tnote{3}   & Skitter        & 1,689,805  & 21,987,076   &  social    \\ \cline{2-6} 
		& WI \tnote{3}   & Wiki-pedia     & 3,333,272  & 200,923,676   &  web   \\ \cline{2-6} 
		& FR \tnote{2} & com-Friendster & 65,608,366 & 3,612,134,270  &  social\\ \hline
	\end{tabular}
[1] http://www.dis.uniroma1.it/challenge9/download.shtml \newline
[2] http://snap.stanford.edu/data     [3] http://konect.cc/
\end{threeparttable}%
\end{table}


\subsection{Experimental Settings} \label{subsec:Experiment_Settings}

\textbf{Datasets.} 
We test on eight real-life datasets (four road networks and four complex networks as shown in Table \ref{table:Dataset}). 
Following \cite{DPSL_zhang2021efficient,DCT_zhou2024Scalable,zhang2021experimental,MCSPs_zhou2024efficient}, the edge weights of complex networks are randomly generated inversely proportional to the highest degree of the endpoints.

\textbf{Partition Methods and Their Performance Metrics.} 
To select the suitable partition method for PSP index, we implement 8 representative methods: \textit{Bubble}~\cite{Bubble_diekmann2000shape}, \textit{RCB}~\cite{simon1991partitioning}, \textit{KaHyPar}~\cite{kahypar_ahss2017alenex}, \textit{PUNCH}~\cite{PUNCH_delling2011graph}, \textit{SCOTCH}~\cite{Scotch18}, \textit{METIS}~\cite{METIS_Karypis98MeTis}, \textit{HEP}~\cite{HEP_mayer2021hybrid}, \textit{CLUGP}~\cite{CLUGP_kong2022clustering}, all of which are planar with PUNCH and RCB only working on road networks. Since core-periphery (core-tree~\cite{DCT_zhou2024Scalable} and Sketch \cite{farhan2022batchhl}) and hierarchical partition (uses METIS \cite{METIS_Karypis98MeTis}) methods are finite, they are selected by default in the corresponding PSP index. We evaluate the performance of a partition method from three aspects: overall border number $|B|=\sum_{i\in [1,k]} |B_i|$, average partition border number $\overline{|B_i|}=\frac{\sum_{i\in [1,k]} |B_i|}{k}$, and partition connectivity $R_C=\frac{\sum_{i\in [1,k]}|C(G_i)|}{k}$, where $C(G_i)$ is the connected component number of subgraph $G_i$.

\textbf{PSP Indexes and Performance Metrics.} We evaluate 5 SOTA PSP indexes combining with the three proposed partition structures: 
a) \textit{FHL} \cite{liu2021efficient,liu2022FHL} and \textit{THop}~\cite{li2019time} from planar partition; 
b) \textit{Core-Tree (CT)} \cite{CT_li2020scaling,zheng2022workload,DCT_zhou2024Scalable} and \textit{Sketch}~\cite{farhan2022batchhl} from core-periphery;
c) \textit{G-Tree} \cite{Gtree_zhong2013g}
from hierarchical partition. 
Besides, we implement our 5 newly proposed PSP indexes: \textit{P-TD-P}, \textit{N-CH-P}, \textit{P-PT-CP}, \textit{N-PC-CP}, and \textit{N-TS-HP}. 
For the SOTA shortest path indexes CH \cite{CH_geisberger2008contraction,DCH_ouyang2020efficient}, TD~\cite{H2H_ouyang2018hierarchy,DH2H_zhang2021dynamic}, and PLL \cite{PLL_akiba2013fast,DPSL_zhang2021efficient}, we implement their partitioned version (named \textit{\textit{PCH}}, \textit{\textit{PTD}}, and \textit{\textit{PPLL}}). We measure the performance from four aspects: index construction time $t_c$, query time $t_q$, update time $t_u$, and index size $s$. We do not report the result ($\infty$) if index time exceeds 24 hours.

\textbf{Query and Update.} 
For each dataset, we randomly generate a set of 10,000 queries and 1,000 edge update instances to evaluate $t_q$ and $t_u$. Following \cite{DH2H_zhang2021dynamic,zhang2021experimental}, for each selected edge $e$, we decrease or increase its weight to $\alpha\cdot |e|$, where $\alpha$ randomly ranges in $(0,2]$.
Besides, we also randomly generate 1000 update instances for each type of structural update (edge/vertex insertion/deletion) for Exp 5.

\textbf{Parameter Setting.} 
We set the default partition number as $32$ for all planar PSP and the bandwidth of \textit{CT}-based PSP as $40$ as per our preliminary experimental results in \cite{FullVersion}. The landmark number of Sketch is $20$ \cite{farhan2022batchhl}. 
For the hierarchical PSP, we follow G-tree~\cite{Gtree_zhong2013g, Gstar_tree_li2019g} and set the fan-out $f$ as $4$ and the leaf node size $\tau$ as 128 in \textit{NY}, 256 in \textit{FL}, 512 in \textit{W} and \textit{US}.
We use \textit{No-boundary} as the default strategy if the PSP strategy is not specified.


\begin{figure*}[t]
	\centering
	\includegraphics[width=1\linewidth]{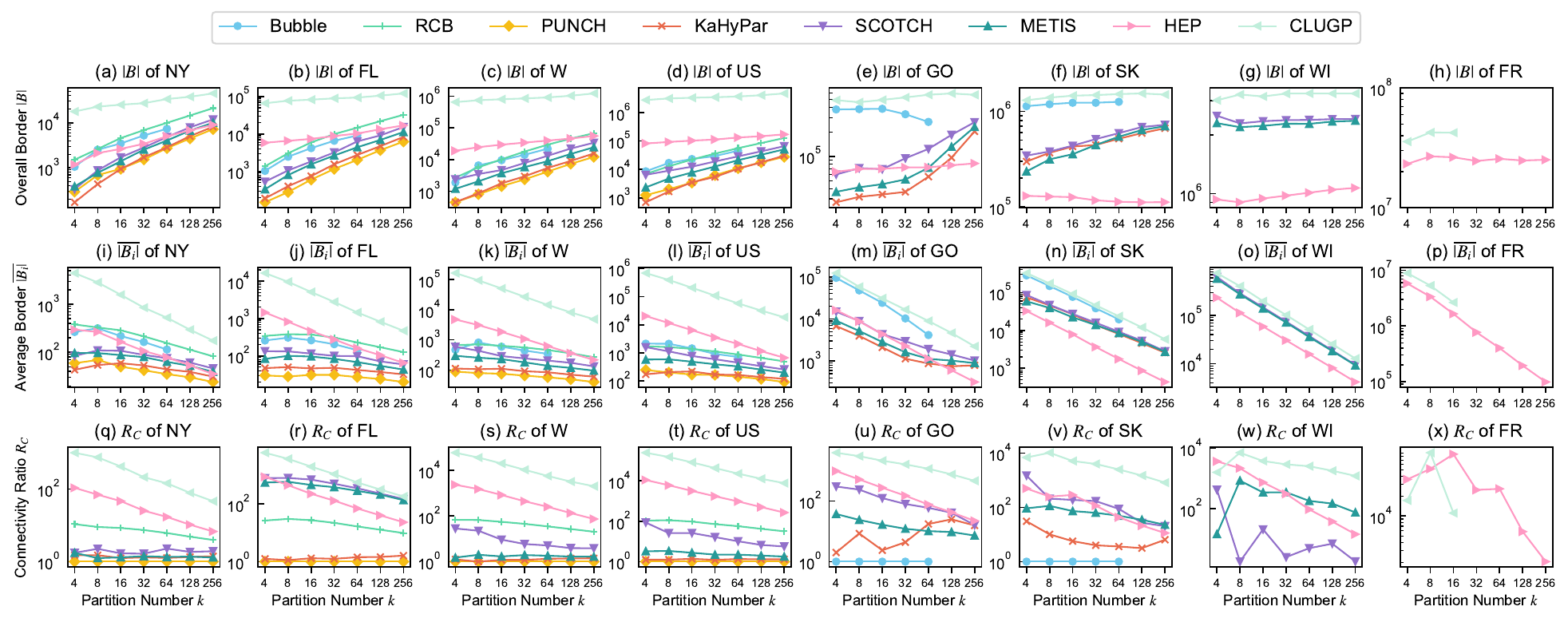}
	\caption{Performance of Partition Methods When Varying Partition Number $k$.}
	\label{fig:partitionInfo}
\end{figure*}

\begin{figure}[t]
	\centering
    \includegraphics[width=0.85\linewidth]{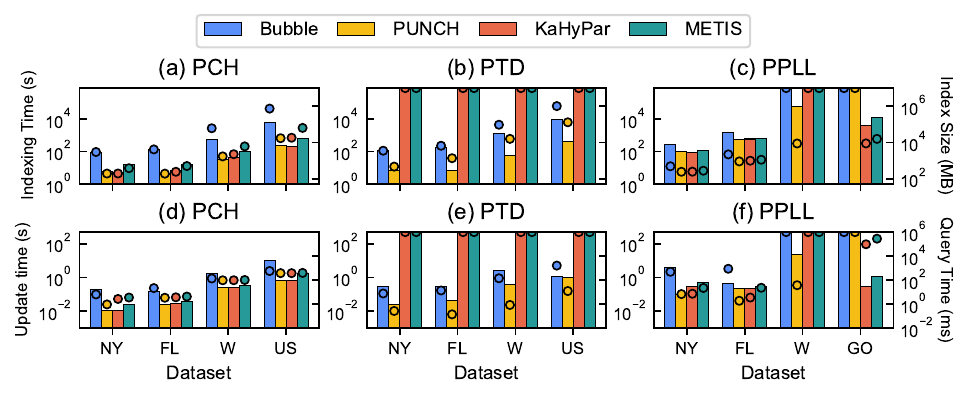}
	\caption{Effect of Partition Methods. Bar: Left, Ball: Right}
	\label{fig:partition_method}
\end{figure}



\subsection{Experiment Result Analysis}\label{subsec:ExperimentResults}

\textbf{Exp 1: Performance of Partition Methods When Varying $k$.} 
As shown in Figure~\ref{fig:partitionInfo} (a)-(p), the overall border number $|B|$ generally increases while the average border number $\overline{|B_i|}$ decreases with the increase of $k$. This indicates that having a larger $k$ can potentially speed up the query processing of the partition index due to a smaller average partition border number. However, it also leads to a larger overlay graph, which can slow down the query processing and construction of the overlay index.
Therefore, there is a trade-off between them.
Furthermore, as per Figure~\ref{fig:partitionInfo} (q)-(x), among the partition methods used, only \textit{PUNCH} and \textit{Bubble} have a partition connectivity ratio $R_C$ of 1. This implies that the partitions generated by other partition methods generally contain multiple unconnected components. Since \textit{PTD} requires the vertices within the same partition to be connected, only \textit{PUNCH} and Bubble suit it.
Overall, \textit{PUNCH} and \textit{KaHyPar} perform best in $|B|$ and $\overline{|B_i|}$ on road networks. When dealing with complex networks, it is advisable to use \textit{KaHyPar}, \textit{METIS}, or \textit{HEP}. In the case of the largest graph FR, only \textit{HEP} and \textit{CLUGP} are capable of completing graph partitioning within 24 hours.



\textbf{Exp 2: Effect of Partition Method on PSP Index.}
Next, we choose \textit{Bubble}, \textit{PUNCH}, \textit{KaHyPar}, and \textit{METIS} as the partition methods according to Exp 1 and evaluate their effect on PSP indexes.
We test \textit{PCH} and \textit{PTD} on road networks NY, FL, W, and US since they are infeasible on complex networks with large treewidth.
We test \textit{PPLL} on road networks and complex network GO and do not present the result on US since \textit{PPLL} cannot finish index construction within 24 hours for any partition methods. 
Figure~\ref{fig:partition_method} shows that \textit{PUNCH} outperforms other partition methods in most cases due to smaller border numbers $|B|$ and $\overline{B_i}$ as evidenced in Exp 1, which reveals that they are the most important factors in the partition method selection. 
For complex networks that \textit{PUNCH} cannot be applied, \textit{KaHyPar} has the best performance.
We have tried to test \textit{PPLL} on larger complex networks. However, it fails to complete the test within 24 hours due to the large and condensed overlay graph, which reveals that the planar PSP index is generally unsuitable for complex networks.
Therefore, we select \textit{PUNCH} and \textit{KaHyPar} as the default planar partition methods for road networks and complex networks, respectively.
Besides, we also evaluate the effect of partition number 
as presented in the appendix~\cite{FullVersion} due to limited space. In short, we set the default partition number as 32 for all datasets since it provided satisfactory performance in most cases.
\begin{table}[]
\footnotesize
\setlength\tabcolsep{1pt}
\caption{Effectiveness of Overlay Optimization}
\label{table:prune}
\begin{tabular}{|lllr|rrr|rrr|rrr|}
\hline
\multicolumn{4}{|c|}{\textbf{PSP   Index}}                                                                                                                                                & \multicolumn{3}{c|}{\textbf{\textit{PCH}}}                                                                                  & \multicolumn{3}{c|}{\textbf{\textit{PTD}}}                                                                                  & \multicolumn{3}{c|}{\textbf{\textit{PPLL}}}                                                                                   \\ \hline
\multicolumn{1}{|l|}{\textbf{Dataset}}             & \multicolumn{2}{c|}{\textbf{Method}}                                                          & \multicolumn{1}{r|}{\textbf{$\tilde{D}$}} & \multicolumn{1}{r|}{\textbf{$t_c$}}   & \multicolumn{1}{r|}{\textbf{$t_q$}}  & \multicolumn{1}{r|}{\textbf{$t_u$}} & \multicolumn{1}{r|}{\textbf{$t_c$}}   & \multicolumn{1}{r|}{\textbf{$t_q$}}  & \multicolumn{1}{r|}{\textbf{$t_u$}} & \multicolumn{1}{r|}{\textbf{$t_c$}}   & \multicolumn{1}{r|}{\textbf{$t_q$}}    & \multicolumn{1}{r|}{\textbf{$t_u$}} \\ \hline
\multicolumn{1}{|l|}{\multirow{4}{*}{NY}} & \multicolumn{1}{l|}{\multirow{2}{*}{Bubble}} & \multicolumn{1}{l|}{w/o opt} & 187.45                             & \multicolumn{1}{r|}{112.8}          & \multicolumn{1}{r|}{18.04}          & 0.20                             & \multicolumn{1}{r|}{134.5}          & \multicolumn{1}{r|}{18.06}          & 0.42                             & \multicolumn{1}{r|}{70.27}           & \multicolumn{1}{r|}{1743}          & 11.77                            \\ \cline{3-13} 
\multicolumn{1}{|l|}{}                             & \multicolumn{1}{l|}{}                                 & \multicolumn{1}{l|}{w/ opt}  & \textbf{163.76}                     & \multicolumn{1}{r|}{\textbf{104.2}} & \multicolumn{1}{r|}{\textbf{16.90}} & \textbf{0.19}                      & \multicolumn{1}{r|}{\textbf{128.0}} & \multicolumn{1}{r|}{\textbf{17.85}} & \textbf{0.38}                      & \multicolumn{1}{r|}{\textbf{68.03}}  & \multicolumn{1}{r|}{\textbf{1561}} & \textbf{10.00}                     \\ \cline{2-13} 
\multicolumn{1}{|l|}{}                             & \multicolumn{1}{l|}{\multirow{2}{*}{PUNCH}}  & \multicolumn{1}{l|}{w/o opt} & 53.53                             & \multicolumn{1}{r|}{4.67}            & \multicolumn{1}{r|}{3.29}           & 0.02                               & \multicolumn{1}{r|}{5.33}            & \multicolumn{1}{r|}{0.32}           & 0.016                               & \multicolumn{1}{r|}{25.94}           & \multicolumn{1}{r|}{9.70}             & 0.19                              \\ \cline{3-13} 
\multicolumn{1}{|l|}{}                             & \multicolumn{1}{l|}{}                                 & \multicolumn{1}{l|}{w/ opt}  & \textbf{52.68}                      & \multicolumn{1}{r|}{\textbf{4.65}}   & \multicolumn{1}{r|}{\textbf{3.21}}  & \textbf{0.01}                      & \multicolumn{1}{r|}{\textbf{5.28}}   & \multicolumn{1}{r|}{\textbf{0.30}}  & \textbf{0.02}                      & \multicolumn{1}{r|}{\textbf{25.87}}  & \multicolumn{1}{r|}{\textbf{9.56}}    & \textbf{0.18}                      \\ \hline
\multicolumn{1}{|l|}{\multirow{4}{*}{FL}} & \multicolumn{1}{l|}{\multirow{2}{*}{Bubble}} & \multicolumn{1}{l|}{w/o opt} & 235.80                             & \multicolumn{1}{r|}{143.6}          & \multicolumn{1}{r|}{44.56}          & 0.06                             & \multicolumn{1}{r|}{195.9}          & \multicolumn{1}{r|}{33.75}          & 0.12                           & \multicolumn{1}{r|}{748.7}          & \multicolumn{1}{r|}{3032}          & 1.28                              \\ \cline{3-13} 
\multicolumn{1}{|l|}{}                             & \multicolumn{1}{l|}{}                                 & \multicolumn{1}{l|}{w/ opt}  & \textbf{211.87}                     & \multicolumn{1}{r|}{\textbf{138.0}} & \multicolumn{1}{r|}{\textbf{38.82}} & \textbf{0.05}                      & \multicolumn{1}{r|}{\textbf{187.8}} & \multicolumn{1}{r|}{\textbf{31.37}} & \textbf{0.09}                      & \multicolumn{1}{r|}{\textbf{742.4}} & \multicolumn{1}{r|}{\textbf{2921}} & \textbf{0.98}                      \\ \cline{2-13} 
\multicolumn{1}{|l|}{}                             & \multicolumn{1}{l|}{\multirow{2}{*}{PUNCH}}  & \multicolumn{1}{l|}{w/o opt} & 37.42                               & \multicolumn{1}{r|}{3.57}            & \multicolumn{1}{r|}{13.98}          & \textbf{0.03}                      & \multicolumn{1}{r|}{5.69}            & \multicolumn{1}{r|}{0.17}           & 0.05                               & \multicolumn{1}{r|}{368.7}          & \multicolumn{1}{r|}{4.83}             & \textbf{0.24}                              \\ \cline{3-13} 
\multicolumn{1}{|l|}{}                             & \multicolumn{1}{l|}{}                                 & \multicolumn{1}{l|}{w/ opt}  & \textbf{37.26}                      & \multicolumn{1}{r|}{\textbf{3.43}}   & \multicolumn{1}{r|}{\textbf{13.81}} & \textbf{0.03}                      & \multicolumn{1}{r|}{\textbf{5.61}}   & \multicolumn{1}{r|}{\textbf{0.16}}  & \textbf{0.04}                      & \multicolumn{1}{r|}{\textbf{365.7}} & \multicolumn{1}{r|}{\textbf{4.34}}    & \textbf{0.24}                      \\ \hline
\end{tabular}

$\tilde{D}$: average vertex degree of $\tilde{G}$; 
$t_c$: indexing time (s); $t_q$: query time (ms); $t_u$: index update time (s).
\end{table}

\begin{figure}[t]
	\centering
	\includegraphics[width=0.85\linewidth]{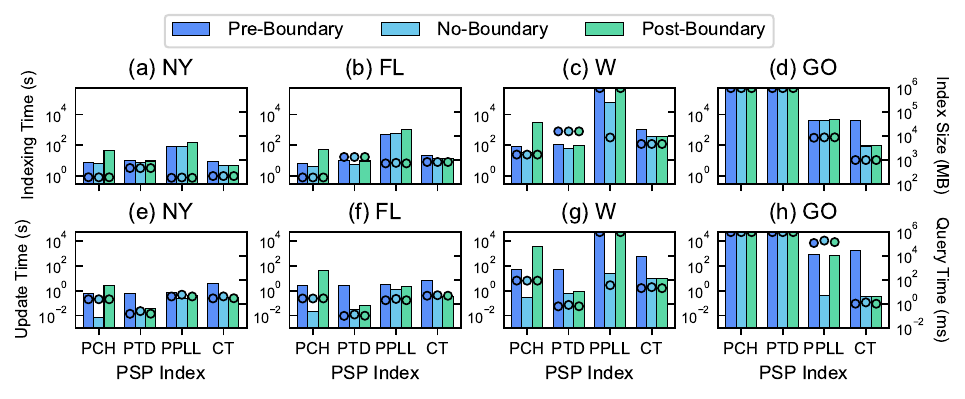}
	\caption{Effectiveness of PSP Strategy. Bar: Left, Ball: Right}
	\label{fig:boundary_strategy}
\end{figure}

\textbf{Exp 3: Effectiveness of Overlay Graph Optimization.}
We evaluate the effectiveness of overlay graph optimization by reporting $t_c$, $t_q$, $t_u$, and average degree $\tilde{D}$ of $\tilde{G}$ on NY and FL.
We do not use large networks since they fail to report the results of \textit{PCH}, \textit{PTD}, and \textit{PPLL} simultaneously.
As shown in Table~\ref{table:prune}, it enhances the index performance in most cases with evident improvement in Bubble-based partitions (achieving $14.6\%$, $8.3\%$, $14.7\%$, and $17.8\%$ speed-up on $\tilde{D}$, $t_c$, $t_q$, and $t_u$). 
The improvement on \textit{PUNCH} is marginal since it already has a great partition result. 
Nevertheless, this optimization improves the index performance by reducing the overlay graph density with a theoretical guarantee of the index correctness and we will use it by default in the following experiments.

\begin{figure}[t]
	\centering
	\includegraphics[width=0.8\linewidth]{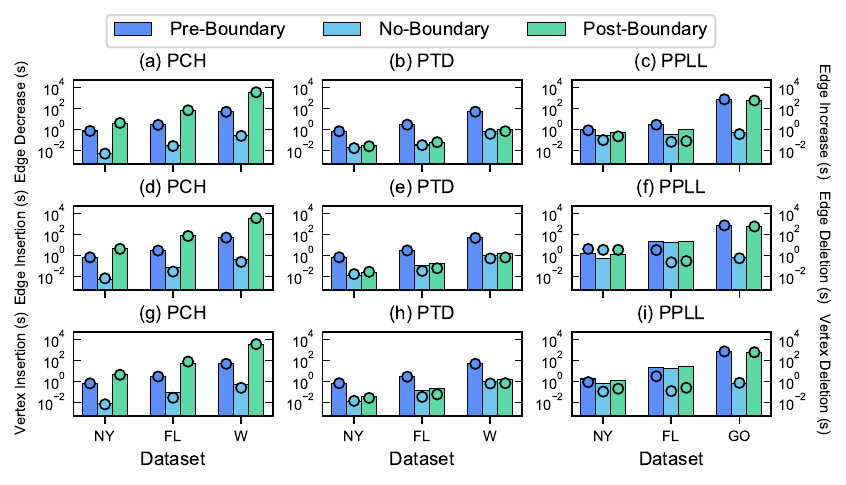}
	\caption{Performance of Different Index Update Types. Bar: Left, Ball: Right}
	\label{fig:indexUpdateType}
\end{figure}

\begin{figure}[t]
	\centering
	\includegraphics[width=0.8\linewidth]{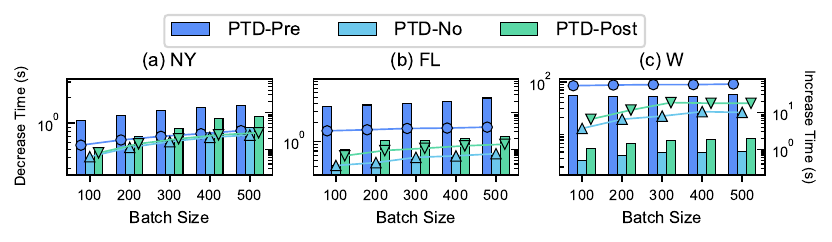}
	\caption{Edge Weight Batch Updates of \textit{PTD}. Bar: Left, Polyline: Right}
	\label{fig:updateVaryBatch}
\end{figure}

\begin{figure}[t]
	\centering
	\includegraphics[width=0.75\linewidth]{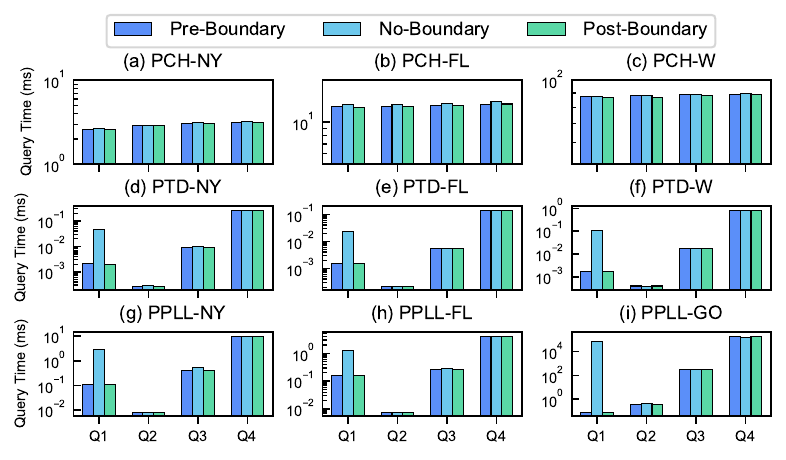}
	\caption{Performance of Different Query Types}
	\label{fig:query_type}
\end{figure}

\textbf{Exp 4: Effectiveness of PSP Strategy.}
We evaluate the effectiveness of the proposed PSP strategies on \textit{PCH}, \textit{PTD}, \textit{PPLL}, and \textit{CT} in Figure~\ref{fig:boundary_strategy}. The no-boundary always the fastest in update, achieving up to $185.2\times$, $91.3\times$, $1759.4\times$, and $5358.8\times$ faster than the pre-boundary for \textit{PCH}, \textit{PTD}, \textit{PPLL}, and \textit{CT}.
Moreover, it also outperforms the pre-boundary in index construction time in \textit{PCH} and \textit{PTD} with up to $2.4\times$ and $1.7\times$ faster.
This is because no-boundary replaces the time-consuming Dijkstra's searches among the boundary vertices with index query.
Nevertheless, it is slow in query since the same-partition queries require concatenating the overlay and partition indexes.
The post-boundary addresses this drawback by repairing the partition indexes, achieving equivalent query time as pre-boundary.
Moreover, since it needs to repair all-boundary-pair distances, it is time-consuming when the query efficiency of $\tilde{L}$ is low.
For example, because the CH query processing is relatively slow, the index construction and maintenance of \textit{PCH} with post-boundary is even slower than the pre-boundary as it involves twice all-pair searches among the boundaries. Nevertheless, the post-boundary could achieve a significant update efficiency improvement ($14.6-59.3\times$ for \textit{PTD}) for \textit{PTD}, \textit{PPLL}, and \textit{CT} that use 2-hop labeling (TD and PLL) as the underlying SP index. 

\textbf{Exp 5: Evaluation of Update Types.}
As shown in Figure \ref{fig:indexUpdateType}, 
\textit{no-boundary} is much faster (up to $185.2\times$, $91.3\times$, $1759.4\times$ for \textit{PCH}, \textit{PTD}, and \textit{PPLL}) than pre-boundary in all update types.
The post-boundary works better than the pre-boundary on \textit{PTD} and \textit{PPLL} since it replaces all-pair searches with index query. Note that the performance of edge/vertex insertion (deletion) is almost equivalent to edge decrease (increase) since this structural update essentially shares a similar update mechanism with the edge weight update. Furthermore, the batch update for \textit{PTD} is tested by varying batch size from 100 to 500. 
As shown in Figure \ref{fig:updateVaryBatch}, \textit{PTD-No} and \textit{PTD-Post} always outperform \textit{PTD-Pre}. 
As the batch size increases, the improvement decreases since the all-pair boundary searches in the pre-boundary could be shared by different edge updates.
Nonetheless, \textit{PTD-No} and \textit{PTD-Post} also share the computation for different edge updates during the all-pair boundary distance check and outperform \textit{PTD-Pre}. 

\textbf{Exp 6: Performance of Different Query Types.}
We classify the queries into four types based on $s$ and $t$ locations: 
Q1 (same partition), Q2 (both boundary), Q3 (one boundary one non-boundary), Q4 (different partitions). 
As shown in Figure~\ref{fig:query_type}, for \textit{PTD} and \textit{PPLL}, Q2 is the fastest since it only uses $\tilde{L}$. By contrast, Q3 and Q4 are much slower due to distance concatenation.
For Q1, the post-boundary is the same with pre-boundary since they both only use the partition index, while the no-boundary has to concatenate $\tilde{L}$ and $\{L_i\}$. Lastly, \textit{PCH} has similar query efficiency on different query types and strategies due to its search manner.


\begin{figure}[t]
	\centering
    \includegraphics[width=0.75\linewidth]{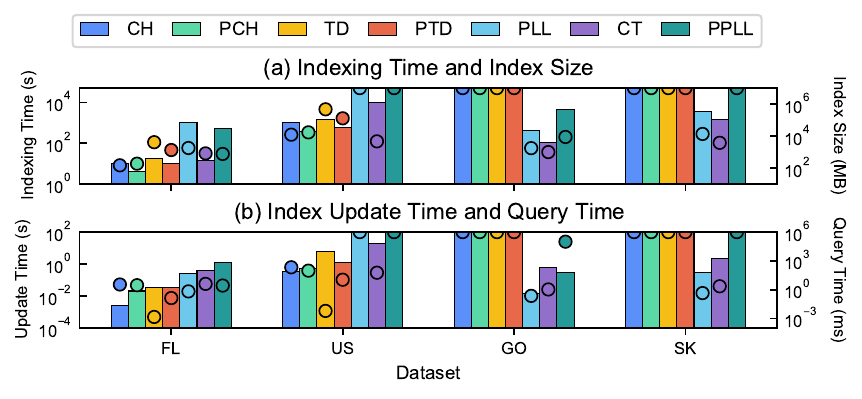}
	\caption{Comparison with Non-partitioned SP Indexes. (Bar: Left axis, Ball: Right axis)}
	\label{fig:nonPartitionedSP}
\end{figure}

\begin{figure}[t]
	\centering
    \includegraphics[width=0.95\linewidth]{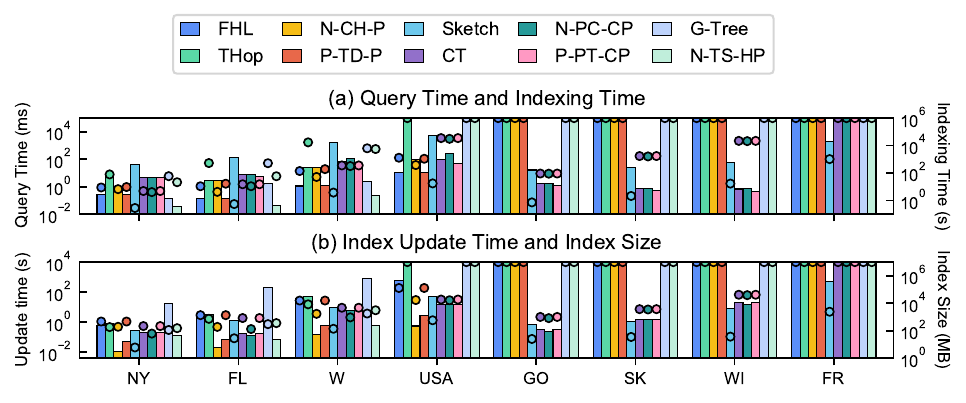}
	\caption{PSP Indexes Comparison. Bar: Left, Ball: Right}
	\label{fig:PSPIndexComparison}
\end{figure}




\textbf{Exp 7: Comparison with Non-partitioned SP Algorithms.} 
We conduct a comparison between PSP indexes (including \textit{PCH}, \textit{PTD}, \textit{PPLL}, and \textit{CT}) and their non-partitioned counterparts (namely CH~\cite{CH_geisberger2008contraction}, TD~\cite{H2H_ouyang2018hierarchy}, PLL~\cite{PLL_akiba2013fast}). 
As depicted in Figure~\ref{fig:nonPartitionedSP}, PSP indexes generally provide faster indexing time, smaller index size, and better scalability than their non-partitioned counterparts.
For instance, \textit{PCH} and \textit{PTD} exhibit up to $2.76\times$ and $2.25\times$ faster index construction compared to CH and TD, respectively. Moreover, \textit{CT} demonstrates $72.3\times$ and $3.58\times$ improvement over PLL in terms of indexing time and index size, respectively. However, PSP indexes generally compromise query processing efficiency and even index update time (except for \textit{PTD}). 
Besides, it can also be seen that planar structure is more suited for road networks, while core-periphery structure is better suited for complex networks, as only \textit{CT} can finish tests on complex networks while \textit{PPLL} cannot.

\textbf{Exp 8: Comparison of All PSP Algorithms.} 
We compare our proposed \textit{PSP} indexes with the existing ones in Figure~\ref{fig:PSPIndexComparison}. 
We first analyze the performance of our proposed indexes:
1) \textit{P-TD-P} has the same query efficiency and index size as \textit{FHL} but is much faster to update with $11.1-202.7\times$ speed-up;
2) \textit{N-CH-P} has the fastest index construction and maintenance efficiency and the smallest index size among all planar PSP indexes, achieving up to $3.4\times$, $1004\times$, and $7.6\times$ speed-up over FHL;
3) \textit{N-PC-CP} has faster index construction and update than \textit{CT} since it utilizes faster CH for tree index. \textit{P-PT-C}P outperforms \textit{CT} in terms of query processing as it is faster in same-partition queries;
4) \textit{N-TS-HP} has faster index construction, smaller index size, faster query and update than G-Tree by leveraging TD for overlay index construction;
5) Sketch can scale up to very large graphs (the only algorithm that can be applied on graph FR), but its query efficiency is much slower than N-PC-CP and P-PT-CP due to the light index.

We next discuss the advantages of different partition structures. Figure~\ref{fig:PSPIndexComparison}-(b) shows that the state-of-the-art planar PSP indexes (\textit{N-CH-P}, \textit{P-TD-P}) and hierarchical PSP indexes (\textit{N-TS-HP}) generally provide better query and index update efficiency for road networks than core-periphery PSP indexes. However, only core-periphery PSP indexes can be used on complex networks, which typically have large treewidth. Therefore, planar and hierarchical partitions are more suitable for road networks, while complex networks work better with core-periphery partitions.
\subsection{PSP Application Guidance}
\label{subsec:ExpeGuidance}
Based on the experimental analysis above, we clarify the misconceptions mentioned in Section~\ref{sec:intro} and conclude the following observations: 1) All PSP indexes are essentially interrelated because they all consist of three dimensions in our PSP index scheme: graph partition method, PSP strategy, and shortest path algorithm. Moreover, we could combine the elements of these three dimensions to create a new PSP index;
2) It is important to choose the right partition structure for different types of graphs. Generally, for road networks, planar and hierarchical PSP indexes are more suitable than core-periphery PSP due to their efficient query and index update efficiency. On the other hand, core-periphery PSP indexes are more suitable for complex networks due to their excellent scalability;
3) Although PSP indexes typically have faster indexing efficiency, smaller index size, and better scalability than their non-partitioned counterparts, they sacrifice query and index update efficiency.

\begin{figure}[t]
	\centering
    \includegraphics[width=0.9\linewidth]{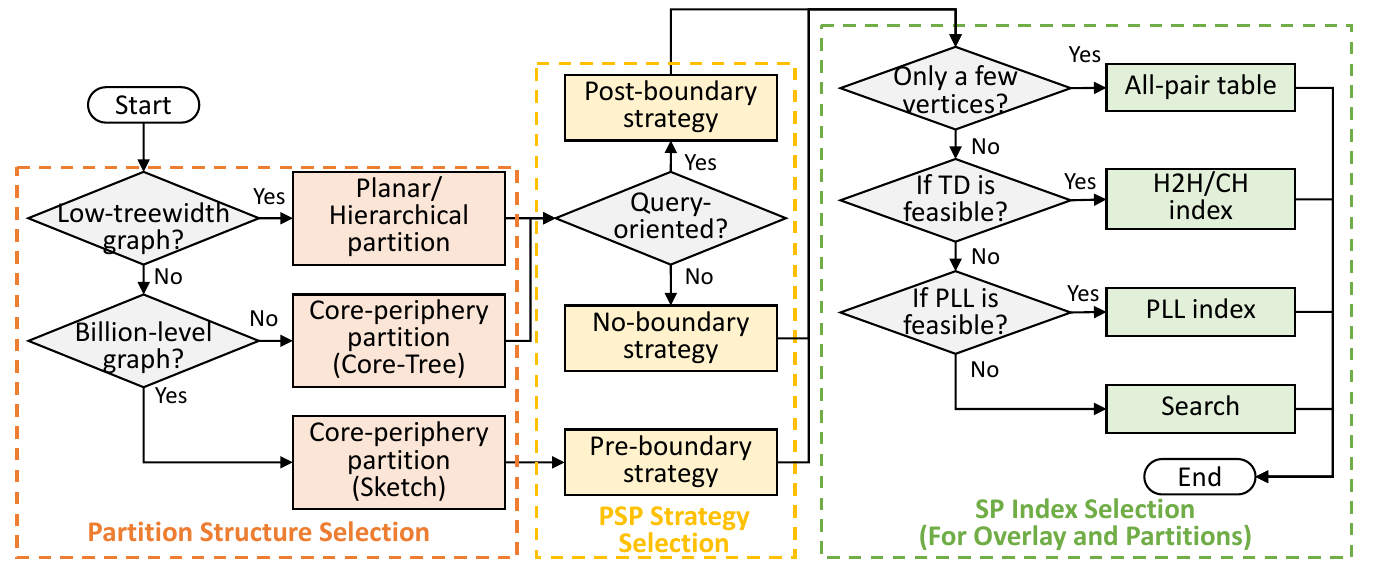}
	\caption{Our Decision Tree for \textit{PSP} Index Design}
	\label{fig:DecisionTree}
\end{figure}

According to our theoretical analysis and experimental results, we further provide our decision tree for PSP index design as illustrated in Figure \ref{fig:DecisionTree}. 
The first step in creating a new PSP index is to select a partition structure that is suitable for the specific application, such as graph type and size. For graphs with low treewidth, such as road networks, we recommend using planar or hierarchical partition structures. For larger graphs, we suggest using the core-periphery structure, with the core-tree structure for million-level graphs and sketch for billion-level graphs. 
The second step is to select the PSP strategy based on the required query processing and index update efficiency. For query-oriented applications, we recommend using the post-boundary strategy, while for update-oriented applications, the no-boundary strategy is more suitable.
The third step is to choose the SP algorithm for the overlay and partitions. For a small number of vertices, we recommend using the most efficient all-pair table. For larger graphs, if tree decomposition is feasible, we suggest using TD or CH to achieve better query/update efficiency. If tree decomposition is infeasible, we recommend selecting PLL or even the direct search for SP algorithm selection.

\section{Future Directions}
\label{sec:FutureDirection}
The insightful structure-based partition classification and universal PSP scheme proposed in this paper provide a systematical way to analyze and design the PSP index. Besides, the novel no-boundary and post-boundary strategies significantly enrich the design space for the PSP index. 
However, there are still some unresolved problems with the PSP index. Next, we will discuss some future directions and open challenges to inspire researchers in this field.

\textbf{1) I/O-efficient and Distributed PSP Algorithms}. There are two general ways to scale up graph algorithms: scale up to external memory on a single machine (out-of-core PSP index), or scale out to a distributed cluster (distributed PSP index). 
Since the PSP indexes naturally decompose a graph into several subgraphs, we could regard each subgraph (or its partition index) as a whole for disk accessory to reduce the I/O numbers or distribute each subgraph to a server in the cluster to reduce communication. Nevertheless, how to partition the graph for fewer I/O numbers or communication workload according to the network structure and algorithm procedure is still an open challenge. Moreover, how to optimize the PSP strategy and underlying SP algorithm simultaneously according to the I/O accessory or communication pattern is also an open challenge.

\textbf{2) Automatic PSP Index Design System}. Although we have propose a universal PSP index scheme and insightful guideline for the PSP index design, they are essentially experience-based. In the era of Artificial Intelligence (AI), how to leverage AI to automatically design the PSP index according to the system requirements is an open challenge. Moreover, how to make the elements of different dimensions in PSP index scheme portable such that the system could react promptly with the evolving system requirements is also an open challenge.

\textbf{3) High Throughput System Based on PSP Index}.
Another future direction is to leverage PSP index for high throughput shortest path query processing. For example, road networks generally incur high throughput queries while the edge weight of road networks (travel time) could dynamically changes due to evolving traffic conditions. Therefore, how to achieve fast index maintenance and query processing simultaneously is important to enhance the system throughput. The PSP index could leverage parallelization among partition indexes to speed up the index update. Besides, adopting 2-hop labeling as the underlying SP algorithm also contributes to a fast query processing, making the PSP index a promising option for building path-finding system with high throughput. Nevertheless, how to design such a PSP index for high throughput query processing on dynamic road networks is an open challenge. 
Besides, partial edge updates in a partition may not influence all queries on other partitions. Thus how to identify the affected queries such that the system could still leverage the unaffected index for fast query processing is another open challenge.


\section{Conclusions}
\label{sec:Conclusion}
In this work, we systematically study the partitioned shortest path index and propose a universal PSP scheme by considering three dimensions: SP algorithm, partition method, and PSP strategy. 
We first briefly review the SP algorithms used in the PSP index and then propose a new structure-based classification to facilitate the selection of the partition method and conduct a comprehensive review of corresponding partition methods and PSP indexes. 
For the PSP strategy, we summarize the classic pre-boundary strategy and then propose two new strategies and pruning-based overlay graph optimization to improve the index performance. We also provide index maintenance methods for all PSP strategies. 
Furthermore, we construct our PSP index scheme by seamlessly coupling those three dimensions. We also propose five new PSP indexes that are either more efficient in query or update than the current state-of-the-art. Systematic and extensive experimental evaluations demonstrate the effectiveness of our proposed techniques and new indexes guided by the insightful PSP index scheme. 
An insightful PSP index design guideline is provided to help future researchers and practitioners design their indexes.
Lastly, the future directions and open challenges for the PSP index are also discussed.

\balance

\bibliographystyle{ACM-Reference-Format}
\bibliography{Reference}


\begin{thebibliography}{127}


\ifx \showCODEN    \undefined \def \showCODEN     #1{\unskip}     \fi
\ifx \showDOI      \undefined \def \showDOI       #1{#1}\fi
\ifx \showISBNx    \undefined \def \showISBNx     #1{\unskip}     \fi
\ifx \showISBNxiii \undefined \def \showISBNxiii  #1{\unskip}     \fi
\ifx \showISSN     \undefined \def \showISSN      #1{\unskip}     \fi
\ifx \showLCCN     \undefined \def \showLCCN      #1{\unskip}     \fi
\ifx \shownote     \undefined \def \shownote      #1{#1}          \fi
\ifx \showarticletitle \undefined \def \showarticletitle #1{#1}   \fi
\ifx \showURL      \undefined \def \showURL       {\relax}        \fi
\providecommand\bibfield[2]{#2}
\providecommand\bibinfo[2]{#2}
\providecommand\natexlab[1]{#1}
\providecommand\showeprint[2][]{arXiv:#2}

\bibitem[Abraham et~al\mbox{.}(2011)]%
        {abraham2011hub}
\bibfield{author}{\bibinfo{person}{Ittai Abraham}, \bibinfo{person}{Daniel Delling}, \bibinfo{person}{Andrew~V Goldberg}, {and} \bibinfo{person}{Renato~F Werneck}.} \bibinfo{year}{2011}\natexlab{}.
\newblock \showarticletitle{A hub-based labeling algorithm for shortest paths in road networks}. In \bibinfo{booktitle}{\emph{Experimental Algorithms: 10th International Symposium, SEA 2011, Kolimpari, Chania, Crete, Greece, May 5-7, 2011. Proceedings 10}}. Springer, \bibinfo{pages}{230--241}.
\newblock


\bibitem[Akhremtsev et~al\mbox{.}(2017)]%
        {kahypar_ahss2017alenex}
\bibfield{author}{\bibinfo{person}{Yaroslav Akhremtsev}, \bibinfo{person}{Tobias Heuer}, \bibinfo{person}{Peter Sanders}, {and} \bibinfo{person}{Sebastian Schlag}.} \bibinfo{year}{2017}\natexlab{}.
\newblock \showarticletitle{Engineering a direct \emph{k}-way Hypergraph Partitioning Algorithm}. In \bibinfo{booktitle}{\emph{19th Workshop on Algorithm Engineering and Experiments, (ALENEX 2017)}}. \bibinfo{pages}{28--42}.
\newblock


\bibitem[Akiba et~al\mbox{.}(2013)]%
        {PLL_akiba2013fast}
\bibfield{author}{\bibinfo{person}{Takuya Akiba}, \bibinfo{person}{Yoichi Iwata}, {and} \bibinfo{person}{Yuichi Yoshida}.} \bibinfo{year}{2013}\natexlab{}.
\newblock \showarticletitle{Fast exact shortest-path distance queries on large networks by pruned landmark labeling}. In \bibinfo{booktitle}{\emph{Proceedings of the 2013 ACM SIGMOD International Conference on Management of Data}}. \bibinfo{pages}{349--360}.
\newblock


\bibitem[Akiba et~al\mbox{.}(2014)]%
        {IncPLL_akiba2014dynamic}
\bibfield{author}{\bibinfo{person}{Takuya Akiba}, \bibinfo{person}{Yoichi Iwata}, {and} \bibinfo{person}{Yuichi Yoshida}.} \bibinfo{year}{2014}\natexlab{}.
\newblock \showarticletitle{Dynamic and historical shortest-path distance queries on large evolving networks by pruned landmark labeling}. In \bibinfo{booktitle}{\emph{Proceedings of the 23rd international conference on World wide web}}. \bibinfo{pages}{237--248}.
\newblock


\bibitem[Barat et~al\mbox{.}(2018)]%
        {Scotch18}
\bibfield{author}{\bibinfo{person}{Remi Barat}, \bibinfo{person}{C{\'e}dric Chevalier}, {and} \bibinfo{person}{Fran{\c c}ois Pellegrini}.} \bibinfo{year}{2018}\natexlab{}.
\newblock \showarticletitle{Multi-criteria Graph Partitioning with Scotch}. In \bibinfo{booktitle}{\emph{2018 Proceedings of the SIAM Workshop on Combinatorial Scientific Computing (CSC)}}. \bibinfo{pages}{66--75}.
\newblock


\bibitem[Barnard and Simon(1994)]%
        {barnard1994fast}
\bibfield{author}{\bibinfo{person}{Stephen~T Barnard} {and} \bibinfo{person}{Horst~D Simon}.} \bibinfo{year}{1994}\natexlab{}.
\newblock \showarticletitle{Fast multilevel implementation of recursive spectral bisection for partitioning unstructured problems}.
\newblock \bibinfo{journal}{\emph{Concurrency: Practice and experience}} \bibinfo{volume}{6}, \bibinfo{number}{2} (\bibinfo{year}{1994}), \bibinfo{pages}{101--117}.
\newblock


\bibitem[Bast et~al\mbox{.}(2016)]%
        {bast2016route}
\bibfield{author}{\bibinfo{person}{Hannah Bast}, \bibinfo{person}{Daniel Delling}, \bibinfo{person}{Andrew Goldberg}, \bibinfo{person}{Matthias M{\"u}ller-Hannemann}, \bibinfo{person}{Thomas Pajor}, \bibinfo{person}{Peter Sanders}, \bibinfo{person}{Dorothea Wagner}, {and} \bibinfo{person}{Renato~F Werneck}.} \bibinfo{year}{2016}\natexlab{}.
\newblock \showarticletitle{Route planning in transportation networks}.
\newblock In \bibinfo{booktitle}{\emph{Algorithm engineering}}. \bibinfo{publisher}{Springer}, \bibinfo{pages}{19--80}.
\newblock


\bibitem[Bast et~al\mbox{.}(2007)]%
        {TNR_bast2007transit}
\bibfield{author}{\bibinfo{person}{Holger Bast}, \bibinfo{person}{Stefan Funke}, \bibinfo{person}{Domagoj Matijevic}, \bibinfo{person}{Peter Sanders}, {and} \bibinfo{person}{Dominik Schultes}.} \bibinfo{year}{2007}\natexlab{}.
\newblock \showarticletitle{In transit to constant time shortest-path queries in road networks}. In \bibinfo{booktitle}{\emph{2007 Proceedings of the Ninth Workshop on Algorithm Engineering and Experiments (ALENEX)}}. SIAM, \bibinfo{pages}{46--59}.
\newblock


\bibitem[Bauer and Delling(2010)]%
        {bauer2010sharc}
\bibfield{author}{\bibinfo{person}{Reinhard Bauer} {and} \bibinfo{person}{Daniel Delling}.} \bibinfo{year}{2010}\natexlab{}.
\newblock \showarticletitle{SHARC: Fast and robust unidirectional routing}.
\newblock \bibinfo{journal}{\emph{Journal of Experimental Algorithmics (JEA)}}  \bibinfo{volume}{14} (\bibinfo{year}{2010}), \bibinfo{pages}{2--4}.
\newblock


\bibitem[Berger and Bokhari(1987)]%
        {BB87RCB}
\bibfield{author}{\bibinfo{person}{Marsha~J Berger} {and} \bibinfo{person}{Shahid~H Bokhari}.} \bibinfo{year}{1987}\natexlab{}.
\newblock \showarticletitle{A Partitioning Strategy for Nonuniform Problems on Multiprocessors}.
\newblock \bibinfo{journal}{\emph{IEEE Trans. Comput.}} \bibinfo{volume}{C-36}, \bibinfo{number}{5} (\bibinfo{year}{1987}), \bibinfo{pages}{570--580}.
\newblock


\bibitem[Berry et~al\mbox{.}(2003)]%
        {MDE_berry2003minimum}
\bibfield{author}{\bibinfo{person}{Anne Berry}, \bibinfo{person}{Pinar Heggernes}, {and} \bibinfo{person}{Genevieve Simonet}.} \bibinfo{year}{2003}\natexlab{}.
\newblock \showarticletitle{The minimum degree heuristic and the minimal triangulation process}. In \bibinfo{booktitle}{\emph{International Workshop on Graph-Theoretic Concepts in Computer Science}}. Springer, \bibinfo{pages}{58--70}.
\newblock


\bibitem[Boccaletti et~al\mbox{.}(2006)]%
        {boccaletti2006complex}
\bibfield{author}{\bibinfo{person}{Stefano Boccaletti}, \bibinfo{person}{Vito Latora}, \bibinfo{person}{Yamir Moreno}, \bibinfo{person}{Martin Chavez}, {and} \bibinfo{person}{D-U Hwang}.} \bibinfo{year}{2006}\natexlab{}.
\newblock \showarticletitle{Complex networks: Structure and dynamics}.
\newblock \bibinfo{journal}{\emph{Physics reports}} \bibinfo{volume}{424}, \bibinfo{number}{4-5} (\bibinfo{year}{2006}), \bibinfo{pages}{175--308}.
\newblock


\bibitem[Borgatti and Everett(2000)]%
        {borgatti2000models}
\bibfield{author}{\bibinfo{person}{Stephen~P Borgatti} {and} \bibinfo{person}{Martin~G Everett}.} \bibinfo{year}{2000}\natexlab{}.
\newblock \showarticletitle{Models of core/periphery structures}.
\newblock \bibinfo{journal}{\emph{Social networks}} \bibinfo{volume}{21}, \bibinfo{number}{4} (\bibinfo{year}{2000}), \bibinfo{pages}{375--395}.
\newblock


\bibitem[Bourse et~al\mbox{.}(2014)]%
        {bourse2014balanced}
\bibfield{author}{\bibinfo{person}{Florian Bourse}, \bibinfo{person}{Marc Lelarge}, {and} \bibinfo{person}{Milan Vojnovic}.} \bibinfo{year}{2014}\natexlab{}.
\newblock \showarticletitle{Balanced graph edge partition}. In \bibinfo{booktitle}{\emph{Proceedings of the 20th ACM SIGKDD international conference on Knowledge discovery and data mining}}. \bibinfo{pages}{1456--1465}.
\newblock


\bibitem[Bulu{\c{c}} et~al\mbox{.}(2016)]%
        {bulucc2016recent}
\bibfield{author}{\bibinfo{person}{Ayd{\i}n Bulu{\c{c}}}, \bibinfo{person}{Henning Meyerhenke}, \bibinfo{person}{Ilya Safro}, \bibinfo{person}{Peter Sanders}, {and} \bibinfo{person}{Christian Schulz}.} \bibinfo{year}{2016}\natexlab{}.
\newblock \showarticletitle{Recent advances in graph partitioning}.
\newblock \bibinfo{journal}{\emph{Algorithm engineering}} (\bibinfo{year}{2016}), \bibinfo{pages}{117--158}.
\newblock


\bibitem[{\c{C}}ataly{\"u}rek and Aykanat(2011)]%
        {atalyrek2011PaToH}
\bibfield{author}{\bibinfo{person}{{\"U}mit {\c{C}}ataly{\"u}rek} {and} \bibinfo{person}{Cevdet Aykanat}.} \bibinfo{year}{2011}\natexlab{}.
\newblock \showarticletitle{PaToH (Partitioning Tool for Hypergraphs)}. In \bibinfo{booktitle}{\emph{Encyclopedia of Parallel Computing}}. \bibinfo{pages}{1479--1487}.
\newblock


\bibitem[Chen et~al\mbox{.}(2021)]%
        {chen2021p2h}
\bibfield{author}{\bibinfo{person}{Zitong Chen}, \bibinfo{person}{Ada Wai-Chee Fu}, \bibinfo{person}{Minhao Jiang}, \bibinfo{person}{Eric Lo}, {and} \bibinfo{person}{Pengfei Zhang}.} \bibinfo{year}{2021}\natexlab{}.
\newblock \showarticletitle{P2h: Efficient distance querying on road networks by projected vertex separators}. In \bibinfo{booktitle}{\emph{Proceedings of the 2021 International Conference on Management of Data}}. \bibinfo{pages}{313--325}.
\newblock


\bibitem[Chondrogiannis and Gamper(2014)]%
        {PbS_chondrogiannis2014exploring}
\bibfield{author}{\bibinfo{person}{Theodoros Chondrogiannis} {and} \bibinfo{person}{Johann Gamper}.} \bibinfo{year}{2014}\natexlab{}.
\newblock \showarticletitle{Exploring graph partitioning for shortest path queries on road networks}. In \bibinfo{booktitle}{\emph{26th GI-Workshop Grundlagen von Datenbanken: GvDB'14}}. \bibinfo{pages}{71--76}.
\newblock


\bibitem[Chondrogiannis and Gamper(2016)]%
        {ParDiSP_chondrogiannis2016pardisp}
\bibfield{author}{\bibinfo{person}{Theodoros Chondrogiannis} {and} \bibinfo{person}{Johann Gamper}.} \bibinfo{year}{2016}\natexlab{}.
\newblock \showarticletitle{ParDiSP: A partition-based framework for distance and shortest path queries on road networks}. In \bibinfo{booktitle}{\emph{2016 17th IEEE International Conference on Mobile Data Management (MDM)}}, Vol.~\bibinfo{volume}{1}. IEEE, \bibinfo{pages}{242--251}.
\newblock


\bibitem[Cohen et~al\mbox{.}(2003)]%
        {cohen2003reachability}
\bibfield{author}{\bibinfo{person}{Edith Cohen}, \bibinfo{person}{Eran Halperin}, \bibinfo{person}{Haim Kaplan}, {and} \bibinfo{person}{Uri Zwick}.} \bibinfo{year}{2003}\natexlab{}.
\newblock \showarticletitle{Reachability and distance queries via 2-hop labels}.
\newblock \bibinfo{journal}{\emph{SIAM J. Comput.}} \bibinfo{volume}{32}, \bibinfo{number}{5} (\bibinfo{year}{2003}), \bibinfo{pages}{1338--1355}.
\newblock


\bibitem[Dan et~al\mbox{.}(2022)]%
        {LG-Tree_dan2022lg}
\bibfield{author}{\bibinfo{person}{Tangpeng Dan}, \bibinfo{person}{Changyin Luo}, \bibinfo{person}{Yanhong Li}, \bibinfo{person}{Zhong Guan}, {and} \bibinfo{person}{Xiaofeng Meng}.} \bibinfo{year}{2022}\natexlab{}.
\newblock \showarticletitle{LG-Tree: An Efficient Labeled Index for Shortest Distance Search on Massive Road Networks}.
\newblock \bibinfo{journal}{\emph{IEEE Transactions on Intelligent Transportation Systems}} \bibinfo{volume}{23}, \bibinfo{number}{12} (\bibinfo{year}{2022}), \bibinfo{pages}{23721--23735}.
\newblock


\bibitem[D'angelo et~al\mbox{.}(2019)]%
        {DecPLL_d2019fully}
\bibfield{author}{\bibinfo{person}{Gianlorenzo D'angelo}, \bibinfo{person}{Mattia D'emidio}, {and} \bibinfo{person}{Daniele Frigioni}.} \bibinfo{year}{2019}\natexlab{}.
\newblock \showarticletitle{Fully dynamic 2-hop cover labeling}.
\newblock \bibinfo{journal}{\emph{Journal of Experimental Algorithmics (JEA)}}  \bibinfo{volume}{24} (\bibinfo{year}{2019}), \bibinfo{pages}{1--36}.
\newblock


\bibitem[Das~Sarma et~al\mbox{.}(2010)]%
        {das2010sketch}
\bibfield{author}{\bibinfo{person}{Atish Das~Sarma}, \bibinfo{person}{Sreenivas Gollapudi}, \bibinfo{person}{Marc Najork}, {and} \bibinfo{person}{Rina Panigrahy}.} \bibinfo{year}{2010}\natexlab{}.
\newblock \showarticletitle{A sketch-based distance oracle for web-scale graphs}. In \bibinfo{booktitle}{\emph{Proceedings of the third ACM international conference on Web search and data mining}}. \bibinfo{pages}{401--410}.
\newblock


\bibitem[Delling et~al\mbox{.}(2011a)]%
        {CRP_delling2011customizable}
\bibfield{author}{\bibinfo{person}{Daniel Delling}, \bibinfo{person}{Andrew~V Goldberg}, \bibinfo{person}{Thomas Pajor}, {and} \bibinfo{person}{Renato~F Werneck}.} \bibinfo{year}{2011}\natexlab{a}.
\newblock \showarticletitle{Customizable route planning}. In \bibinfo{booktitle}{\emph{Experimental Algorithms: 10th International Symposium, SEA 2011, Kolimpari, Chania, Crete, Greece, May 5-7, 2011. Proceedings 10}}. Springer, \bibinfo{pages}{376--387}.
\newblock


\bibitem[Delling et~al\mbox{.}(2017)]%
        {CRP_delling2017customizable}
\bibfield{author}{\bibinfo{person}{Daniel Delling}, \bibinfo{person}{Andrew~V Goldberg}, \bibinfo{person}{Thomas Pajor}, {and} \bibinfo{person}{Renato~F Werneck}.} \bibinfo{year}{2017}\natexlab{}.
\newblock \showarticletitle{Customizable route planning in road networks}.
\newblock \bibinfo{journal}{\emph{Transportation Science}} \bibinfo{volume}{51}, \bibinfo{number}{2} (\bibinfo{year}{2017}), \bibinfo{pages}{566--591}.
\newblock


\bibitem[Delling et~al\mbox{.}(2011b)]%
        {PUNCH_delling2011graph}
\bibfield{author}{\bibinfo{person}{Daniel Delling}, \bibinfo{person}{Andrew~V. Goldberg}, \bibinfo{person}{Ilya Razenshteyn}, {and} \bibinfo{person}{Renato~F. Werneck}.} \bibinfo{year}{2011}\natexlab{b}.
\newblock \showarticletitle{Graph Partitioning with Natural Cuts}. In \bibinfo{booktitle}{\emph{2011 IEEE International Parallel Distributed Processing Symposium}}. \bibinfo{pages}{1135--1146}.
\newblock


\bibitem[Diekmann et~al\mbox{.}(2000)]%
        {Bubble_diekmann2000shape}
\bibfield{author}{\bibinfo{person}{Ralf Diekmann}, \bibinfo{person}{Robert Preis}, \bibinfo{person}{Frank Schlimbach}, {and} \bibinfo{person}{Chris Walshaw}.} \bibinfo{year}{2000}\natexlab{}.
\newblock \showarticletitle{Shape-optimized mesh partitioning and load balancing for parallel adaptive FEM}.
\newblock \bibinfo{journal}{\emph{Parallel Comput.}} \bibinfo{volume}{26}, \bibinfo{number}{12} (\bibinfo{year}{2000}), \bibinfo{pages}{1555--1581}.
\newblock


\bibitem[Dijkstra(1959)]%
        {dijkstra1959note}
\bibfield{author}{\bibinfo{person}{Edsger~W Dijkstra}.} \bibinfo{year}{1959}\natexlab{}.
\newblock \showarticletitle{A note on two problems in connexion with graphs}.
\newblock \bibinfo{journal}{\emph{Numerische mathematik}} \bibinfo{volume}{1}, \bibinfo{number}{1} (\bibinfo{year}{1959}), \bibinfo{pages}{269--271}.
\newblock


\bibitem[Ding et~al\mbox{.}(2008)]%
        {ding2008finding}
\bibfield{author}{\bibinfo{person}{Bolin Ding}, \bibinfo{person}{Jeffrey~Xu Yu}, {and} \bibinfo{person}{Lu Qin}.} \bibinfo{year}{2008}\natexlab{}.
\newblock \showarticletitle{Finding time-dependent shortest paths over large graphs}. In \bibinfo{booktitle}{\emph{Proceedings of the 11th international conference on Extending database technology: Advances in database technology}}. \bibinfo{pages}{205--216}.
\newblock


\bibitem[Elliott et~al\mbox{.}(2020)]%
        {elliott2020core}
\bibfield{author}{\bibinfo{person}{Andrew Elliott}, \bibinfo{person}{Angus Chiu}, \bibinfo{person}{Marya Bazzi}, \bibinfo{person}{Gesine Reinert}, {and} \bibinfo{person}{Mihai Cucuringu}.} \bibinfo{year}{2020}\natexlab{}.
\newblock \showarticletitle{Core--periphery structure in directed networks}.
\newblock \bibinfo{journal}{\emph{Proceedings of the Royal Society A}} \bibinfo{volume}{476}, \bibinfo{number}{2241} (\bibinfo{year}{2020}), \bibinfo{pages}{20190783}.
\newblock


\bibitem[Farhan et~al\mbox{.}(2022)]%
        {farhan2022batchhl}
\bibfield{author}{\bibinfo{person}{Muhammad Farhan}, \bibinfo{person}{Qing Wang}, {and} \bibinfo{person}{Henning Koehler}.} \bibinfo{year}{2022}\natexlab{}.
\newblock \showarticletitle{BatchHL: Answering Distance Queries on Batch-Dynamic Networks at Scale}. In \bibinfo{booktitle}{\emph{Proceedings of the 2022 International Conference on Management of Data}}. \bibinfo{pages}{2020--2033}.
\newblock


\bibitem[Farhan et~al\mbox{.}(2018)]%
        {farhan2018highly}
\bibfield{author}{\bibinfo{person}{Muhammad Farhan}, \bibinfo{person}{Qing Wang}, \bibinfo{person}{Yu Lin}, {and} \bibinfo{person}{Brendan Mckay}.} \bibinfo{year}{2018}\natexlab{}.
\newblock \showarticletitle{A highly scalable labelling approach for exact distance queries in complex networks}.
\newblock \bibinfo{journal}{\emph{EDBT}} (\bibinfo{year}{2018}).
\newblock


\bibitem[Fiduccia and Mattheyses(1982)]%
        {FiducciaMattheyses82}
\bibfield{author}{\bibinfo{person}{C.M. Fiduccia} {and} \bibinfo{person}{R.M. Mattheyses}.} \bibinfo{year}{1982}\natexlab{}.
\newblock \showarticletitle{A Linear-Time Heuristic for Improving Network Partitions}. In \bibinfo{booktitle}{\emph{Proceedings of the 19th Design Automation Conference}}. \bibinfo{pages}{175--181}.
\newblock


\bibitem[Ford and Fulkerson(1956)]%
        {ford1956maximal}
\bibfield{author}{\bibinfo{person}{Lester~Randolph Ford} {and} \bibinfo{person}{Delbert~R Fulkerson}.} \bibinfo{year}{1956}\natexlab{}.
\newblock \showarticletitle{Maximal flow through a network}.
\newblock \bibinfo{journal}{\emph{Canadian journal of Mathematics}}  \bibinfo{volume}{8} (\bibinfo{year}{1956}), \bibinfo{pages}{399--404}.
\newblock


\bibitem[Geisberger et~al\mbox{.}(2008)]%
        {CH_geisberger2008contraction}
\bibfield{author}{\bibinfo{person}{Robert Geisberger}, \bibinfo{person}{Peter Sanders}, \bibinfo{person}{Dominik Schultes}, {and} \bibinfo{person}{Daniel Delling}.} \bibinfo{year}{2008}\natexlab{}.
\newblock \showarticletitle{Contraction hierarchies: Faster and simpler hierarchical routing in road networks}. In \bibinfo{booktitle}{\emph{International workshop on experimental and efficient algorithms}}. Springer, \bibinfo{pages}{319--333}.
\newblock


\bibitem[Geisberger et~al\mbox{.}(2012)]%
        {geisberger2012exact}
\bibfield{author}{\bibinfo{person}{Robert Geisberger}, \bibinfo{person}{Peter Sanders}, \bibinfo{person}{Dominik Schultes}, {and} \bibinfo{person}{Christian Vetter}.} \bibinfo{year}{2012}\natexlab{}.
\newblock \showarticletitle{Exact routing in large road networks using contraction hierarchies}.
\newblock \bibinfo{journal}{\emph{Transportation Science}} \bibinfo{volume}{46}, \bibinfo{number}{3} (\bibinfo{year}{2012}), \bibinfo{pages}{388--404}.
\newblock


\bibitem[Gilbert et~al\mbox{.}(1998)]%
        {gilbert1998geometric}
\bibfield{author}{\bibinfo{person}{John~R Gilbert}, \bibinfo{person}{Gary~L Miller}, {and} \bibinfo{person}{Shang-Hua Teng}.} \bibinfo{year}{1998}\natexlab{}.
\newblock \showarticletitle{Geometric mesh partitioning: Implementation and experiments}.
\newblock \bibinfo{journal}{\emph{SIAM Journal on Scientific Computing}} \bibinfo{volume}{19}, \bibinfo{number}{6} (\bibinfo{year}{1998}), \bibinfo{pages}{2091--2110}.
\newblock


\bibitem[GitHub({[n.\,d.]})]%
        {SourceCode}
\bibfield{author}{\bibinfo{person}{GitHub}.} \bibinfo{year}{[n.\,d.]}\natexlab{}.
\newblock \bibinfo{title}{EPSP}.
\newblock \bibinfo{howpublished}{\url{https://github.com/ZXJ-DSA/EPSP}}.
\newblock


\bibitem[Goehring and Saad(1994)]%
        {goehring1994heuristic}
\bibfield{author}{\bibinfo{person}{Todd Goehring} {and} \bibinfo{person}{Yousef Saad}.} \bibinfo{year}{1994}\natexlab{}.
\newblock \bibinfo{booktitle}{\emph{Heuristic algorithms for automatic graph partitioning}}.
\newblock \bibinfo{type}{{T}echnical {R}eport}. \bibinfo{institution}{Citeseer}.
\newblock


\bibitem[Goldberg and Harrelson(2005)]%
        {ALT_goldberg2005computing}
\bibfield{author}{\bibinfo{person}{Andrew~V Goldberg} {and} \bibinfo{person}{Chris Harrelson}.} \bibinfo{year}{2005}\natexlab{}.
\newblock \showarticletitle{Computing the shortest path: A search meets graph theory.}. In \bibinfo{booktitle}{\emph{SODA}}, Vol.~\bibinfo{volume}{5}. Citeseer, \bibinfo{pages}{156--165}.
\newblock


\bibitem[Goldberg and Tarjan(1988)]%
        {goldberg1988new}
\bibfield{author}{\bibinfo{person}{Andrew~V Goldberg} {and} \bibinfo{person}{Robert~E Tarjan}.} \bibinfo{year}{1988}\natexlab{}.
\newblock \showarticletitle{A new approach to the maximum-flow problem}.
\newblock \bibinfo{journal}{\emph{Journal of the ACM (JACM)}} \bibinfo{volume}{35}, \bibinfo{number}{4} (\bibinfo{year}{1988}), \bibinfo{pages}{921--940}.
\newblock


\bibitem[Goldschmidt and Hochbaum(1994)]%
        {Goldschmidt94}
\bibfield{author}{\bibinfo{person}{Olivier Goldschmidt} {and} \bibinfo{person}{Dorit~S. Hochbaum}.} \bibinfo{year}{1994}\natexlab{}.
\newblock \showarticletitle{A Polynomial Algorithm for the k-Cut Problem for Fixed k}.
\newblock \bibinfo{journal}{\emph{Mathematics of Operations Research}} \bibinfo{volume}{19}, \bibinfo{number}{1} (\bibinfo{year}{1994}), \bibinfo{pages}{24--37}.
\newblock


\bibitem[Gong et~al\mbox{.}(2016)]%
        {gong2016efficient}
\bibfield{author}{\bibinfo{person}{Maoguo Gong}, \bibinfo{person}{Guanjun Li}, \bibinfo{person}{Zhao Wang}, \bibinfo{person}{Lijia Ma}, {and} \bibinfo{person}{Dayong Tian}.} \bibinfo{year}{2016}\natexlab{}.
\newblock \showarticletitle{An efficient shortest path approach for social networks based on community structure}.
\newblock \bibinfo{journal}{\emph{CAAI Transactions on Intelligence Technology}} \bibinfo{volume}{1}, \bibinfo{number}{1} (\bibinfo{year}{2016}), \bibinfo{pages}{114--123}.
\newblock


\bibitem[Gonzalez et~al\mbox{.}(2012)]%
        {gonzalez2012powergraph}
\bibfield{author}{\bibinfo{person}{Joseph~E Gonzalez}, \bibinfo{person}{Yucheng Low}, \bibinfo{person}{Haijie Gu}, \bibinfo{person}{Danny Bickson}, {and} \bibinfo{person}{Carlos Guestrin}.} \bibinfo{year}{2012}\natexlab{}.
\newblock \showarticletitle{$\{$PowerGraph$\}$: Distributed $\{$Graph-Parallel$\}$ Computation on Natural Graphs}. In \bibinfo{booktitle}{\emph{10th USENIX symposium on operating systems design and implementation (OSDI 12)}}. \bibinfo{pages}{17--30}.
\newblock


\bibitem[Gubichev et~al\mbox{.}(2010)]%
        {gubichev2010fast}
\bibfield{author}{\bibinfo{person}{Andrey Gubichev}, \bibinfo{person}{Srikanta Bedathur}, \bibinfo{person}{Stephan Seufert}, {and} \bibinfo{person}{Gerhard Weikum}.} \bibinfo{year}{2010}\natexlab{}.
\newblock \showarticletitle{Fast and accurate estimation of shortest paths in large graphs}. In \bibinfo{booktitle}{\emph{Proceedings of the 19th ACM international conference on Information and knowledge management}}. \bibinfo{pages}{499--508}.
\newblock


\bibitem[Hart et~al\mbox{.}(1968)]%
        {Astar_hart1968formal}
\bibfield{author}{\bibinfo{person}{Peter~E Hart}, \bibinfo{person}{Nils~J Nilsson}, {and} \bibinfo{person}{Bertram Raphael}.} \bibinfo{year}{1968}\natexlab{}.
\newblock \showarticletitle{A formal basis for the heuristic determination of minimum cost paths}.
\newblock \bibinfo{journal}{\emph{IEEE transactions on Systems Science and Cybernetics}} \bibinfo{volume}{4}, \bibinfo{number}{2} (\bibinfo{year}{1968}), \bibinfo{pages}{100--107}.
\newblock


\bibitem[Hendrickson and Leland(1995a)]%
        {hendrickson1995improved}
\bibfield{author}{\bibinfo{person}{Bruce Hendrickson} {and} \bibinfo{person}{Robert Leland}.} \bibinfo{year}{1995}\natexlab{a}.
\newblock \showarticletitle{An improved spectral graph partitioning algorithm for mapping parallel computations}.
\newblock \bibinfo{journal}{\emph{SIAM Journal on Scientific Computing}} \bibinfo{volume}{16}, \bibinfo{number}{2} (\bibinfo{year}{1995}), \bibinfo{pages}{452--469}.
\newblock


\bibitem[Hendrickson and Leland(1995b)]%
        {HL95}
\bibfield{author}{\bibinfo{person}{Bruce Hendrickson} {and} \bibinfo{person}{Robert Leland}.} \bibinfo{year}{1995}\natexlab{b}.
\newblock \showarticletitle{A Multilevel Algorithm for Partitioning Graphs}. In \bibinfo{booktitle}{\emph{Proceedings of the 1995 ACM/IEEE Conference on Supercomputing}}. \bibinfo{pages}{28--–es}.
\newblock


\bibitem[Hilger et~al\mbox{.}(2009)]%
        {ArcFlag_hilger2009fast}
\bibfield{author}{\bibinfo{person}{Moritz Hilger}, \bibinfo{person}{Ekkehard K{\"o}hler}, \bibinfo{person}{Rolf~H M{\"o}hring}, {and} \bibinfo{person}{Heiko Schilling}.} \bibinfo{year}{2009}\natexlab{}.
\newblock \showarticletitle{Fast point-to-point shortest path computations with arc-flags}.
\newblock \bibinfo{journal}{\emph{The Shortest Path Problem: Ninth DIMACS Implementation Challenge}}  \bibinfo{volume}{74} (\bibinfo{year}{2009}), \bibinfo{pages}{41--72}.
\newblock


\bibitem[Holtgrewe et~al\mbox{.}(2010)]%
        {kappa2010}
\bibfield{author}{\bibinfo{person}{Manuel Holtgrewe}, \bibinfo{person}{Peter Sanders}, {and} \bibinfo{person}{Christian Schulz}.} \bibinfo{year}{2010}\natexlab{}.
\newblock \showarticletitle{Engineering a scalable high quality graph partitioner}. In \bibinfo{booktitle}{\emph{2010 IEEE International Symposium on Parallel \& Distributed Processing (IPDPS)}}. \bibinfo{pages}{1--12}.
\newblock


\bibitem[Holzer et~al\mbox{.}(2009)]%
        {holzer2009engineering}
\bibfield{author}{\bibinfo{person}{Martin Holzer}, \bibinfo{person}{Frank Schulz}, {and} \bibinfo{person}{Dorothea Wagner}.} \bibinfo{year}{2009}\natexlab{}.
\newblock \showarticletitle{Engineering multilevel overlay graphs for shortest-path queries}.
\newblock \bibinfo{journal}{\emph{Journal of Experimental Algorithmics (JEA)}}  \bibinfo{volume}{13} (\bibinfo{year}{2009}), \bibinfo{pages}{2--5}.
\newblock


\bibitem[Huang et~al\mbox{.}(2021)]%
        {huang2021learning}
\bibfield{author}{\bibinfo{person}{Shuai Huang}, \bibinfo{person}{Yong Wang}, \bibinfo{person}{Tianyu Zhao}, {and} \bibinfo{person}{Guoliang Li}.} \bibinfo{year}{2021}\natexlab{}.
\newblock \showarticletitle{A Learning-based Method for Computing Shortest Path Distances on Road Networks}. In \bibinfo{booktitle}{\emph{2021 IEEE 37th International Conference on Data Engineering (ICDE)}}. IEEE, \bibinfo{pages}{360--371}.
\newblock


\bibitem[Huang et~al\mbox{.}(1996)]%
        {huang1996effective}
\bibfield{author}{\bibinfo{person}{Yun-Wu Huang}, \bibinfo{person}{Ning Jing}, {and} \bibinfo{person}{Elke~A Rundensteiner}.} \bibinfo{year}{1996}\natexlab{}.
\newblock \showarticletitle{Effective graph clustering for path queries in digital map databases}. In \bibinfo{booktitle}{\emph{Proceedings of the fifth international conference on Information and knowledge management}}. \bibinfo{pages}{215--222}.
\newblock


\bibitem[ich Lauther(2006)]%
        {ArcFlag_ich2006extremely}
\bibfield{author}{\bibinfo{person}{Ulr ich Lauther}.} \bibinfo{year}{2006}\natexlab{}.
\newblock \showarticletitle{An extremely fast, exact algorithm for finding shortest paths in static networks with geographical background}.
\newblock  (\bibinfo{year}{2006}).
\newblock


\bibitem[Jin et~al\mbox{.}(2020)]%
        {jin2020parallelizing}
\bibfield{author}{\bibinfo{person}{Ruoming Jin}, \bibinfo{person}{Zhen Peng}, \bibinfo{person}{Wendell Wu}, \bibinfo{person}{Feodor Dragan}, \bibinfo{person}{Gagan Agrawal}, {and} \bibinfo{person}{Bin Ren}.} \bibinfo{year}{2020}\natexlab{}.
\newblock \showarticletitle{Parallelizing pruned landmark labeling: dealing with dependencies in graph algorithms}. In \bibinfo{booktitle}{\emph{Proceedings of the 34th ACM International Conference on Supercomputing}}. \bibinfo{pages}{1--13}.
\newblock


\bibitem[Jung and Pramanik(1996)]%
        {jung1996hiti}
\bibfield{author}{\bibinfo{person}{Sungwon Jung} {and} \bibinfo{person}{Sakti Pramanik}.} \bibinfo{year}{1996}\natexlab{}.
\newblock \showarticletitle{HiTi graph model of topographical road maps in navigation systems}. In \bibinfo{booktitle}{\emph{Proceedings of the Twelfth International Conference on Data Engineering}}. IEEE, \bibinfo{pages}{76--84}.
\newblock


\bibitem[Jung and Pramanik(2002)]%
        {jung2002efficient}
\bibfield{author}{\bibinfo{person}{Sungwon Jung} {and} \bibinfo{person}{Sakti Pramanik}.} \bibinfo{year}{2002}\natexlab{}.
\newblock \showarticletitle{An efficient path computation model for hierarchically structured topographical road maps}.
\newblock \bibinfo{journal}{\emph{IEEE Transactions on Knowledge and Data Engineering}} \bibinfo{volume}{14}, \bibinfo{number}{5} (\bibinfo{year}{2002}), \bibinfo{pages}{1029--1046}.
\newblock


\bibitem[Karypis and Kumar(1998)]%
        {METIS_Karypis98MeTis}
\bibfield{author}{\bibinfo{person}{George Karypis} {and} \bibinfo{person}{Vipin Kumar}.} \bibinfo{year}{1998}\natexlab{}.
\newblock \showarticletitle{A fast and high quality multilevel scheme for partitioning irregular graphs}.
\newblock \bibinfo{journal}{\emph{SIAM JOURNAL ON SCIENTIFIC COMPUTING}} \bibinfo{volume}{20}, \bibinfo{number}{1} (\bibinfo{year}{1998}), \bibinfo{pages}{359--392}.
\newblock


\bibitem[Kernighan and Lin(1970)]%
        {KernighanLin70}
\bibfield{author}{\bibinfo{person}{B.~W. Kernighan} {and} \bibinfo{person}{S. Lin}.} \bibinfo{year}{1970}\natexlab{}.
\newblock \showarticletitle{An efficient heuristic procedure for partitioning graphs}.
\newblock \bibinfo{journal}{\emph{The Bell System Technical Journal}} \bibinfo{volume}{49}, \bibinfo{number}{2} (\bibinfo{year}{1970}), \bibinfo{pages}{291--307}.
\newblock


\bibitem[Kolaczyk et~al\mbox{.}(2009)]%
        {kolaczyk2009group}
\bibfield{author}{\bibinfo{person}{Eric~D Kolaczyk}, \bibinfo{person}{David~B Chua}, {and} \bibinfo{person}{Marc Barth{\'e}lemy}.} \bibinfo{year}{2009}\natexlab{}.
\newblock \showarticletitle{Group betweenness and co-betweenness: Inter-related notions of coalition centrality}.
\newblock \bibinfo{journal}{\emph{Social Networks}} \bibinfo{volume}{31}, \bibinfo{number}{3} (\bibinfo{year}{2009}), \bibinfo{pages}{190--203}.
\newblock


\bibitem[Kong et~al\mbox{.}(2022)]%
        {CLUGP_kong2022clustering}
\bibfield{author}{\bibinfo{person}{Deyu Kong}, \bibinfo{person}{Xike Xie}, {and} \bibinfo{person}{Zhuoxu Zhang}.} \bibinfo{year}{2022}\natexlab{}.
\newblock \showarticletitle{Clustering-based partitioning for large web graphs}. In \bibinfo{booktitle}{\emph{2022 IEEE 38th International Conference on Data Engineering (ICDE)}}. IEEE, \bibinfo{pages}{593--606}.
\newblock


\bibitem[Lanczos(1950)]%
        {lanczos1950iteration}
\bibfield{author}{\bibinfo{person}{Cornelius Lanczos}.} \bibinfo{year}{1950}\natexlab{}.
\newblock \showarticletitle{An iteration method for the solution of the eigenvalue problem of linear differential and integral operators}.
\newblock  (\bibinfo{year}{1950}).
\newblock


\bibitem[Lee et~al\mbox{.}(2009)]%
        {lee2009fast}
\bibfield{author}{\bibinfo{person}{Ken~CK Lee}, \bibinfo{person}{Wang-Chien Lee}, {and} \bibinfo{person}{Baihua Zheng}.} \bibinfo{year}{2009}\natexlab{}.
\newblock \showarticletitle{Fast object search on road networks}. In \bibinfo{booktitle}{\emph{Proceedings of the 12th International Conference on Extending Database Technology: Advances in Database Technology}}. \bibinfo{pages}{1018--1029}.
\newblock


\bibitem[Lee et~al\mbox{.}(2010)]%
        {lee2010road}
\bibfield{author}{\bibinfo{person}{Ken~CK Lee}, \bibinfo{person}{Wang-Chien Lee}, \bibinfo{person}{Baihua Zheng}, {and} \bibinfo{person}{Yuan Tian}.} \bibinfo{year}{2010}\natexlab{}.
\newblock \showarticletitle{ROAD: A new spatial object search framework for road networks}.
\newblock \bibinfo{journal}{\emph{IEEE transactions on knowledge and data engineering}} \bibinfo{volume}{24}, \bibinfo{number}{3} (\bibinfo{year}{2010}), \bibinfo{pages}{547--560}.
\newblock


\bibitem[Li et~al\mbox{.}(2017)]%
        {li2017minimal}
\bibfield{author}{\bibinfo{person}{Lei Li}, \bibinfo{person}{Wen Hua}, \bibinfo{person}{Xingzhong Du}, {and} \bibinfo{person}{Xiaofang Zhou}.} \bibinfo{year}{2017}\natexlab{}.
\newblock \showarticletitle{Minimal on-road time route scheduling on time-dependent graphs}.
\newblock \bibinfo{journal}{\emph{Proceedings of the VLDB Endowment}} \bibinfo{volume}{10}, \bibinfo{number}{11} (\bibinfo{year}{2017}), \bibinfo{pages}{1274--1285}.
\newblock


\bibitem[Li et~al\mbox{.}(2019c)]%
        {li2019time}
\bibfield{author}{\bibinfo{person}{Lei Li}, \bibinfo{person}{Sibo Wang}, {and} \bibinfo{person}{Xiaofang Zhou}.} \bibinfo{year}{2019}\natexlab{c}.
\newblock \showarticletitle{Time-dependent hop labeling on road network}. In \bibinfo{booktitle}{\emph{2019 IEEE 35th International Conference on Data Engineering (ICDE)}}. IEEE, \bibinfo{pages}{902--913}.
\newblock


\bibitem[Li et~al\mbox{.}(2020b)]%
        {li2020fastest}
\bibfield{author}{\bibinfo{person}{Lei Li}, \bibinfo{person}{Sibo Wang}, {and} \bibinfo{person}{Xiaofang Zhou}.} \bibinfo{year}{2020}\natexlab{b}.
\newblock \showarticletitle{Fastest path query answering using time-dependent hop-labeling in road network}.
\newblock \bibinfo{journal}{\emph{IEEE Transactions on Knowledge and Data Engineering}} (\bibinfo{year}{2020}).
\newblock


\bibitem[Li et~al\mbox{.}(2020c)]%
        {li2020fast}
\bibfield{author}{\bibinfo{person}{Lei Li}, \bibinfo{person}{Mengxuan Zhang}, \bibinfo{person}{Wen Hua}, {and} \bibinfo{person}{Xiaofang Zhou}.} \bibinfo{year}{2020}\natexlab{c}.
\newblock \showarticletitle{Fast query decomposition for batch shortest path processing in road networks}. In \bibinfo{booktitle}{\emph{2020 IEEE 36th International Conference on Data Engineering (ICDE)}}. IEEE, \bibinfo{pages}{1189--1200}.
\newblock


\bibitem[Li et~al\mbox{.}(2018)]%
        {li2018go}
\bibfield{author}{\bibinfo{person}{Lei Li}, \bibinfo{person}{Kai Zheng}, \bibinfo{person}{Sibo Wang}, \bibinfo{person}{Wen Hua}, {and} \bibinfo{person}{Xiaofang Zhou}.} \bibinfo{year}{2018}\natexlab{}.
\newblock \showarticletitle{Go slow to go fast: minimal on-road time route scheduling with parking facilities using historical trajectory}.
\newblock \bibinfo{journal}{\emph{The VLDB Journal}}  \bibinfo{volume}{27} (\bibinfo{year}{2018}), \bibinfo{pages}{321--345}.
\newblock


\bibitem[Li et~al\mbox{.}(2019b)]%
        {PSL_li2019scaling}
\bibfield{author}{\bibinfo{person}{Wentao Li}, \bibinfo{person}{Miao Qiao}, \bibinfo{person}{Lu Qin}, \bibinfo{person}{Ying Zhang}, \bibinfo{person}{Lijun Chang}, {and} \bibinfo{person}{Xuemin Lin}.} \bibinfo{year}{2019}\natexlab{b}.
\newblock \showarticletitle{Scaling distance labeling on small-world networks}. In \bibinfo{booktitle}{\emph{Proceedings of the 2019 International Conference on Management of Data}}. \bibinfo{pages}{1060--1077}.
\newblock


\bibitem[Li et~al\mbox{.}(2020a)]%
        {CT_li2020scaling}
\bibfield{author}{\bibinfo{person}{Wentao Li}, \bibinfo{person}{Miao Qiao}, \bibinfo{person}{Lu Qin}, \bibinfo{person}{Ying Zhang}, \bibinfo{person}{Lijun Chang}, {and} \bibinfo{person}{Xuemin Lin}.} \bibinfo{year}{2020}\natexlab{a}.
\newblock \showarticletitle{Scaling up distance labeling on graphs with core-periphery properties}. In \bibinfo{booktitle}{\emph{Proceedings of the 2020 ACM SIGMOD International Conference on Management of Data}}. \bibinfo{pages}{1367--1381}.
\newblock


\bibitem[Li et~al\mbox{.}(2019a)]%
        {Gstar_tree_li2019g}
\bibfield{author}{\bibinfo{person}{Zijian Li}, \bibinfo{person}{Lei Chen}, {and} \bibinfo{person}{Yue Wang}.} \bibinfo{year}{2019}\natexlab{a}.
\newblock \showarticletitle{G*-tree: An efficient spatial index on road networks}. In \bibinfo{booktitle}{\emph{2019 IEEE 35th International Conference on Data Engineering (ICDE)}}. IEEE, \bibinfo{pages}{268--279}.
\newblock


\bibitem[Liu et~al\mbox{.}(2021)]%
        {liu2021efficient}
\bibfield{author}{\bibinfo{person}{Ziyi Liu}, \bibinfo{person}{Lei Li}, \bibinfo{person}{Mengxuan Zhang}, \bibinfo{person}{Wen Hua}, \bibinfo{person}{Pingfu Chao}, {and} \bibinfo{person}{Xiaofang Zhou}.} \bibinfo{year}{2021}\natexlab{}.
\newblock \showarticletitle{Efficient constrained shortest path query answering with forest hop labeling}. In \bibinfo{booktitle}{\emph{2021 IEEE 37th International Conference on Data Engineering (ICDE)}}. IEEE, \bibinfo{pages}{1763--1774}.
\newblock


\bibitem[Liu et~al\mbox{.}(2022)]%
        {liu2022FHL}
\bibfield{author}{\bibinfo{person}{Ziyi Liu}, \bibinfo{person}{Lei Li}, \bibinfo{person}{Mengxuan Zhang}, \bibinfo{person}{Wen Hua}, {and} \bibinfo{person}{Xiaofang Zhou}.} \bibinfo{year}{2022}\natexlab{}.
\newblock \showarticletitle{FHL-Cube: Multi-Constraint Shortest Path Querying with Flexible Combination of Constraints}.
\newblock \bibinfo{journal}{\emph{Proceedings of the VLDB Endowment}}  \bibinfo{volume}{15} (\bibinfo{year}{2022}).
\newblock


\bibitem[Liu et~al\mbox{.}(2023)]%
        {liu2023multi}
\bibfield{author}{\bibinfo{person}{Ziyi Liu}, \bibinfo{person}{Lei Li}, \bibinfo{person}{Mengxuan Zhang}, \bibinfo{person}{Wen Hua}, {and} \bibinfo{person}{Xiaofang Zhou}.} \bibinfo{year}{2023}\natexlab{}.
\newblock \showarticletitle{Multi-constraint shortest path using forest hop labeling}.
\newblock \bibinfo{journal}{\emph{The VLDB Journal}} \bibinfo{volume}{32}, \bibinfo{number}{3} (\bibinfo{year}{2023}), \bibinfo{pages}{595--621}.
\newblock


\bibitem[Maehara et~al\mbox{.}(2014a)]%
        {CTD_maehara2014computing}
\bibfield{author}{\bibinfo{person}{Takanori Maehara}, \bibinfo{person}{Takuya Akiba}, \bibinfo{person}{Yoichi Iwata}, {and} \bibinfo{person}{Ken-ichi Kawarabayashi}.} \bibinfo{year}{2014}\natexlab{a}.
\newblock \showarticletitle{Computing personalized pagerank quickly by exploiting graph structures}.
\newblock \bibinfo{journal}{\emph{Proceedings of the VLDB Endowment}} \bibinfo{volume}{7}, \bibinfo{number}{12} (\bibinfo{year}{2014}), \bibinfo{pages}{1023--1034}.
\newblock


\bibitem[Maehara et~al\mbox{.}(2014b)]%
        {maehara2014computing}
\bibfield{author}{\bibinfo{person}{Takanori Maehara}, \bibinfo{person}{Takuya Akiba}, \bibinfo{person}{Yoichi Iwata}, {and} \bibinfo{person}{Ken-ichi Kawarabayashi}.} \bibinfo{year}{2014}\natexlab{b}.
\newblock \showarticletitle{Computing personalized pagerank quickly by exploiting graph structures}.
\newblock \bibinfo{journal}{\emph{Proceedings of the VLDB Endowment}} \bibinfo{volume}{7}, \bibinfo{number}{12} (\bibinfo{year}{2014}), \bibinfo{pages}{1023--1034}.
\newblock


\bibitem[Mayer and Jacobsen(2021)]%
        {HEP_mayer2021hybrid}
\bibfield{author}{\bibinfo{person}{Ruben Mayer} {and} \bibinfo{person}{Hans-Arno Jacobsen}.} \bibinfo{year}{2021}\natexlab{}.
\newblock \showarticletitle{Hybrid Edge Partitioner: Partitioning Large Power-Law Graphs under Memory Constraints}. In \bibinfo{booktitle}{\emph{Proceedings of the 2021 International Conference on Management of Data}}. \bibinfo{pages}{1289--1302}.
\newblock


\bibitem[Meyerhenke et~al\mbox{.}(2006)]%
        {meyerhenke2006accelerating}
\bibfield{author}{\bibinfo{person}{Henning Meyerhenke}, \bibinfo{person}{Burkhard Monien}, {and} \bibinfo{person}{Stefan Schamberger}.} \bibinfo{year}{2006}\natexlab{}.
\newblock \showarticletitle{Accelerating shape optimizing load balancing for parallel FEM simulations by algebraic multigrid}. In \bibinfo{booktitle}{\emph{Proceedings 20th IEEE International Parallel \& Distributed Processing Symposium}}. IEEE, \bibinfo{pages}{10--pp}.
\newblock


\bibitem[Miller et~al\mbox{.}(1993)]%
        {miller1993automatic}
\bibfield{author}{\bibinfo{person}{GL Miller}, \bibinfo{person}{SH Teng}, \bibinfo{person}{W Thurston}, {and} \bibinfo{person}{SA Vavasis}.} \bibinfo{year}{1993}\natexlab{}.
\newblock \bibinfo{title}{Automatic mesh partitioning, in “Sparse Matrix Computations: Graph Theory Issues and Algorithms,” An IMA Workshop Volume,(A. George, JR Gilbert, and JW-H. Liu, Eds.)}.
\newblock
\newblock


\bibitem[Miller et~al\mbox{.}(1998)]%
        {miller1998geometric}
\bibfield{author}{\bibinfo{person}{Gary~L Miller}, \bibinfo{person}{Shang-Hua Teng}, \bibinfo{person}{William Thurston}, {and} \bibinfo{person}{Stephen~A Vavasis}.} \bibinfo{year}{1998}\natexlab{}.
\newblock \showarticletitle{Geometric separators for finite-element meshes}.
\newblock \bibinfo{journal}{\emph{SIAM Journal on Scientific Computing}} \bibinfo{volume}{19}, \bibinfo{number}{2} (\bibinfo{year}{1998}), \bibinfo{pages}{364--386}.
\newblock


\bibitem[M{\"o}hring et~al\mbox{.}(2007)]%
        {mohring2007partitioning}
\bibfield{author}{\bibinfo{person}{Rolf~H M{\"o}hring}, \bibinfo{person}{Heiko Schilling}, \bibinfo{person}{Birk Sch{\"u}tz}, \bibinfo{person}{Dorothea Wagner}, {and} \bibinfo{person}{Thomas Willhalm}.} \bibinfo{year}{2007}\natexlab{}.
\newblock \showarticletitle{Partitioning graphs to speedup Dijkstra's algorithm}.
\newblock \bibinfo{journal}{\emph{Journal of Experimental Algorithmics (JEA)}}  \bibinfo{volume}{11} (\bibinfo{year}{2007}), \bibinfo{pages}{2--8}.
\newblock


\bibitem[Opsahl et~al\mbox{.}(2010)]%
        {opsahl2010node}
\bibfield{author}{\bibinfo{person}{Tore Opsahl}, \bibinfo{person}{Filip Agneessens}, {and} \bibinfo{person}{John Skvoretz}.} \bibinfo{year}{2010}\natexlab{}.
\newblock \showarticletitle{Node centrality in weighted networks: Generalizing degree and shortest paths}.
\newblock \bibinfo{journal}{\emph{Social networks}} \bibinfo{volume}{32}, \bibinfo{number}{3} (\bibinfo{year}{2010}), \bibinfo{pages}{245--251}.
\newblock


\bibitem[Ouyang et~al\mbox{.}(2018)]%
        {H2H_ouyang2018hierarchy}
\bibfield{author}{\bibinfo{person}{Dian Ouyang}, \bibinfo{person}{Lu Qin}, \bibinfo{person}{Lijun Chang}, \bibinfo{person}{Xuemin Lin}, \bibinfo{person}{Ying Zhang}, {and} \bibinfo{person}{Qing Zhu}.} \bibinfo{year}{2018}\natexlab{}.
\newblock \showarticletitle{When hierarchy meets 2-hop-labeling: Efficient shortest distance queries on road networks}. In \bibinfo{booktitle}{\emph{Proceedings of the 2018 International Conference on Management of Data}}. \bibinfo{pages}{709--724}.
\newblock


\bibitem[Ouyang et~al\mbox{.}(2020)]%
        {DCH_ouyang2020efficient}
\bibfield{author}{\bibinfo{person}{Dian Ouyang}, \bibinfo{person}{Long Yuan}, \bibinfo{person}{Lu Qin}, \bibinfo{person}{Lijun Chang}, \bibinfo{person}{Ying Zhang}, {and} \bibinfo{person}{Xuemin Lin}.} \bibinfo{year}{2020}\natexlab{}.
\newblock \showarticletitle{Efficient shortest path index maintenance on dynamic road networks with theoretical guarantees}.
\newblock \bibinfo{journal}{\emph{Proceedings of the VLDB Endowment}} \bibinfo{volume}{13}, \bibinfo{number}{5} (\bibinfo{year}{2020}), \bibinfo{pages}{602--615}.
\newblock


\bibitem[Pellegrini and Roman(1996)]%
        {pellegrini1996scotch}
\bibfield{author}{\bibinfo{person}{Fran{\c{c}}ois Pellegrini} {and} \bibinfo{person}{Jean Roman}.} \bibinfo{year}{1996}\natexlab{}.
\newblock \showarticletitle{Scotch: A software package for static mapping by dual recursive bipartitioning of process and architecture graphs}. In \bibinfo{booktitle}{\emph{International Conference on High-Performance Computing and Networking}}. Springer, \bibinfo{pages}{493--498}.
\newblock


\bibitem[Petroni et~al\mbox{.}(2015)]%
        {petroni2015hdrf}
\bibfield{author}{\bibinfo{person}{Fabio Petroni}, \bibinfo{person}{Leonardo Querzoni}, \bibinfo{person}{Khuzaima Daudjee}, \bibinfo{person}{Shahin Kamali}, {and} \bibinfo{person}{Giorgio Iacoboni}.} \bibinfo{year}{2015}\natexlab{}.
\newblock \showarticletitle{Hdrf: Stream-based partitioning for power-law graphs}. In \bibinfo{booktitle}{\emph{Proceedings of the 24th ACM international on conference on information and knowledge management}}. \bibinfo{pages}{243--252}.
\newblock


\bibitem[Pilkington and Baden(1994)]%
        {pilkington1994partitioning}
\bibfield{author}{\bibinfo{person}{John~R Pilkington} {and} \bibinfo{person}{Scott~B Baden}.} \bibinfo{year}{1994}\natexlab{}.
\newblock \showarticletitle{Partitioning with spacefilling curves}.
\newblock  (\bibinfo{year}{1994}).
\newblock


\bibitem[Potamias et~al\mbox{.}(2009)]%
        {potamias2009fast}
\bibfield{author}{\bibinfo{person}{Michalis Potamias}, \bibinfo{person}{Francesco Bonchi}, \bibinfo{person}{Carlos Castillo}, {and} \bibinfo{person}{Aristides Gionis}.} \bibinfo{year}{2009}\natexlab{}.
\newblock \showarticletitle{Fast shortest path distance estimation in large networks}. In \bibinfo{booktitle}{\emph{Proceedings of the 18th ACM conference on Information and knowledge management}}. \bibinfo{pages}{867--876}.
\newblock


\bibitem[Pothen et~al\mbox{.}(1990)]%
        {Pothen90SB}
\bibfield{author}{\bibinfo{person}{Alex Pothen}, \bibinfo{person}{Horst~D. Simon}, {and} \bibinfo{person}{Kang-Pu Liou}.} \bibinfo{year}{1990}\natexlab{}.
\newblock \showarticletitle{Partitioning Sparse Matrices with Eigenvectors of Graphs}.
\newblock \bibinfo{journal}{\emph{SIAM J. Matrix Anal. Appl.}} \bibinfo{volume}{11}, \bibinfo{number}{3} (\bibinfo{year}{1990}), \bibinfo{pages}{430--452}.
\newblock


\bibitem[Qi et~al\mbox{.}(2013)]%
        {qi2013toward}
\bibfield{author}{\bibinfo{person}{Zichao Qi}, \bibinfo{person}{Yanghua Xiao}, \bibinfo{person}{Bin Shao}, {and} \bibinfo{person}{Haixun Wang}.} \bibinfo{year}{2013}\natexlab{}.
\newblock \showarticletitle{Toward a distance oracle for billion-node graphs}.
\newblock \bibinfo{journal}{\emph{Proceedings of the VLDB Endowment}} \bibinfo{volume}{7}, \bibinfo{number}{1} (\bibinfo{year}{2013}), \bibinfo{pages}{61--72}.
\newblock


\bibitem[Qiao et~al\mbox{.}(2012)]%
        {qiao2012approximate}
\bibfield{author}{\bibinfo{person}{Miao Qiao}, \bibinfo{person}{Hong Cheng}, \bibinfo{person}{Lijun Chang}, {and} \bibinfo{person}{Jeffrey~Xu Yu}.} \bibinfo{year}{2012}\natexlab{}.
\newblock \showarticletitle{Approximate shortest distance computing: A query-dependent local landmark scheme}.
\newblock \bibinfo{journal}{\emph{IEEE Transactions on Knowledge and Data Engineering}} \bibinfo{volume}{26}, \bibinfo{number}{1} (\bibinfo{year}{2012}), \bibinfo{pages}{55--68}.
\newblock


\bibitem[Qin et~al\mbox{.}(2017)]%
        {IUPLL_qin2017efficient}
\bibfield{author}{\bibinfo{person}{Yongrui Qin}, \bibinfo{person}{Quan~Z Sheng}, \bibinfo{person}{Nickolas~JG Falkner}, \bibinfo{person}{Lina Yao}, {and} \bibinfo{person}{Simon Parkinson}.} \bibinfo{year}{2017}\natexlab{}.
\newblock \showarticletitle{Efficient computation of distance labeling for decremental updates in large dynamic graphs}.
\newblock \bibinfo{journal}{\emph{World Wide Web}}  \bibinfo{volume}{20} (\bibinfo{year}{2017}), \bibinfo{pages}{915--937}.
\newblock


\bibitem[Robertson and Seymour(1984)]%
        {robertson1984graph}
\bibfield{author}{\bibinfo{person}{Neil Robertson} {and} \bibinfo{person}{Paul~D Seymour}.} \bibinfo{year}{1984}\natexlab{}.
\newblock \showarticletitle{Graph minors. III. Planar tree-width}.
\newblock \bibinfo{journal}{\emph{Journal of Combinatorial Theory, Series B}} \bibinfo{volume}{36}, \bibinfo{number}{1} (\bibinfo{year}{1984}), \bibinfo{pages}{49--64}.
\newblock


\bibitem[Rombach et~al\mbox{.}(2014)]%
        {rombach2014core}
\bibfield{author}{\bibinfo{person}{M~Puck Rombach}, \bibinfo{person}{Mason~A Porter}, \bibinfo{person}{James~H Fowler}, {and} \bibinfo{person}{Peter~J Mucha}.} \bibinfo{year}{2014}\natexlab{}.
\newblock \showarticletitle{Core-periphery structure in networks}.
\newblock \bibinfo{journal}{\emph{SIAM Journal on Applied mathematics}} \bibinfo{volume}{74}, \bibinfo{number}{1} (\bibinfo{year}{2014}), \bibinfo{pages}{167--190}.
\newblock


\bibitem[Sagan(2012)]%
        {sagan2012space}
\bibfield{author}{\bibinfo{person}{Hans Sagan}.} \bibinfo{year}{2012}\natexlab{}.
\newblock \bibinfo{booktitle}{\emph{Space-filling curves}}.
\newblock \bibinfo{publisher}{Springer Science \& Business Media}.
\newblock


\bibitem[Sanders and Schulz(2011)]%
        {sanders2011engineering}
\bibfield{author}{\bibinfo{person}{Peter Sanders} {and} \bibinfo{person}{Christian Schulz}.} \bibinfo{year}{2011}\natexlab{}.
\newblock \showarticletitle{Engineering multilevel graph partitioning algorithms}. In \bibinfo{booktitle}{\emph{European Symposium on Algorithms}}. Springer, \bibinfo{pages}{469--480}.
\newblock


\bibitem[Schamberger(2004)]%
        {schamberger2004partitioning}
\bibfield{author}{\bibinfo{person}{Stefan Schamberger}.} \bibinfo{year}{2004}\natexlab{}.
\newblock \showarticletitle{On partitioning FEM graphs using diffusion}. In \bibinfo{booktitle}{\emph{18th International Parallel and Distributed Processing Symposium, 2004. Proceedings.}} IEEE, \bibinfo{pages}{277}.
\newblock


\bibitem[Scheideler(2002)]%
        {scheideler2002models}
\bibfield{author}{\bibinfo{person}{Christian Scheideler}.} \bibinfo{year}{2002}\natexlab{}.
\newblock \showarticletitle{Models and techniques for communication in dynamic networks}. In \bibinfo{booktitle}{\emph{Annual Symposium on Theoretical Aspects of Computer Science}}. Springer, \bibinfo{pages}{27--49}.
\newblock


\bibitem[Schloegel et~al\mbox{.}(2000)]%
        {schloegel2000graph}
\bibfield{author}{\bibinfo{person}{Kirk Schloegel}, \bibinfo{person}{George Karypis}, {and} \bibinfo{person}{Vipin Kumar}.} \bibinfo{year}{2000}\natexlab{}.
\newblock \showarticletitle{Graph partitioning for high performance scientific simulations}.
\newblock  (\bibinfo{year}{2000}).
\newblock


\bibitem[Simon(1991)]%
        {simon1991partitioning}
\bibfield{author}{\bibinfo{person}{Horst~D Simon}.} \bibinfo{year}{1991}\natexlab{}.
\newblock \showarticletitle{Partitioning of unstructured problems for parallel processing}.
\newblock \bibinfo{journal}{\emph{Computing systems in engineering}} \bibinfo{volume}{2}, \bibinfo{number}{2-3} (\bibinfo{year}{1991}), \bibinfo{pages}{135--148}.
\newblock


\bibitem[Sommer(2014)]%
        {sommer2014shortest}
\bibfield{author}{\bibinfo{person}{Christian Sommer}.} \bibinfo{year}{2014}\natexlab{}.
\newblock \showarticletitle{Shortest-path queries in static networks}.
\newblock \bibinfo{journal}{\emph{ACM Computing Surveys (CSUR)}} \bibinfo{volume}{46}, \bibinfo{number}{4} (\bibinfo{year}{2014}), \bibinfo{pages}{1--31}.
\newblock


\bibitem[Ta et~al\mbox{.}(2017)]%
        {ta2017efficient}
\bibfield{author}{\bibinfo{person}{Na Ta}, \bibinfo{person}{Guoliang Li}, \bibinfo{person}{Tianyu Zhao}, \bibinfo{person}{Jianhua Feng}, \bibinfo{person}{Hanchao Ma}, {and} \bibinfo{person}{Zhiguo Gong}.} \bibinfo{year}{2017}\natexlab{}.
\newblock \showarticletitle{An efficient ride-sharing framework for maximizing shared route}.
\newblock \bibinfo{journal}{\emph{IEEE Transactions on Knowledge and Data Engineering}} \bibinfo{volume}{30}, \bibinfo{number}{2} (\bibinfo{year}{2017}), \bibinfo{pages}{219--233}.
\newblock


\bibitem[Teng and Points(1991)]%
        {teng1991unified}
\bibfield{author}{\bibinfo{person}{SH Teng} {and} \bibinfo{person}{Spheres Points}.} \bibinfo{year}{1991}\natexlab{}.
\newblock \showarticletitle{Unified geometric approach to graph separators}. In \bibinfo{booktitle}{\emph{Proc. 31st Ann. Symp. Foundations of Computer Science}}. \bibinfo{pages}{538--547}.
\newblock


\bibitem[Tretyakov et~al\mbox{.}(2011)]%
        {tretyakov2011fast}
\bibfield{author}{\bibinfo{person}{Konstantin Tretyakov}, \bibinfo{person}{Abel Armas-Cervantes}, \bibinfo{person}{Luciano Garc{\'\i}a-Ba{\~n}uelos}, \bibinfo{person}{Jaak Vilo}, {and} \bibinfo{person}{Marlon Dumas}.} \bibinfo{year}{2011}\natexlab{}.
\newblock \showarticletitle{Fast fully dynamic landmark-based estimation of shortest path distances in very large graphs}. In \bibinfo{booktitle}{\emph{Proceedings of the 20th ACM international conference on Information and knowledge management}}. \bibinfo{pages}{1785--1794}.
\newblock


\bibitem[Wang et~al\mbox{.}(2016)]%
        {wang2016effective}
\bibfield{author}{\bibinfo{person}{Sibo Wang}, \bibinfo{person}{Xiaokui Xiao}, \bibinfo{person}{Yin Yang}, {and} \bibinfo{person}{Wenqing Lin}.} \bibinfo{year}{2016}\natexlab{}.
\newblock \showarticletitle{Effective indexing for approximate constrained shortest path queries on large road networks}.
\newblock \bibinfo{journal}{\emph{Proceedings of the VLDB Endowment}} \bibinfo{volume}{10}, \bibinfo{number}{2} (\bibinfo{year}{2016}), \bibinfo{pages}{61--72}.
\newblock


\bibitem[Wang et~al\mbox{.}(2019)]%
        {wang2019querying}
\bibfield{author}{\bibinfo{person}{Yong Wang}, \bibinfo{person}{Guoliang Li}, {and} \bibinfo{person}{Nan Tang}.} \bibinfo{year}{2019}\natexlab{}.
\newblock \showarticletitle{Querying shortest paths on time dependent road networks}.
\newblock \bibinfo{journal}{\emph{Proceedings of the VLDB Endowment}} \bibinfo{volume}{12}, \bibinfo{number}{11} (\bibinfo{year}{2019}), \bibinfo{pages}{1249--1261}.
\newblock


\bibitem[Wang et~al\mbox{.}(2021)]%
        {QbS_wang2021query}
\bibfield{author}{\bibinfo{person}{Ye Wang}, \bibinfo{person}{Qing Wang}, \bibinfo{person}{Henning Koehler}, {and} \bibinfo{person}{Yu Lin}.} \bibinfo{year}{2021}\natexlab{}.
\newblock \showarticletitle{Query-by-sketch: Scaling shortest path graph queries on very large networks}. In \bibinfo{booktitle}{\emph{Proceedings of the 2021 International Conference on Management of Data}}. \bibinfo{pages}{1946--1958}.
\newblock


\bibitem[Wei et~al\mbox{.}(2020)]%
        {wei2020architecture}
\bibfield{author}{\bibinfo{person}{Victor~Junqiu Wei}, \bibinfo{person}{Raymond Chi-Wing Wong}, {and} \bibinfo{person}{Cheng Long}.} \bibinfo{year}{2020}\natexlab{}.
\newblock \showarticletitle{Architecture-intact oracle for fastest path and time queries on dynamic spatial networks}. In \bibinfo{booktitle}{\emph{Proceedings of the 2020 ACM SIGMOD International Conference on Management of Data}}. \bibinfo{pages}{1841--1856}.
\newblock


\bibitem[Xu et~al\mbox{.}(2005)]%
        {MDETD_xu2005tree}
\bibfield{author}{\bibinfo{person}{Jinbo Xu}, \bibinfo{person}{Feng Jiao}, {and} \bibinfo{person}{Bonnie Berger}.} \bibinfo{year}{2005}\natexlab{}.
\newblock \showarticletitle{A tree-decomposition approach to protein structure prediction}. In \bibinfo{booktitle}{\emph{2005 IEEE Computational Systems Bioinformatics Conference (CSB'05)}}. IEEE, \bibinfo{pages}{247--256}.
\newblock


\bibitem[Yu et~al\mbox{.}(2024)]%
        {yu2024distributed}
\bibfield{author}{\bibinfo{person}{Ziqiang Yu}, \bibinfo{person}{Xiaohui Yu}, \bibinfo{person}{Nick Koudas}, \bibinfo{person}{Yueting Chen}, {and} \bibinfo{person}{Yang Liu}.} \bibinfo{year}{2024}\natexlab{}.
\newblock \showarticletitle{A Distributed Solution for Efficient K Shortest Paths Computation Over Dynamic Road Networks}.
\newblock \bibinfo{journal}{\emph{IEEE Transactions on Knowledge and Data Engineering}} (\bibinfo{year}{2024}).
\newblock


\bibitem[Yu et~al\mbox{.}(2020)]%
        {yu2020distributed}
\bibfield{author}{\bibinfo{person}{Ziqiang Yu}, \bibinfo{person}{Xiaohui Yu}, \bibinfo{person}{Nick Koudas}, \bibinfo{person}{Yang Liu}, \bibinfo{person}{Yifan Li}, \bibinfo{person}{Yueting Chen}, {and} \bibinfo{person}{Dingyu Yang}.} \bibinfo{year}{2020}\natexlab{}.
\newblock \showarticletitle{Distributed processing of k shortest path queries over dynamic road networks}. In \bibinfo{booktitle}{\emph{Proceedings of the 2020 ACM SIGMOD International Conference on Management of Data}}. \bibinfo{pages}{665--679}.
\newblock


\bibitem[Zhang et~al\mbox{.}(2017)]%
        {zhang2017graph}
\bibfield{author}{\bibinfo{person}{Chenzi Zhang}, \bibinfo{person}{Fan Wei}, \bibinfo{person}{Qin Liu}, \bibinfo{person}{Zhihao~Gavin Tang}, {and} \bibinfo{person}{Zhenguo Li}.} \bibinfo{year}{2017}\natexlab{}.
\newblock \showarticletitle{Graph edge partitioning via neighborhood heuristic}. In \bibinfo{booktitle}{\emph{Proceedings of the 23rd ACM SIGKDD International Conference on Knowledge Discovery and Data Mining}}. \bibinfo{pages}{605--614}.
\newblock


\bibitem[Zhang et~al\mbox{.}(2022)]%
        {zhang2022shortest}
\bibfield{author}{\bibinfo{person}{Junhua Zhang}, \bibinfo{person}{Wentao Li}, \bibinfo{person}{Long Yuan}, \bibinfo{person}{Lu Qin}, \bibinfo{person}{Ying Zhang}, {and} \bibinfo{person}{Lijun Chang}.} \bibinfo{year}{2022}\natexlab{}.
\newblock \showarticletitle{Shortest-path queries on complex networks: experiments, analyses, and improvement}.
\newblock \bibinfo{journal}{\emph{Proceedings of the VLDB Endowment}} \bibinfo{volume}{15}, \bibinfo{number}{11} (\bibinfo{year}{2022}), \bibinfo{pages}{2640--2652}.
\newblock


\bibitem[Zhang et~al\mbox{.}(2021c)]%
        {DH2H_zhang2021dynamic}
\bibfield{author}{\bibinfo{person}{Mengxuan Zhang}, \bibinfo{person}{Lei Li}, \bibinfo{person}{Wen Hua}, \bibinfo{person}{Rui Mao}, \bibinfo{person}{Pingfu Chao}, {and} \bibinfo{person}{Xiaofang Zhou}.} \bibinfo{year}{2021}\natexlab{c}.
\newblock \showarticletitle{Dynamic hub labeling for road networks}. In \bibinfo{booktitle}{\emph{2021 IEEE 37th International Conference on Data Engineering (ICDE)}}. IEEE, \bibinfo{pages}{336--347}.
\newblock


\bibitem[Zhang et~al\mbox{.}(2020)]%
        {zhang2020stream}
\bibfield{author}{\bibinfo{person}{Mengxuan Zhang}, \bibinfo{person}{Lei Li}, \bibinfo{person}{Wen Hua}, {and} \bibinfo{person}{Xiaofang Zhou}.} \bibinfo{year}{2020}\natexlab{}.
\newblock \showarticletitle{Stream processing of shortest path query in dynamic road networks}.
\newblock \bibinfo{journal}{\emph{IEEE Transactions on Knowledge and Data Engineering}} (\bibinfo{year}{2020}).
\newblock


\bibitem[Zhang et~al\mbox{.}(2021b)]%
        {DPSL_zhang2021efficient}
\bibfield{author}{\bibinfo{person}{Mengxuan Zhang}, \bibinfo{person}{Lei Li}, \bibinfo{person}{Wen Hua}, {and} \bibinfo{person}{Xiaofang Zhou}.} \bibinfo{year}{2021}\natexlab{b}.
\newblock \showarticletitle{Efficient 2-hop labeling maintenance in dynamic small-world networks}. In \bibinfo{booktitle}{\emph{2021 IEEE 37th International Conference on Data Engineering (ICDE)}}. IEEE, \bibinfo{pages}{133--144}.
\newblock


\bibitem[Zhang et~al\mbox{.}(2023)]%
        {zhang2023parallel}
\bibfield{author}{\bibinfo{person}{Mengxuan Zhang}, \bibinfo{person}{Lei Li}, \bibinfo{person}{Goce Trajcevski}, \bibinfo{person}{Anderas Zufle}, {and} \bibinfo{person}{Xiaofang Zhou}.} \bibinfo{year}{2023}\natexlab{}.
\newblock \showarticletitle{Parallel Hub Labeling Maintenance with High Efficiency in Dynamic Small-World Networks}.
\newblock \bibinfo{journal}{\emph{IEEE Transactions on Knowledge and Data Engineering}} (\bibinfo{year}{2023}).
\newblock


\bibitem[Zhang et~al\mbox{.}(2021a)]%
        {zhang2021experimental}
\bibfield{author}{\bibinfo{person}{Mengxuan Zhang}, \bibinfo{person}{Lei Li}, {and} \bibinfo{person}{Xiaofang Zhou}.} \bibinfo{year}{2021}\natexlab{a}.
\newblock \showarticletitle{An experimental evaluation and guideline for path finding in weighted dynamic network}.
\newblock \bibinfo{journal}{\emph{Proceedings of the VLDB Endowment}} \bibinfo{volume}{14}, \bibinfo{number}{11} (\bibinfo{year}{2021}), \bibinfo{pages}{2127--2140}.
\newblock


\bibitem[Zhang et~al\mbox{.}({[n.\,d.]})]%
        {FullVersion}
\bibfield{author}{\bibinfo{person}{Mengxuan Zhang}, \bibinfo{person}{Xinjie Zhou}, \bibinfo{person}{Lei Li}, \bibinfo{person}{Ziyi Liu}, \bibinfo{person}{Goce Trajcevski}, \bibinfo{person}{Yan Huang}, {and} \bibinfo{person}{Xiaofang Zhou}.} \bibinfo{year}{[n.\,d.]}\natexlab{}.
\newblock \bibinfo{title}{A Universal Scheme for Dynamic Partitioned Shortest Path Index: Survey, Improvement, and Experiments (Full Version)}.
\newblock \bibinfo{howpublished}{\url{https://hkustconnect-my.sharepoint.com/:f:/g/personal/xzhouby_connect_ust_hk/EhWRISAsqhFBiOLw3SAQ72cBHHQ60G5IurNtwzH6bAuq0Q?e=jNV71L}}.
\newblock


\bibitem[Zhang and Yu(2022)]%
        {zhang2022relative}
\bibfield{author}{\bibinfo{person}{Yikai Zhang} {and} \bibinfo{person}{Jeffrey~Xu Yu}.} \bibinfo{year}{2022}\natexlab{}.
\newblock \showarticletitle{Relative subboundedness of contraction hierarchy and hierarchical 2-hop index in dynamic road networks}. In \bibinfo{booktitle}{\emph{Proceedings of the 2022 International Conference on Management of Data}}. \bibinfo{pages}{1992--2005}.
\newblock


\bibitem[Zheng et~al\mbox{.}(2022)]%
        {zheng2022workload}
\bibfield{author}{\bibinfo{person}{Bolong Zheng}, \bibinfo{person}{Jingyi Wan}, \bibinfo{person}{Yongyong Gao}, \bibinfo{person}{Yong Ma}, \bibinfo{person}{Kai Huang}, \bibinfo{person}{Xiaofang~Zhou Zhou}, {and} \bibinfo{person}{Christian S.~Jensen}.} \bibinfo{year}{2022}\natexlab{}.
\newblock \showarticletitle{Workload-Aware Shortest Path Distance Querying in Road Networks}. In \bibinfo{booktitle}{\emph{2022 IEEE 38th International Conference on Data Engineering (ICDE)}}. IEEE.
\newblock


\bibitem[Zhong et~al\mbox{.}(2013)]%
        {Gtree_zhong2013g}
\bibfield{author}{\bibinfo{person}{Ruicheng Zhong}, \bibinfo{person}{Guoliang Li}, \bibinfo{person}{Kian-Lee Tan}, {and} \bibinfo{person}{Lizhu Zhou}.} \bibinfo{year}{2013}\natexlab{}.
\newblock \showarticletitle{G-tree: An efficient index for knn search on road networks}. In \bibinfo{booktitle}{\emph{Proceedings of the 22nd ACM international conference on Information \& Knowledge Management}}. \bibinfo{pages}{39--48}.
\newblock


\bibitem[Zhong et~al\mbox{.}(2015)]%
        {zhong2015g}
\bibfield{author}{\bibinfo{person}{Ruicheng Zhong}, \bibinfo{person}{Guoliang Li}, \bibinfo{person}{Kian-Lee Tan}, \bibinfo{person}{Lizhu Zhou}, {and} \bibinfo{person}{Zhiguo Gong}.} \bibinfo{year}{2015}\natexlab{}.
\newblock \showarticletitle{G-tree: An efficient and scalable index for spatial search on road networks}.
\newblock \bibinfo{journal}{\emph{IEEE Transactions on Knowledge and Data Engineering}} \bibinfo{volume}{27}, \bibinfo{number}{8} (\bibinfo{year}{2015}), \bibinfo{pages}{2175--2189}.
\newblock


\bibitem[Zhou et~al\mbox{.}(2024a)]%
        {MCSPs_zhou2024efficient}
\bibfield{author}{\bibinfo{person}{Xinjie Zhou}, \bibinfo{person}{Kai Huang}, \bibinfo{person}{Lei Li}, \bibinfo{person}{Mengxuan Zhang}, {and} \bibinfo{person}{Xiaofang Zhou}.} \bibinfo{year}{2024}\natexlab{a}.
\newblock \showarticletitle{I/O-Efficient Multi-Criteria Shortest Paths Query Processing on Large Graphs}.
\newblock \bibinfo{journal}{\emph{IEEE Transactions on Knowledge and Data Engineering}} (\bibinfo{year}{2024}).
\newblock


\bibitem[Zhou et~al\mbox{.}(2024b)]%
        {DCT_zhou2024Scalable}
\bibfield{author}{\bibinfo{person}{Xinjie Zhou}, \bibinfo{person}{Mengxuan Zhang}, \bibinfo{person}{Lei Li}, {and} \bibinfo{person}{Xiaofang Zhou}.} \bibinfo{year}{2024}\natexlab{b}.
\newblock \showarticletitle{Scalable Distance Labeling Maintenance and Construction for Dynamic Small-World Networks}. In \bibinfo{booktitle}{\emph{2024 IEEE 40th International Conference on Data Engineering (ICDE)}}. IEEE.
\newblock


\bibitem[Zhou et~al\mbox{.}(2025)]%
        {HTSP_zhou2025High}
\bibfield{author}{\bibinfo{person}{Xinjie Zhou}, \bibinfo{person}{Mengxuan Zhang}, \bibinfo{person}{Lei Li}, {and} \bibinfo{person}{Xiaofang Zhou}.} \bibinfo{year}{2025}\natexlab{}.
\newblock \showarticletitle{High Throughput Shortest Distance Query Processing on Large Dynamic Road Networks}. In \bibinfo{booktitle}{\emph{2025 IEEE 41th International Conference on Data Engineering (ICDE)}}. IEEE.
\newblock


\end{thebibliography}

\newpage
\section{Appendix}
\label{sec:appendix}

\subsection{Query Correctness Proof of Pre-boundary Strategy}

We formally prove the query correctness of the \textit{Pre-boundary} PSP strategy in the following two lemmas.

\begin{lemma}
The same-partition query can be processed correctly.
 	\label{lemma:NaiveSameParti}
\end{lemma}

\begin{proof}
We prove this by dividing all cases into two sub-cases:
	
\textit{Sub-Case 1:} $p_G(s,t)$ does not go outside of $G_i$, \ie $\forall v\in p_G(s,t), v\in G_i$, then $d_G(s,t)=d_{G_i}(s,t)=d_{L_i}(s,t)$;
	
\textit{Sub-Case 2:} $p_G(s,t)$ passes outs of $G_i$ ($\exists v\in p_G(s,t), v\in G_j, j\neq i$). Suppose the first and last vertex on $p_G(s,t)$ which are not in $G_i$ is $v_1$ and $v_2$ ($v_1\notin G_i, v_2\notin G_i$, and $v_1,v_2$ could be the same vertex). Then we denote the vertex right before $v_1$ and right after $v_2$ along $p_G(s,t)$ are $v_{b1}$ and $v_{b2}$. Both $v_{b1}$ and $v_{b2}$ are boundary vertices with $v_{b1}, v_{b2}\in B_i$. We can decompose the length of $p_G(s,t)$ as $d_G(s,t)=d_G(s,v_{b1})+d_G(v_{b1},v_{b2})+d_G(v_{b2},t)$, and  $p_{G_i}(s,t)$ on $G_i$ as $d_{G_i}(s,v_{b1})+d_{G_i}(v_{b1},v_{b2})+d_{G_i}(v_{b2},t)$. Since $d_G(s,v_{b1})=d_{G_i}(s,v_{b1})$ and $d_G(v_{b2},t)=d_{G_i}(v_{b2},t)$ by referring to the above \textit{sub-case 1}, and $d_G(v_{b1},v_{b2})=d_{G_i}(v_{b1},v_{b2})$ as indicated in the \textit{Step 1} of \textit{Index Construction}. So $d_G(s,t)=d_{G_i}(s,t)=d_{L_i}(s,t)$ holds.
\end{proof}

\begin{lemma}
The cross-partition query can be processed correctly.
\end{lemma}
\begin{proof}
We discuss those 4 sub-cases one by one:
	
\textit{Sub-Case 1:} Both $s$ and $t$ are boundary vertices. Suppose that the concise path of $p_G(s,t)$ by extracting the boundary vertices is $p=\left<s=b_0,b_1,\dots, b_l=t\right>$ with $\forall b_h\in p (0\le h\le l), b_h\in \tilde{G}$. Then $d_G(s,t)=\sum_{h=0}^{l-1}d_G(b_h,b_{h+1})$. If $b_h$ and $b_{h+1}$ are in the same partition, then $d_G(b_h,b_{h+1})=e_{\tilde{G}}(b_h,b_{h+1})$ with $(b_h,b_{h+1})\in \tilde{G}$ by referring to the \textit{Index Construction}. If $b_h$ and $b_{h+1}$ are in different partitions, then $(b_h,b_{h+1})\in E_{inter}$ and $d_G(b_h,b_{h+1})=e_{\tilde{G}}(b_h,b_{h+1})$. So it can be proved that $d_G(s,t)=d_{\tilde{G}}(s,t)=d_{\tilde{L}}(s,t)$;  
	
\textit{Sub-Case 2:} $s$ is a boundary vertex and $t$ is an inner vertex of $G_j$. Suppose the last boundary vertex on $p_G(s,t)$ is $b$. Then $b\in B_j$ since $p_G(s,t)$ would reach $t$ by entering $G_j$ through any vertex in $B_j$. The shortest distance can be calculated as $d_G(s,t)=d_G(s,b)+d_G(b,t)$. We can get $d_G(s,b)=d_{\tilde{G}}(s,b)$ by referring to the \textit{Sub-Case 1} and $d_G(b,t)=d_{\tilde{G}}(b,t)$ according to Lemma \ref{lemma:NaiveSameParti}, such that $d_G(s,t)=d_{\tilde{G}}(s,b)+d_{G_j}(b,t)=d_{\tilde{L}}(s,b)+d_{L_j}(b,t)$;
	
\textit{Sub-Case 3:} $s$ is a inner vertex and $t$ is a boundary vertex. This is the reverse version of \textit{Sub-Case 2}; 
	
\textit{Sub-Case 4:} Neither $s$ nor $t$ is boundary vertex. It can be proved by extending \textit{Sub-Case 2}. Therefore, both \textit{sub-case 3} and \textit{sub-case 4} can be proved similarly as \textit{sub-case 2}.
\end{proof}

	



\begin{algorithm}[b]
	\caption{N-CH-P Index Construction}
	\label{algo:PCHBuild}
    \footnotesize
	\LinesNumbered
    \KwIn{Graph $G=\{V,E\}$}
	\KwOut{N-CH-P index $L=\{\tilde{L},\{L_i\}\}$}
    \SetKwBlock{DoParallel}{parallel\_for}{end}
     $\{G_i|1\leq i\leq k\}\gets$ partition result of $G$ by PUNCH~\cite{PUNCH_delling2011graph}\label{pchb:1}\\
     // Step 1: Partition Index Construction\label{pchb:2}\\
    \DoParallel($i\in [1,k]$)
    {
        $L_i\gets$ \textsc{CHIndexing}($G_i$); \label{pchb:4}\\
    }
    // Step 2: Overlay Graph Construction\label{pchb:5}\\
    $\tilde{G}\gets E_{inter}$; \Comment{Get the initial inter-partition edges}\label{pchb:6}\\
    \DoParallel($i\in [1,k]$)
    {
        \For{$v\in B_i$ and $\forall u\in L_i(v)$}
        {
            $\tilde{G}\gets \tilde{G}\cup L_i(v,u)$; // $L_i(v,u)$ is the shortcut among $v$ and $u$\label{pchb:9}\\
        }
    }
    // Step 3: Overlay Index Construction\label{pchb:10}\\
    $\tilde{L}\gets$ \textsc{CHIndexing}($\tilde{G}$);\label{pchb:11}\\
    \Return $L=\{\tilde{L},\{L_i\}\}$;\\
    \SetKwProg{Fn}{Function}{:}{end}
    \Fn{\textsc{CHIndexing}($G$)}
    {
        $L(v)\gets\phi, \forall v\in V$;\label{pchb:14}\\
        \For{$v\in V$ in increasing vertex order}{
            $L(v)\gets v$'s adjacent edges in $G$;\label{pchb:16}\\
             insert/update the all-pair clique among $N_G(v)$ to $G$;\label{pchb:17}\\
            $V\gets V\setminus v$; remove $v$'s adjacent edges from $G$;\label{pchb:18} \\
        }
        \Return $L$;\label{pchb:19}\\
    }
\end{algorithm}

\subsection{Conversion of Vertex-Cut Partitions}
Though graph partition can be categorized into \textit{edge-cut} and \textit{vertex-cut}, all of our proposed techniques are illustrated in the form of \textit{edge-cut} for clear presentation since 
the vertex-cut partitions can be converted to edge-cut partitions.
Specifically, we generalize the \textit{vertex-cut} to \textit{edge-cut} by duplicating those vertices that are cut into different partitions and connecting the duplicated vertices with their original vertices through an edge of zero weight. Suppose in a graph $G$, a vertex $x$ is cut into $l+1$ different partitions $\{G_{0}, G_{1}, \dots, G_{l}\}$ in a \textit{vertex-cut partition} with $x$ and its neighbors $\{x_{n_i}\} (0\le i\le l)$ belonging to subgraph $G_{i}$. To obtain its equivalent \textit{edge-cut partition}, we transform $G$ to $G'$: $\forall$ $x\in X$ (\textit{cut vertex} set), keep the connection between $x$ and its partial neighbors $\{x_{n_0}\}$, duplicate $x$ as $x_i$ connecting neighbors $x_{n_i}$, and connect $x$ with $x_i$ by an edge with $e(x,x_i)=0$. Then we partition $G'$ by using \textit{edge-cut partition}: cut each added edge $(x,x_i)$, which leads to Lemma~\ref{lemma:partitionEquivalent}.

\begin{lemma}\label{lemma:partitionEquivalent}
	The edge-cut partition of $G'$ by cutting those added edges $(x,x_i)$ is equivalent to the vertex-cut partition of $G$ by cutting the vertices in $X$.
\end{lemma}

Following by, we verify that the partition equivalence has no effect on the shortest distance computation.

\begin{theorem}
	$\forall s,t\in V, d_G(s,t)=d_{G'}(s,t)$.
\end{theorem}

\begin{proof}
	We classify all the scenarios into two cases:
	
	Case 1: $p_G(s,t)$ ($p_{s,t}$ for short) does not pass through any $x\in X$. It indicates that $p_{s,t}$ is totally within one partition $G_i (1\le i\le k)$. Since the vertex-cut partition of $G$ and the edge-cut partition of $G'$ are equivalent, $d_{G_i}(s,t)=d_G(s,t)=d_{G'}(s,t)$ holds.
	
	Case 2: $p_{s,t}$ pass through $x\in X$. We suppose that $p_{s,t}=<s,\dots,x_f$,$x,x_b,\dots,t>$ with $x_f\in G_{i}, x_b\in G_{j}$, then it can correspond to $p_{s,t}'=<s,\dots,x_f,x_i,x,x_j,x_b,\dots,t>$ in $G'$. Since $e(x_i,x)=e(x,x_j)=0$, $l(p_{s,t})=l(p_{s,t}')$ holds which prove that $d_G(s,t)=d_{G'}(s,t)$ is correct as well.
\end{proof}

\begin{algorithm}[h]
	\caption{N-CH-P Query Processing}
	\label{algo:PCHQuery}
    \footnotesize
	\LinesNumbered
    \KwIn{Query $Q(s,t)$ with $s\in G_p, t\in G_q$, N-CH-P index $L=\{\tilde{L},\{L_i\}\}$}
	\KwOut{Shortest distance of $Q(s,t)$}
    $d\gets \infty$;\label{pchq:1}\\
    \If{$s\in \tilde{G}$ and $t\in \tilde{G}$}{
        $d\gets$ \textsc{QueryCH}($s,t,\tilde{L}$); \Comment{Query on overlay index $\tilde{L}$}\label{pchq:3}\\
    }\Else{
        \textbf{if} $s\notin\tilde{G}$ and $t\notin\tilde{G}$ \textbf{then} $d\gets$ \textsc{QueryCH}($s,t,\tilde{L}\cup L_p\cup L_q$);\label{pchq:5}\\
        \textbf{if} $s\notin\tilde{G}$ and $t\in\tilde{G}$ \textbf{then} $d\gets$ \textsc{QueryCH}($s,t,\tilde{L}\cup L_p$);\label{pchq:6}\\
        \textbf{if} $s\in\tilde{G}$ and $t\notin\tilde{G}$ \textbf{then} $d\gets$ \textsc{QueryCH}($s,t,\tilde{L}\cup L_q$);\label{pchq:7}\\
    }
    \Return $d$;\\
    \SetKwProg{Fn}{Function}{:}{end}
    \Fn{\textsc{QueryCH}($s,t,L$)}
    {
        initialize min-priority queues $PQ_f$ and $PQ_b$ with $\left<s,0\right>$ and $\left<t,0\right>$;\label{pchq:10}\\
        $d\gets\infty$; $F_f, F_b\gets false$; $C_f[v], C_b[v]\gets false, \forall v\in V$;\label{pchq:11}\\
        $D_f[v], D_b[v]\gets \infty, \forall v\in V$; $D_f[s]\gets 0$, $D_b[t]\gets 0$;\label{pchq:12}\\
        \While{$PQ_f$ is not empty and $PQ_b$ is not empty}
        {
            \textbf{if} $F_f=true$ and $F_b=true$ \textbf{then} break;\label{pchq:14}\\
            \textbf{if} $F_f=true$ and $PQ_b$ is empty \textbf{then} break;\label{pchq:15}\\
            \textbf{if} $F_b=true$ and $PQ_f$ is empty \textbf{then} break;\label{pchq:16}\\
            // Forward Search\label{pchq:17}\\
            \If{$PQ_f$ is not empty and $F_f=false$}
            {
                $\left<v,dis\right>\gets PQ_f$.Top(); $PQ_f$.Pop(); $C_f[v]\gets true$;\label{pchq:19}\\
                \textbf{if} $dis>d$ \textbf{then} $F_f\gets true$;\label{pchq:20}\\
                \If{$C_b[v]=true$ and $dis+D_b[v]<d$}
                {
                    $d\gets dis+D_b[v]$;\label{pchq:22}\\
                }
                \For{$u\in L(v)$}
                {
                    \If{$C_f[u]=false$ and $dis+L(v,u)<D_f[u]$}
                    {
                        $D_f[u]\gets dis+L(v,u)$; $PQ_f$.Update($u,D_f[u]$);\label{pchq:25}\\
                    }
                }
            }
            // Backward Search. Similar to the forward search, omit here.\label{pchq:26}\\
        }
        \Return $d$;\label{pchq:27}\\
    }
\end{algorithm} 

\begin{algorithm}[t]
	\caption{N-CH-P Index Maintenance}
	\label{algo:PCHUpdate}
    \footnotesize
	\LinesNumbered
    \KwIn{Batch update set $U$}
	\KwOut{Updated N-CH-P index $L$}
    \SetKwBlock{DoParallel}{parallel\_for}{end}
    $\tilde{U},\{U_i,...,U_k\}\gets$ \textsc{UpdateClassify}($U$);\Comment{Classify updates by distribution}\label{pchu:1}\\
     // Step 1: Partition Index Maintenance\label{pchu:2}\\
     $u_i\gets\phi, \forall i\in[1,k]$; // $u_i$ stores the the updated shortcuts for $G_i$\\
    \DoParallel($i\in [1,k]$ and $U_i\neq\phi$)
    {
        $u_i\gets$ \textsc{CHIndexUpdate}($U_i,L_i$); // $u_i$ is the updated shortcut set\label{pchu:4}\\
    }
    // Step 2: Get Overlay Graph Updates \label{pchu:5}\\
    \DoParallel($i\in [1,k]$ and $u_i\neq\phi$)
    {
        \For{$e(u,v)\in u_i$}{
        \If{$e(u,v)\in \tilde{G}$ and $|e(u,v)|\neq |e_{\tilde{G}}(u,v)|$}
        {
            $\tilde{U}\gets \tilde{U}\cup e(u,v)$;\label{pchu:9}\\
        }
        
        }
    }
    // Step 3: Overlay Index Maintenance\label{pchu:10}\\
    \textsc{CHIndexUpdate}($\tilde{U},\tilde{L}$); \Comment{Overlay index update} \label{pchu:11}\\
    \Return $L$;\\
\end{algorithm}

\subsection{Details of Update-oriented N-CH-P Index}
We take the N-CH-P index as an example to introduce our proposed PSP indexes. The N-CH-P adopts the \textit{no-boundary} strategy for index construction, query processing, and index update. Besides, \textit{CH} is chosen as the SP algorithm for both the overlay and partition indexes since it is fast to update and has much better query efficiency than \textit{direct search}. In what follows, we introduce the index construction, query processing, and index maintenance of N-CH-P, respectively.

\textbf{Index Construction.} Algorithm~\ref{algo:PCHBuild} demonstrates the N-CH-P index construction. In particular, we first leverage the planar partition method such as \textit{PUNCH}~\cite{PUNCH_delling2011graph} to obtain the partition result of $G$ (Line~\ref{pchb:1}). After that, we parallelly construct the partition indexes $\{L_i\}$ by the \textsc{CHIndexing} function~\cite{DCH_ouyang2020efficient,zhang2021experimental} (Lines 2-\ref{pchb:4}). The key idea of \textsc{CHIndexing} function is to iteratively contract the least important vertex $v$ to obtain the distance-preserving shortcuts (Lines 15-\ref{pchb:18}). Step 2 builds the overlay graph according to the initial inter-partition edges and the contraction results of partition indexes. It is worth noting that the \textit{no-boundary} strategy entails constructing an all-pair clique among the boundary vertices of $B_i$ by querying on $\{L_i\}$ to build the overlay graph. However, this approach could be time-consuming if the number of boundaries is large as the CH query efficiency is low. The final step is to build the overlay index on $\tilde{G}$, which is also conducted by running the \textsc{CHIndexing} function (Lines 10-\ref{pchb:11}).

\textbf{Query processing.} It is worth noting that the no-boundary strategy has the same process for cross-partition and same-partition query processing (both rely on the distance concatenation on the overlay index and partition index). Nevertheless, we can prune the search space of the N-CH-P index by only searching on the overlay index or partial partition index. Algorithm~\ref{algo:PCHQuery} presents the pseudo-code of N-CH-P query processing. In particular, when both $s$ and $t$ are in the overlay graph, we only need to conduct the CH search on the overlay index $\tilde{L}$ according to Theorem~\ref{the:BoundaryLabel}. The CH search is implemented in \textsc{QueryCH} function (Lines 9-\ref{pchq:27}). 
Specifically, for the forward search, we assign a min-priority queue $PQ_f$, a search termination flag $F_f$, a flag vector $C_f$ indicating whether a vertex has been popped from $PQ_f$, a distance vector $D_f$ storing the distance to the source vertex $s$, and initialize $PQ_f$ with $\left<s,0\right>$ and $D_f[s]$ with 0 (the same goes for backward search but with subscript $b$, Lines \ref{pchq:10}-\ref{pchq:12}). When both $PQ_f$ and $PQ_b$ are non-empty, and the termination condition lines \ref{pchq:14}-\ref{pchq:16} are not met, we conduct forward and backward searches iteratively on the CH index $L$ (Lines 13-\ref{pchb:27}).
If both of them do not belong to $\tilde{G}$, we conduct the CH search on the combination of $\tilde{L}$, $L_p$, and $L_q$, where $L_p$ and $L_q$ are the partition indexes involving $s$ and $t$. Otherwise, in the case that only $s$ (or $t$) belongs to $\tilde{G}$, we only conduct the CH search on $\tilde{L}$ and $L_p$ (or $L_q$). In short, the N-CH-P is essentially equivalent to the CH index~\cite{ouyang2020efficient,zhang2021experimental} in terms of query processing.

\textbf{Index maintenance.} 
Algorithm~\ref{algo:PCHUpdate} presents the pseudo-code of N-CH-P index update. Given a batch of update $U$, we first classify it into inter-partition updates $\tilde{U}$, and in-partition updates $U_1,...U_k$ corresponding to the updates in $G_1,...,G_k$ (Line \ref{pchu:1}). 
Similar to the index construction procedure, the N-CH-P index maintenance also has three main steps. The first step maintains the partition indexes for all affected partitions in parallel (Lines \ref{pchu:2}-\ref{pchu:5}). This is implemented by running \textsc{CHIndexUpdate} function~\cite{DCH_ouyang2020efficient}. For each affected partition, such as $G_i$, it outputs the updated shortcuts set $u_i$, which is further fed to step 2 for obtaining the final overlay graph updates $\tilde{U}$. Given $\tilde{U}$ as input, step 3 also maintains $\tilde{L}$ by \textsc{CHIndexUpdate} function.

\subsection{Additional Experimental Results}

\begin{figure}[t]
	\centering
    \includegraphics[width=0.75\linewidth]{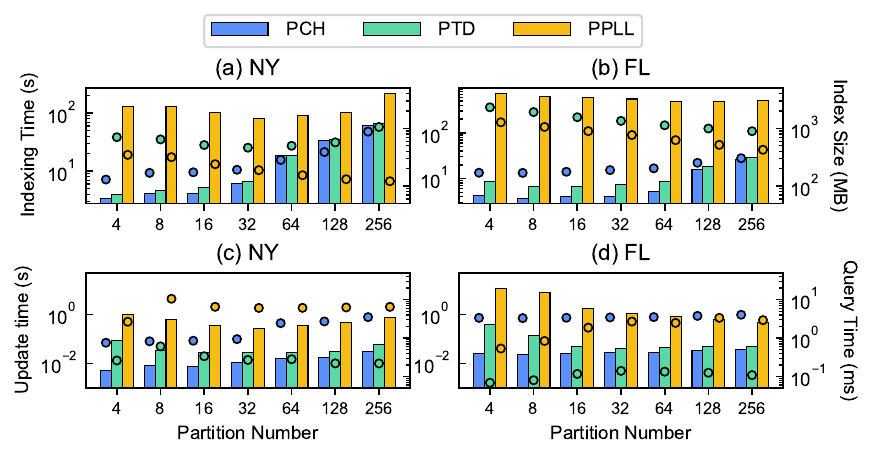}
	\caption{Effect of Partition Number. Bar: Left, Ball: Right}
	\label{fig:partitionNumber}
\end{figure}

\begin{figure}[t!]
	\centering
    \includegraphics[width=0.75\linewidth]{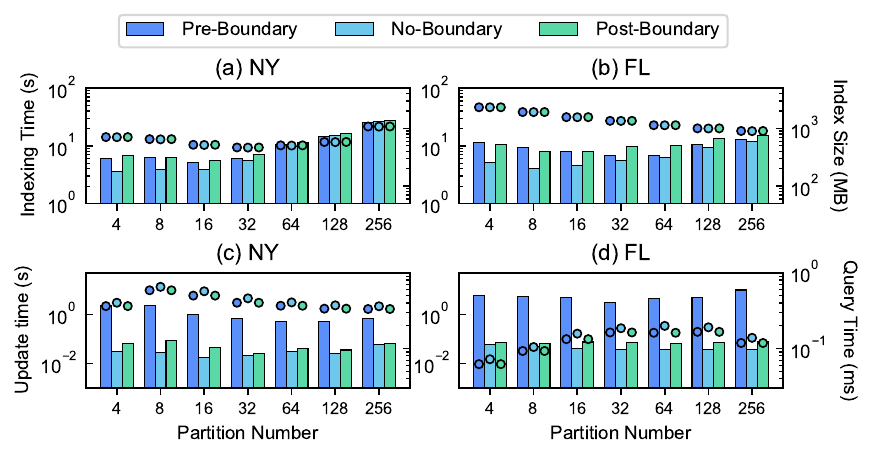}
	\caption{Effect of Partition Number on Different PSP Strategies. Bar: Left axis, Ball: Right axis}
	\label{fig:partiNumber_PSP}
\end{figure}

\textbf{Exp 10: Effect of Partition Number.}
We next evaluate the effect of partition number and present the results on NY and FL in Figure~\ref{fig:partitionNumber}.
In particular, the indexing time of PCH and PTD increased as $k$ increased from 4 to 256. This is because a larger $k$ could lead to a larger overlay graph that cannot leverage thread parallelization, thus resulting in higher indexing time. By contrast, PPLL's indexing time first decreases and then increases in NY while tending to decrease in FL because a larger $k$ could also lead to a smaller $\overline{|B_i|}$, thus reducing the overlay graph construction time.
In Figure~\ref{fig:partitionNumber} (c)-(d), we observe that a smaller $k$ generally improves the query time but leads to worse index update efficiency. Therefore, it is crucial to balance the query and update efficiency by selecting the appropriate $k$. We set the default partition number as 32 for all datasets since it provided satisfactory performance in most cases.

\textbf{Exp 11: Effect of Partition Number on Different PSP Strategies.}
We use PTD as the PSP index and evaluate the effect of the partition number $k$ on different PSP strategies by varying $k$ from 4 to 256. Figure~\ref{fig:partiNumber_PSP} presents the result on NY and FL.
Different PSP strategies share a similar tendency with the increase of partition number, which is in accordance with the observation discussed in Exp 10. In particular, a partition number that is too large or too small could lead to longer index maintenance and construction times. This is because a larger $k$ lead to better parallelization among partition indexes but result in a large overlay graph and longer overlay index construction/maintenance that cannot be paralleled. There is a trade-off between the parallelization and overlay graph processing. Since $k=32$ generally produces satisfactory performance, we set the default partition number as 32 for all datasets.

\begin{figure}[t]
	\centering
	\includegraphics[width=0.75\linewidth]{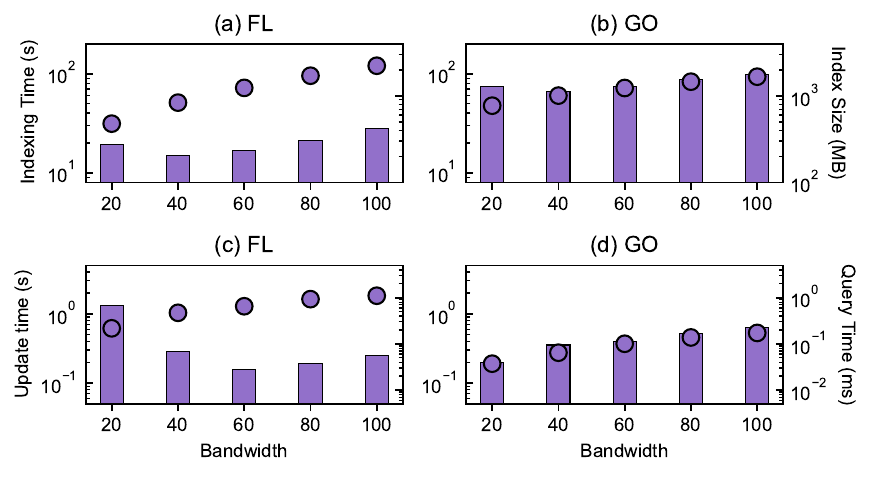}
	\caption{Effect of the Bandwidth.}
	\label{fig:bandwidth}
\end{figure}

\textbf{Exp 12: Effect of the Bandwidth.} 
We leverage CT to evaluate the effect of bandwidth. 
As shown in Figure~\ref{fig:bandwidth}, the indexing time first decreases and then increases with the increase of bandwidth on FL and GO. The index update time also has such a tendency on FL. In other cases, a smaller treewidth leads to a smaller index size and query time. This is because large treewidth can result in a dense core for CT, thus dramatically enlarging the pruning point records that are necessitated by the PLL update~\cite{DCT_zhou2024Scalable}. We set the default treewidth as 40 since it has satisfactory performance.

\begin{figure}[t]
	\centering
    \includegraphics[width=0.75\linewidth]{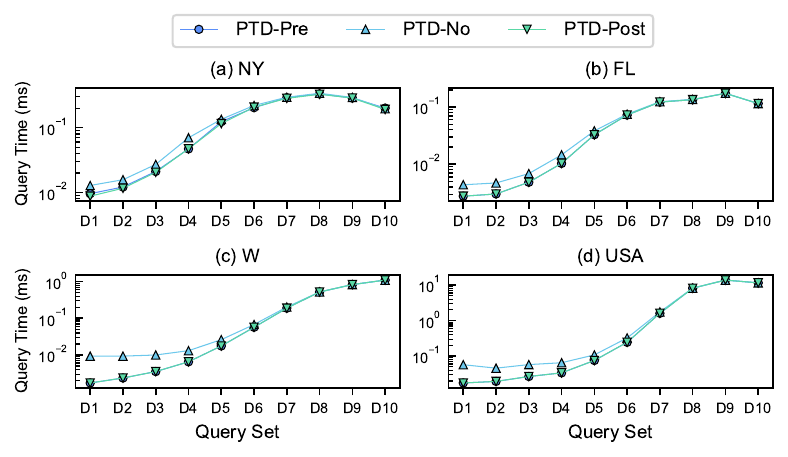}
	\caption{Performance of Different Query Distances}
	\label{fig:query_distance}
\end{figure}

\begin{figure}[t]
	\centering
	\includegraphics[width=0.75\linewidth]{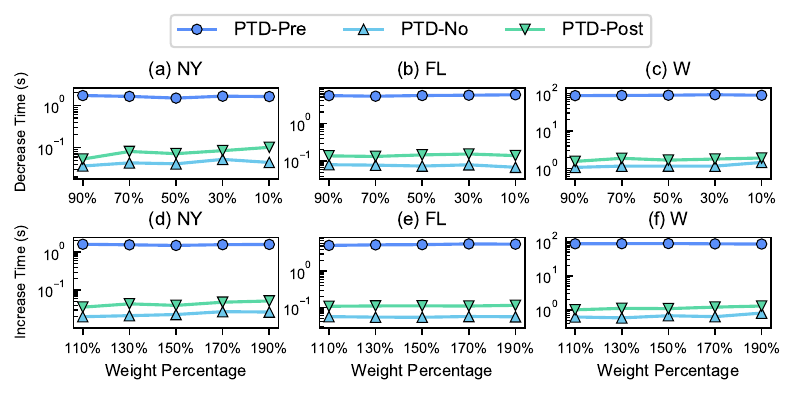}
	\caption{Varying the Percentage of Edge Weight Change}
	\label{fig:updateVaryRatio}
\end{figure}

\begin{figure*}[t]
	\centering
	\includegraphics[width=1\linewidth]{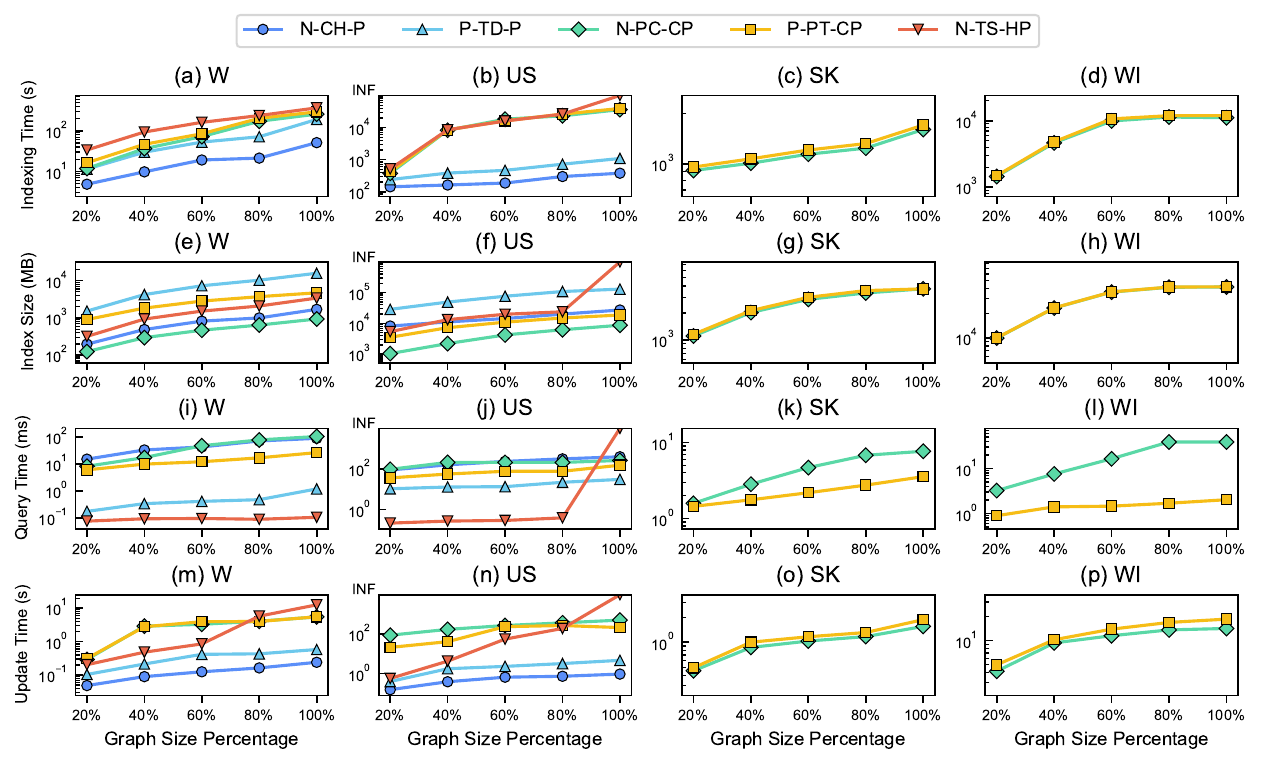}
	\caption{The Test of Scalability}
	\label{fig:scalabilityTest}
\end{figure*}

\textbf{Exp 13: Effect of Query Distance.}
We also evaluate the effect of query distance on PTD for road networks NY, FL, W, and US. In particular, we generate 10 groups of queries $D1$, $D2$,..., $D10$.
Each query set $Di, 1\leq i\leq 10$ contains 10,000 random queries, in which the distance of the source and target vertices for each query fall in the range $(l_{min} \cdot x^{i-1},l_{min}\cdot x^i]$, where $x = (l_{max} /l_{min})^{1/10}$, $l_{min}$ is set as 1 kilometer and $l_{max}$ is the maximum distance of any pair of vertices in the map.
As shown in Figure~\ref{fig:query_distance}, with the increase in distance, the query time tends to increase first but decreases when it comes to query set $D10$. It is observed that partitions on the outskirts of road networks typically have smaller boundary vertex numbers compared to central partitions due to fewer adjacent partitions. Consequently, even though $D10$ involves more cross-partition queries, the number of boundary vertices involved in query processing may be less than those less-distance queries, as the partitions that $D10$'s queries tend to be located on the outskirts of the network.
No matter what queries are involved, the post-boundary PTD (PTD-Post) and pre-boundary PTD (PTD-Pre) always outperform no-boundary PTD (PTD-No) since the former ones have better same-partition query efficiency and equivalent efficiency for other types of queries as evidenced in Exp 5. Moreover, the improvement of the post-boundary strategy is more obvious when it comes to queries with closer distances.

\textbf{Exp 14: Varying the Percentage of Edge Weight Change.}
We also vary the percentage of edge weight change for the edge weight decrease and increase updates on PTD. In particular, for each update edge $e$, we change its weight to $\alpha\cdot |e|$, where $\alpha$ is set as $0.9, 0.7, 0.5, 0.3, 0.1$ for decrease update while $1.1, 1.3, 1.5, 1.7, 1.9$ for the increase update, respectively. The result is presented in Figure~\ref{fig:updateVaryRatio}. The pre-boundary PTD (PTD-Pre) always consumes the most time in all cases while no-boundary PTD (PTD-No) and post-boundary PTD (PTD-Post) are about $47\times$ and $32\times$ faster than PTD-Pre. This is because no-boundary and post-boundary strategies replace the time-consuming Dijkstra's searches among the boundary vertices with efficient H2H query processing for overlay graph construction/maintenance. Besides, the index maintenance efficiency remains stable when varying weight change percentages, reflecting the robustness of our PSP strategies on PTD.

\textbf{Exp 15: The Test of Scalability.}
Lastly, we test the scalability of our proposed PSP indexes. We randomly divided the
vertices of a graph into 5 equally sized vertex groups and created 5 graphs for the cases of $20\%$, $40\%$, $60\%$, $80\%$, and $100\%$: the $i$-th graph is the induced subgraph on the first $i$ node groups. Figure~\ref{fig:scalabilityTest} presents the performance of our proposed PSP indexes in terms of index construction time, index size, query time, and index update time. Note that we omit the results of N-CH-P, P-TD-P, and N-TS-HP on datasets SK and WI since they are infeasible on complex networks.
Figure~\ref{fig:scalabilityTest} (a)-(h) shows that the indexing time and index size grow smoothly with the graph size for all methods. N-CH-P generally has the fastest index construction on road networks since its overlay and partition indexes are CH which is the fastest SP index for index construction. 
It is worth noting that we do not present the result of N-TS-HP since it cannot finish the test within 24 hours.
Regarding index size, N-PC-CP outperforms other methods since its core index is canonical 2-hop labeling (PLL) and the tree index is light index CH.
Note that N-PC-CP and P-PT-CP have almost the same indexing time and index size because the large core index (PLL) dominates the partition index concerning index construction time and index size.
Figure~\ref{fig:scalabilityTest} (i)-(p) shows that the query time and index update time also grow smoothly with the graph size for all methods. 
In particular, for road networks, N-TS-HP has the fastest query efficiency since it uses all-pair tables for the leaf node (the bottom-level partition) and its overlay index is efficient TD. Nevertheless, it has poor scalability for very large road networks ($100\%$ of US). For complex networks, P-PT-CP has better query efficiency and the advantages over N-PC-CP tend to increase with the growth of graph size.
For index update time, N-CH-P outperforms other methods on road networks since its overlay and partition indexes (both are CH) have the most efficient index maintenance efficiency. For complex networks, N-PC-CP has better index update efficiency than P-PT-CP since its tree index is faster CH.

\end{document}